# Review on magnetic and related properties of *RTX* compounds

Sachin Gupta and K. G. Suresh*

Indian Institute of Technology Bombay, Mumbai-400076, India

# Abstract

*RTX* (*R*=rare earths, *T*= 3d/4d/5d, transition metals such as Sc, Ti, Mn, Fe, Co, Ni, Cu, Ru, Rh, Pd, Ag, Os, Ir, Pt, Au, and *X*=p-block elements such as Al, Ga, In, Si, Ge, Sn, As, Sb, Bi) series is a huge family of intermetallics compounds. These compounds crystallize in different crystal structures depending on the constituents. Though these compounds have been known for a long time, they came to limelight recently in view of the large magnetocaloric effect (MCE) and magnetoresistance (MR) shown by many of them. Most of these compounds crystallize in hexagonal and tetragonal crystal structures. Some of them show crystal structure modification with annealing temperature; while a few of them show iso-structural transition in the paramagnetic regime. Their magnetic ordering temperatures vary from very low temperatures to temperatures well above room temperature (~510 K). Depending on the crystal structure, they show a variety of magnetic and electrical properties. These compounds have been characterized by means of a variety of techniques/measurements such as x-ray diffraction, neutron diffraction, magnetic properties, heat capacity, magnetocaloric properties, electrical resistivity, magnetoresistance, thermoelectric power, thermal expansion, Hall effect, optical properties, XPS, Mössbauer spectroscopy, ESR, μSR, NMR, NQR etc. Some amount of work on theoretical calculations on electronic structure, crystal field interaction and exchange interactions has also been reported. The interesting aspect of this series is that they show a variety of physical properties such as Kondo effect, heavy fermion behavior, spin glass state, intermediate valence, superconductivity, multiple magnetic transitions, metamagnetism, large MCE, large positive as well as negative MR, spin orbital compensation, magnetic polaronic behavior, pseudo gap effect etc. Except Mn, no other transition metal in these compounds possesses considerable magnetic moments. Because of this *R*MnX compounds in general have high ordering temperatures. Interstitial modification using hydrogen and nitrogen is found to alter their crystal structures and magnetic properties considerably. *RTX* compounds also show interesting pressure effects on their structural and magnetic properties. In summary, these compounds show variety of physical properties over a wide range of temperatures. This review is intended to cover all the important results obtained in this family, particularly in the last few years.

*Corresponding author email: suresh@phy.iitb.ac.in (K. G. Suresh)

Email: gsachin55@gmail.com (Sachin Gupta)

# Contents

**List of Symbols and abbreviations**

| | |
|---|---|
| R | Rare earth |
| T | Transition metal |
| X | *p*-block element |
| CEF | crystal electric field |
| MR | Magnetoresistance |
| MCE | Magnetocaloric effect |
| RT | Room temperature |
| ZFC | Zero field cooled |
| FC | Field cooled |
| μSR | Muon spin relaxation |
| VVP | Van Vleck paramagnetism |
| KI | Kondo insulator |
| SC | Super conductor |
| TMI | Thermo-magnetic irreversibility |
| HT | High temperature |
| HP | High pressure |
| FIM | Ferrimagnetic |
| TSW | Transverse sine wave structure |
| RC | Refrigerant capacity |
| SF | Spin fluctuation |
| CMR | Colossal magnetoresistance |
| LTP | Low temperature phase |
| HTP | High temperature phase |
| XRD | X-ray diffraction |
| DOS | Density of state |
| AFM | Antiferromagnetic |
| FM | ferromagnetic |
| SG | Spin glass |
| RSG | Reentrant spin glass |
| IV | Intermediate valence |
| NCW | Non-Curie Weiss |
| HTM | High temperature phase |
| LTM | Low temperature phase |
| PPM | Pauli paramagnetic |
| TEP | Thermoelectric power |
| XPS | X-ray photoemission spectroscopy |
| VF | Valence fluctuator |
| $T_C$ | Curie temperature |
| $T_N$ | Néel temperature |
| $T_K$ | Kondo temperature |
| $T_{SR}$ | Spin reorientation temperature |
| $T_t$ | Magnetic transition temperature |
| $n$ | exponent |
| $\theta_p$ | Paramagnetic Curie temperature |
| $H_C$ | Critical field |
| $E_F$ | Fermi energy |
| $T_K$ | Kondo temperature |
| $T_{SG}$ | Spin glass temperature |
| $T_{Comp}$ | Compensation temperature |
| $\mu_{eff}$ | Effective magnetic moment |

# 1. Introduction

In the field of magnetic materials research, rare earth ($R$) – transition metal ($T$) intermetallic compounds have always attracted a special interest. The research activities in the field of magnetism and magnetic materials are growing day-by-day, which have influenced the industry and daily life of common man significantly. The history of magnetism reveals that it is closely related to practical applications. Magnetic materials form the most vital components in many applications such as memory devices, permanent magnets, transformer cores, magneto-mechanical devices, magneto-electronic devices, magneto-optical devices etc. Recent additions to this list include magnetic refrigeration and spintronics. Magnetic refrigeration, which is the process of cooling a material, is an eco-friendly technique and is being seriously considered as an alternative to conventional gas compression/expansion technology. Similarly, spintronics has revolutionized the magnetic recording industry in a great manner. In many of the above mentioned applications, $R$-$T$ intermetallics play a crucial role by virtue of the attractive magnetic properties of the constituent $R$ and the $T$ components.

It is well known that in rare earths, the magnetism originates from the partially filled $4f$ shell electrons. Since these electrons are well localized, their magnetic moments are large. They are also characterized by strong single-ion magnetocrystalline anisotropy and low ordering temperatures. The first half of the rare earth series is referred to as light rare earth, while the right half is called the heavy rare earths. On the other hand, the magnetism exhibited by transition metals is itinerant, which gives rise to lower magnetic moments and higher ordering temperatures, compared to those of rare earths. A combination of rare earths and transition metals often gives rise to very interesting magnetic and related properties. This is true even when the transition metal is weakly magnetic or nonmagnetic. There are a large number of families comprising of rare earth based intermetallic compounds. Over the years, many of these compounds have dominated the applied magnetic materials family. Among the $R$-$T$ intermetallics compounds, $RTX$ ($X$ is a $p$-block element) family consists of many compounds with interesting fundamental properties and application potential. Of particular interest is the variety of structural, magnetic and transport properties exhibited by them. The number of $RTX$ compounds is very large and could not be completely covered in this review.

Here we present the overview of the studies reported in some well-known compounds of this family. This encompasses all the rare earths, *T*= Sc, Ti, Mn, Fe, Co, Ni, Cu, Ru, Rh, Pd, Ag, Os, Ir, Pt and Au and *X*= Al, Ga, In, Si, Ge, Sn, As, Sb and Bi. We have tried to cover the results reported on single crystalline, polycrystalline and amorphous forms, wherever they are available.

The compounds in *RTX* series with different *R*, *T*, and *X* elements crystallize in different crystal structures. Owing to different crystal structures, these compounds show versatile magnetic and electrical properties such as Kondo effect, complex magnetic structure, valence fluctuation, unconventional superconductivity, heavy fermion behavior, magnetic polaronic behavior, non-Fermi liquid behavior, metamagnetism, spin glass, memory effect, crystal electric field (CEF), exchange-bias, magnetoresistance (MR), magnetocaloric effect (MCE) etc. Due the localized moments in the 4*f* shell in rare earth atoms, the Rudermen-Kittel-Kasuya-Yosida (RKKY) interaction is dominant in these compounds, especially when *T* is a nonmagnetic element. Except Mn, all other transition elements possess nearly zero moment in these compounds, It has been observed that the magnetic ordering temperatures (Néel temperature, $T_N$ or Curie temperature, $T_C$) in these compounds vary from very low to high (compared to the room temperature, RT) and enable the series to cover a very large range of temperatures for certain applications. Some compounds of this series are found to show multiple magnetic transitions. One of the most important differences seen in *RTX* compounds compared to many other intermetallic groups is that in the *RTX* case, the exchange energy and the anisotropy energy are comparable. Experimental probes that have been used to study these compounds include dc magnetization (both zero field cooled, ZFC and the field cooled, FC), ac magnetic susceptibility, thermoelectric power (TEP), Hall effect, electrical resistivity, heat capacity, neutron diffraction, Mössbauer spectroscopy, muon spin relaxation (μSR) etc. Several reports are also available on these compounds subjected to chemical pressure (by hydrogenation) and hydrostatic pressure.

## 2. Crystal structure

The compounds of *RTX* series show a variety of crystal structures. In most of the cases, the compounds with same *T* and *X* atoms, but with different *R* ions show the same crystal structure. In some compounds, there is crystal modification which depends upon the annealing condition. It is also seen that in some cases crystal structure is same; there is a change of space group. The crystal structure with their space groups for different members of *RTX* family are shown in the Table I.

The compounds of *R*ScSi and *R*ScGe show two types of crystal structures depending upon its annealing temperature [1, 2, 3, 4, 5]. One structure is CeScSi, which is derived from La$_2$Sb type structure and the other is Ti$_5$Ga$_4$ type hexagonal structure. EuScGe is the first compound to be synthesized in Eu*T*Ge series and it crystallize in tetragonal structure [6]. Pöttgen et al. [7] published a review on Eu*TX* compounds which consists of results of structure, physical properties and [151]Eu Mössbauer spectroscopy. Compounds in *R*TiSi (*R*=Y, Gd-Tm, Lu) [8, 9] and *R*TiGe (*R*=Y, La-Nd, Sm, Gd-Tm, Lu) [10, 11] crystallize in CeFeSi type tetragonal structure. No compound with light rare earth (La-Nd) has been found in *R*TiSi series. The CeFeSi structure may be considered as made up of sheets with each sheet consists of five layers with sequence *R-X-T$_2$–X-R*, perpendicular to the *c* direction [12, 13]. Some compounds such as CeTiGe, GdTiGe and TbTiGe shows crystal structure modifications with temperature [2, 14]. The CeScSi type structure in GdTiGe was seen by different authors [15,16, 17]. Tencé et al. [18] reported that high-temperature modification (HTM) TbTiGe crystallizes in the tetragonal CeScSi-type, while low-temperature modification (LTM) TbTiGe forms the tetragonal CeFeSi-type structure. The crystal structures of *R*TiSi and *R*TiGe play a key role in its magnetic properties, which is discussed in magnetic properties section. The XRD data shows that GdTiSb crystallizes in CeFeSi type tetragonal structure [19].

From Table I, it can be seen that in *R*MnSi compounds, the crystal structure of compounds containing light rare earths (*R*=La-Sm, Gd) is different from the one containing heavy rare earths (Tb-Er). The former compounds form the CeFeSi type tetragonal structure and the latter form in the Co$_2$Si type orthorhombic structure [20]. *R*MnGe (*R*=Gd-Tm) crystallize in TiNiSi type orthorhombic crystal structure. Klosek et al. [21] reported that TmMnGe of this series crystallizes in both TiNiSi and ZrNiAl type structures depending upon heat treatment. They concluded that the two structural forms (ZrNiAl at high temperatures and TiNiSi at low temperatures) of TmMnGe arise due to the fact that the atomic radius of Tm is close to the critical R size for the TiNiSi → ZrNiAl structural transition. $R$=Yb compounds in $RTX$ family show interesting structural and magnetic properties. An overview of structural and related properties of euiatomic Yb$TX$ compounds has been reported by Pöttgen et al. [22].

The high temperature modification of TbMnSi, DyMnSi and NdMnGe show TiNiSi type orthorhombic structure [23]. It has been observed that samples of TbMnSi and NdMnGe after melting followed by annealing at 1273 K show CeFeSi type structure, however melting and quenching show TiNiSi type structure [23]. $R$MnAl ($R$=Ce, Nd, Gd) crystallize in cubic structure [24, 25, 26]. $R$MnGa compounds are found to crystallize in two forms [27, 28]. In this series, light rare earth compounds and heavy rare earth compounds at higher temperatures show cubic Laves phase structure, while the one with heavy rare earth elements, which are annealed at lower temperatures, were found to crystallize in Fe$_2$P type hexagonal structure [28]. Compounds of $R$FeAl shows two types of structures. Light rare earth compounds except R=La, which is single phase, show two phase C15 type structure, the second phase in these compound is unidentified, while heavy rare earth compounds show MgZn$_2$ type hexagonal structure [29].

The light rare earth $R$CoAl compounds are multiphase, while the heavy rare earth counterparts crystallize in the MgZn$_2$ (C14) hexagonal structure [30]. Two types of crystal structures are reported for TbCoSi [31, 32]. Compounds of $R$NiAl series show hexagonal [33] structure together with iso-structural transition in some of them [34, 35, 36, 37]. Merlo et al. [34] reported that GdNiAl shows an iso-structural transition around 205 K. The compound shows same hexagonal cell i.e., ZrNiAl but the lattice constants $a$ and $c$ show a jump at the transition. The $a$ lattice parameter decreases with decrease in temperature while $c$ lattice parameter shows the opposite trend. This change is also reflected in the resistivity data. Jarosz et al. [35] studied single crystal of GdNiAl and reported that there are two different crystallographic phases of ZrNiAl-type around 220 K. The iso-structural transition around 220 K shows significant changes in Gd-Gd and Gd-Ni interatomic distances [35]. TbNiAl shows an abrupt transition in the lattice parameters around 110 K and is accompanied by a sudden change in resistivity while the space group is conserved [36, 37]. Chevalier et al. [38] reported that CeNiGa has two crystal structure phases (polymorphism) depending on the annealing conditions. CeNiGa is the only compound in the $R$NiGa series, which shows two phases with different crystal structures. $R$NiGa ($R$= Gd-Tm) crystallize in orthorhombic structure [39]. $R$NiSi compounds with $R$=La, Ce, Nd crystallize in tetragonal structure, while those with $R$= Gd-Lu crystallize in orthorhombic structure [40, 41, 42].

Light rare earth $R$NiSb compounds crystallize in the hexagonal ZrBeSi type structure, while the heavy rare earth compounds crystallize in the MgAgAs type cubic structure [43, 44]. It is worth to mention here that ZrBeSi is a super structure of AlB$_2$ type structure, in which $c$ parameter get doubled [43]. GdNiSb crystallizes in two crystal structures [44, 45]. In $R$CuGa series, we found the report on only CeCuGa compound, which crystallizes in orthorhombic structure [46]. CeRuAl and CeRhAl compounds crystallize in LaNiAl type orthorhombic structure [47]. Crystal structure of $R$CuSi compounds also depends on the heat treatment. It has been observed that compounds crystallize in two types of crystal structures. The high temperature phase adopts the AlB$_2$-type structure (*P6/mmm)*, while the low-temperature phase forms in the Ni$_2$In-type structure (*P6$_3$/mmc)* [48, 49, 50]. Similar to $R$CuSi series, structural change with annealing temperature was also found in $R$CuGe series. It has been observed that the compounds with $R$= La-Lu crystallize in AlB$_2$ type hexagonal crystal structure in the as cast-form [51, 52]. After an annealing for a long time at 750 °C, the compounds with $R$=La-Gd (excluding Eu) show AlB$_2$ type structure while $R$=Tb-Lu shows CaIn$_2$ type structure [52, 53]. In case of $R$CuSn, the authors have reported two types of crystal structures such as CaIn$_2$ and LiGaGe type hexagonal structures [54]. Crystal structure of CeCuSn is reported differently by different authors. Dwight et al. [54] and Rianai et al. [55] proposed a CaIn$_2$-type hexagonal structure with a statistical distribution of Cu and Sn atoms on the same 4f site while Yang et al. [56] and Adroja et al. [57] reported that these atoms occupy two different sites in a LiGaGe-type structure. The neutron diffraction data reported by Weill et al. [58] is in favor of the ordered distribution between Cu and Sn. EuCuSn of this series shows crystal structure, which is different from that of the other compounds. It crystallizes in CeCu$_2$ type orthorhombic structure [59]. The crystal structure of EuCuGa is same as that of EuCuSn [60]. In both the compounds, Eu is divalent, while EuCuSb crystallizes in ZrBeSi type hexagonal structure [61]. EuCuAs and YbCuBi crystallize in hexagonal structure [62, 63].

$RTX$ compounds with $T$=4$d$/5$d$ metal also show a variety of structures. It has been observed that in $R$RuSi and $R$RuGe series, compounds with light rare earths are found to crystallize in CeFeSi type tetragonal structures while compounds with heavy rare earths were found to show different structures [64, 65, 66, 67]. In *R*RuSi (*R*=Y, Tb-Tm), heavy rare earths excluding Gd show $Co_2Si$ type orthorhombic structure, while in *R*RuGe (*R*=Sm, Gd-Tm) they crystallize in TiNiSi type orthorhombic structure. To the best of our knowledge only CeRuSn compound is reported in *R*RuSn series, which crystallize with a superstructure of monoclinic CeCoAl type [68]. Such type of structure has two independent Ce sites, which can be fixed to trivalent Ce2 and intermediate valent Ce1 on the basis of interatomic distances [68]. The important feature observed in the structure of CeRuSn is the shortest interatomic distances such as Ce1-Ru1 (233 pm) and Ce1-Ru2 (246 pm), which is shorter than the sum of the covalent radii (289 pm) [68 and ref. therein].

The compounds of *R*RhAl (*R*=La-Nd, Gd, Ho, Tm) series show different types of structures but the space group is same. LaRhAl and CeRhAl adopt Pd(Pd,Mn)Ge-type structure while others show TiNiSi type structure [69]. It has been observed that Ce in CeRhAl, shows variable valency (between 3 and 4) at room temperature in this series. Compounds in *R*RhGa (*R*=Ce-Nd, Gd, Tb, Er-Lu, Y) series crystallize in TiNiSi type orthorhombic structure, which is an ordered derivative of $Co_2Si$ type structure [67, 70, 71]. The Ce shows valence close to +4 in CeRhGa [70]. In *R*RhIn, compounds with *R*=La-Nd, Sm, Gd-Tm crystallize in ZrNiAl type hexagonal structure, while compounds with *R*=Eu, Yb and Lu crystallize in TiNiSi type orthorhombic structure [72, 73]. Katoh et al. [74] reported that YbRhIn and LuRhIn crystallize in MgAgAs type cubic structure. *R*RhSi (*R*=Y, Gd-Er) compounds show TiNiSi type orthorhombic structure [75]. LaRhSi in this series is found to crystallize in cubic structure [76]. *R*RhGe (*R*=Ce-Nd, Sm, Gd-Yb, Y) compounds crystallize in TiNiSi type orthorhombic structure [67, 77]. Recently it is observed that Ce in CeRhGe shows a valence change on temperature modifications. Addition to this the compound shows first order structural transition at 520 K upon heating [78]. *R*RhSn (*R*=La-Nd, Sm, Gd-Lu) compounds were found to crystallize in ZrNiAl type hexagonal structure [79]. Recently, we observed that GdRhSn is the only member of *R*RhSn, which shows an iso-structural transition at about 245 K [80]. The lattice parameters (*a* and *c*) in GdRhSn show discontinous change around 245 K. Compounds with *R*RhSb (*R*=La, Pr, Sm, Gd-Tm) crystallize in TiNiSi type orthorhombic crystal structure [81, 82].

Dwight [83] reported that *R*PdAl (*R*=Sm, Gd-Tm, Y) crystallize in TiNiSi type orthorhombic structure. Later, Hulliger [84] did experiments on this series and observed that the samples, which were kept at 750 ºC for seven days after rapid cooling turned out to be hexagonal whereas annealed samples were orthorhombic. GdPdAl and TbPdAl show an iso-structural transition in paramagnetic regime from HTM I phase to HTM II phase at 180 K and 106 K, respectively [85, 86]. Similar to other compounds, showing iso-structural transition, the lattice parameter *a* decreases and *c* increases on cooling in these compounds.

*R*PdGa (*R*=La-Nd, Sm, Gd-Tm, Lu, Y) compounds crystallize in TiNiSi type orthorhombic structure [67], while *R*PdIn (*R*=La-Sm, Gd-Lu, Y,) in the ZrNiAl type structure [87, 88]. EuPdIn crystallizes in TiNiSi type structure [89]. *R*PdSi (*R*=La-Nd, Sm-Lu, Y) compounds crystallize in TiNiSi type orthorhombic structure [90]. Later, Prots' et al. reported that *R*PdSi (*R*=La, Pr, Ce) crystallize in a new monoclinic structure type [91]. Adroja et al. [92] reported LaIrSi type cubic structure in EuPd(Pt)Si compounds. *R*PdGe (*R*=La-Nd, Sm, Gd-Tm, Y) crystallize in orthorhombic structure while EuPdGe crystallizes in monoclinic structure [67,93]. *R*PdSn (*R*=Ce-Sm, Gd-Dy) compounds show TiNiSi type orthorhombic structure, while those with *R*=Er, Tm, Lu crystallize in $Fe_2P$ type hexagonal structure [94, 95]. Among all the compounds, HoPdSn crystallizes in any of the two structure types [94]. The authors [95, 96] reported TiNiSi type orthorhombic crystal structure for as-cast *R*PdSn (*R*=Ce-Nd, Sm-Yb) compounds. They found that on annealing at 950 ºC, ErPdSn and TmPdSn compounds adopt hexagonal $Fe_2P$- type structure. EuPdAs crystallizes in hexagonal structure [97]. Marazza et al. [98] reported that *R*PdSb (*R*=La-Sm, Gd, Tb) crystallize in $CaIn_2$ type hexagonal structure, while compounds with *R*=Dy to Yb crystallize in MgAgAs type cubic structure. Later it was reported that EuPdSi crystallizes in TiNiSi type orthorhombic and HoPdSb, ErPdSb and TmPdSb in MgAgAs type cubic structures [99]. *R*PdBi (*R*=La-Nd, Sm, Gd-Lu) crystallizes in MgAgAs type cubic structure [82 and ref. therein]. It has been observed that compounds in *R*PdSb and *R*PdBi with MgAgAs structure are also known as half Heusler phases [100 and ref. therein], which is drawing considerable attention these days. Many of striking properties of Heusler alloys are result of strong structure-magnetic property correlation.

Compounds in $R$AgAl ($R$=La-Nd, Sm, Gd- Er, Y) [101, 102, 103] and $R$AgGa ($R$=La-Nd, Gd-Lu, Y) [104, 105] crystallize in CeCu$_2$ type orthorhombic structure. YbAgAl was reported to crystallize in MgZn$_2$ type structure [101]. $R$AgGe ($R$=La, Ce) [106] and $R$AgSn (R=La, Ce, Sm, Gd-Er, Yb) are reported to crystallize in CaIn$_2$ and $R$AgGe ($R$=Gd-Lu) [107, 108] together with $R$AgSi ($R$=Sm, Gd, Dy -Lu, Y) compounds [109] crystallize in ZrNiAl type hexagonal structures. Mazzone et al. [110] reported CaIn$_2$ type hexagonal structure in $R$AgSn compounds. EuAgGe crystallizes in CeCu$_2$ type orthorhombic structure [111]. $R$AgSn ($R$=Ce-Nd, Gd-Er) compounds were found to crystallize in two types of structure; either CaIn$_2$ (disordered) or LiGeGa (ordered) type hexagonal structures [112]. Between these two types of structures, LiGaGe type structure is more stable. Some authors studied high temperature (HT) and high pressure (HP) phases of some compounds of $R$AgSn series [113, 114]. It has been reported that HT-LuAgSn crystallizes in NdPtSb type structure [114]. $R$IrAl ($R$=Ce-Nd, Sm, Gd-Tm, Lu, Y) [115], $R$IrGa ($R$=La-Nd, Gd-Tm, Y) [67, 70] $R$IrGe ($R$=Ce, Gd-Er) [67, 116, 117], CeIrSb [118] and YbIrGe [119] compounds crystallize in TiNiSi type structure. Like CeRhAl, Ce in CeIrAl is reported to be in the tetravalent state [115]. Ce shows valence between +3 and +4 in CeIrGa at room temperature [70]. In $R$IrSi, compounds with $R$=La and Nd crystallize in ZrOS type cubic structure while compounds with $R$= Tb-Er crystallize in TiNiSi type orthorhombic structure [76, 120]. $R$IrSn ($R$=La-Nd, Sm, Gd, Tb, Ho-Lu) compounds crystallize in Fe$_2$P hexagonal structure [121].

$R$PtAl ($R$=Sm, Gd-Tm, Lu, Y) [83], $R$PtGa ($R$=La-Nd, Sm, Gd-Lu, Y), $R$PtSi ($R$=Tb-Tm, Lu, Y) and $R$PtGe ($R$=Sm, Gd-Tm, Y) compounds crystallize in TiNiSi type orthorhombic structure, while $R$PtGe ($R$=Ce-Nd) crystallizes in CeCu$_2$ type orthorhombic crystal structure [67]. $R$PtIn ($R$=La-Nd, Sm, Gd- Lu, Y) compounds crystallize in ZrNiAl type hexagonal structure [122,123, 124, 125 and ref. therein] and $R$PtSi ($R$=La-Nd, Sm, Gd) compounds crystallize in LaPtSi tetragonal structure [67]. EuPtIn crystallizes in TiNiSi type orthorhombic structure [126]. EuPtGe crystallizes in LaIrSi type cubic structure [127]. $R$PtSn (R=Gd-Lu) compounds crystallize in Fe$_2$P type hexagonal structure [128]. The orthorhombic structure in CePtSn shows considerable temperature dependence of lattice parameters [129]. It has been observed that at $T_{N1}$ (7.5 K), the lattice parameters and cell volume exhibit discontinuities while at 5.2 K (second magnetic transition), the effect on lattice parameters and cell volume is small [129]. The high pressure modification of NdPtSn and CePtSn show change of crystal structure from TiNiSi type orthorhombic to ZrNiAl type hexagonal structure [130, 131]. YbPtSn crystallizes in ZrNiAl type hexagonal structure [132]. $R$PtSb series shows three types of structures in which compounds with $R$=La-Sm show CaIn$_2$ type hexagonal structure, EuPtSb shows TiNiSi type orthorhombic structure and compounds with $R$=Gd-Yb show MgAgAs type cubic structure [133].

$R$AuAl (La, Ce, Nd) [134] and $R$AuGa ($R$=La-Nd, Sm-Lu, Y) [105] compounds crystallize in TiNiSi and CeCu$_2$ type orthorhombic structure, respectively, while $R$AuIn ($R$=Ce-Sm, Gd-Tm, Lu] compounds crystallize in ZrNiAl type hexagonal structure [135, 136]. EuAuIn crystallizes in TiNiSi type orthorhombic structure [137]. LiGaGe and CaIn$_2$ type hexagonal structure has been observed in $R$AuGe ($R$=La-Nd, Sm, Gd-Eu, Y) [138, 139] and $R$AuSn compounds [140, 141, 142]. EuAuGe crystallizes in CeCu$_2$ type orthorhombic structure [143]. ErAuSn has a polymorphic character [140, 144]. High-temperature, hexagonal CaIn$_2$ - type modification of ErAuSn can be obtained by arc-melting and rapid quenching in the furnace which can be easily converted into well known cubic MgAsAg-type form after a proper heat-treatment [140].

## 3. Magnetic properties

The nature of magnetic ordering and the magnetic transition temperatures of $RTX$ compounds are given in Table II. From the table, one can see that the compounds order in a range of temperatures (i.e from low temperatures to room temperature and above). Some of them order ferromagnetically, while some others order antiferromagnetically. Apart from these, complex magnetic ordering is also seen in many compounds. The highest ordering temperature observed in this family is 510 K for TbMnGe.

In the following, the important magnetic properties derived mainly from bulk magnetization measurements, neutron diffraction data and heat capacity studies of the $RTX$ family is discussed by categorizing on the basis of $T$ constituent, in the order Sc, Ti, Mn, Fe, Co, Ni, Cu, Ru, Rh, Pd, Ag, Os, Ir, Pt and Au. This has been done for a better readability.

## 3.1 $R$Sc$X$ compounds

The $R$ScSi and $R$ScGe compounds show high magnetic ordering temperatures. Nikitin et al. [145] performed magnetometric measurements in the temperature range 77-750 K and found that GdScSi, GdScGe and TbScGe show FM behavior, while TbScSi shows AFM nature. The transition temperatures of other compounds are not reported due to the absence of low temperature data. Later, Sing et al. [3] reported the magnetization data of some light rare earth compounds. They found that PrScGe orders ferromagnetically along with multiple spin reorientation transitions at different temperatures. SmScSi and SmScGe are FM and show huge coercive fields at low temperatures while PrScSi shows AFM ordering. There are various reports on the magnetic properties of CeScSi and CeScGe. Canfield et al. [146] reported that CeScSi and CeScGe show FM transitions at 26 and 46 K, respectively. On the other hand, Uwatoko et al. [147] reported AFM transition for CeScGe at 46 K. Later Uwatoko et al. [148] reported the magnetic properties of single crystals of CeScSi and CeScGe. The report shows a magnetic transition at 46 K along $a$ and $c$-axes for CeScSi and broad maximum around 27 and 25 K along these two axes, respectively. Later, Singh et al. [4] reported that the compounds CeScSi and CeScGe order antiferromagnetically below 26 and 46 K. Neutron diffraction study on these compounds shows that TbScGe and NdScGe have collinear FM structures [149]. Tencé et al. [150] reported that TbScGe orders ferromagnetically below 250 K. Some other interesting properties were also observed in this series of compounds. Recently, Kulkarni et al. [151] reported that SmScGe and NdScGe are found to show near-zero net magnetization when substituted by 6-9 at. % of Nd and 25 at. % of Gd, respectively. The observation of an appreciable conduction electron polarization and a tunable exchange bias field concomitant with the negligible stray field and high ordering temperatures, make these materials potential for applications in spintronics [151].

## 3.2 $R$Ti$X$ compounds

As it was mentioned earlier, in $R$TiSi series, only heavy rare earth compounds exist. YTiSi and LuTiSi are Pauli paramagnets, while others show AFM ordering. Neutron diffraction data show that no local moment is found in Ti atoms. Similar to $R$ScSi and $R$ScGe, these compounds also show high ordering temperature. It is very surprising that though Ti and Si are non-magnetic, the ordering temperature is very high. To understand this behavior Nikitin et al. [152] substituted Ti in place of Mn in GdMnSi compound. It has been observed that substitution of Ti at Mn place, increases the $T_C$ and the paramagnetic Curie temperature ($\theta_p$) of the compounds. The increase in $T_C$ in these compounds on the substitution of Ti has been explained by assuming itinerant magnetism. In CeFeSi-type structure, atoms can be arranged in the sequence $R$-$X$-$T_2$-$X$-$R$, which shows that $R$ atom in such type of structure is separated by $X$ and $T$ atoms. Thus the hybridization between $p$ and $d$ states of $X$ and $T$ elements, respectively, try to fill $3d$ up and $3d$ down sub-bands [152]. In the study of CeFeSi-type structure, the authors have considered three $R$–$R$ exchange interactions, these interactions are: $J_0$ within the $R$ planes, $J_1$ between $R$ planes separated by $T$ and $X$ planes (BaAl$_4$ block) and $J_2$ between nearest $R$ planes (W block) [8]. The nature and magnitude of magnetic ordering temperature in these compounds depends upon the sign of these interactions [8 and ref. therein].

$R$TiGe compounds have similar magnetic properties as those of $R$TiSi compounds because of the same crystal structure. Most of compounds in this series are AFM with high ordering temperatures. La, Lu and YTiGe compounds show weak Pauli paramagnetic behavior, while CeTiGe is paramagnetic down to 0.4 K. At low temperatures, CeTiGe shows the properties expected for Fermi liquid [153]. The HTM-CeTiGe reported in Ref. [14] revealed that there is decrease of hybridization $J_{cf}$ between $4f$ (Ce) and conduction electrons when one goes from LTM to HTM of CeTiGe. Deppe et al. [154] reported the first order metamagnetic transition in CeTiGe. Magnetic structure of these compounds can be characterized by FM W slabs, where $R$ moments are aligned in basal plane, while moments in different W slabs are coupled antiferromagnetically [13]. Hence, the magnetic structures of these compounds combine both FM and AFM exchange interactions and yields positive $\theta_p$ values [13]. Due to coexistence of FM and AFM interactions, most of the compounds in this series show field induced metamagnetic transition.

According to neutron diffraction report, PrTiGe and NdTiGe show sine modulated magnetic structure [155]. Below 80 K, the magnetic structure of NdTiGe transforms into commensurate arrangement, in which FM (0 0 1) Nd layers are coupled antiferromagntically along the c direction in the sequence ++-- [155]. The magnetic structure shown by TbTiGe,

DyTiGe, HoTiGe and ErTiGe below 312, 185, 124, and 36 K, respectively is also commensurate. Except ErTiGe, the moment in these compounds is aligned along $c$-axis, while in ErTiGe it lies in the basal plane [155]. The high field magnetization study shows forced ferromagnetic alignment of the Dy moments, which is achieved by two metamagnetic transitions and the field larger than 20 T [156]. It is reported that the CeScSi type phase of GdTiGe shows FM ordering, while CeFeSi-type structure shows AFM transition [16]. The magnetic behavior in TbTiGe is affected by the structural transition (HTM-CeScSi to LTM-CeFeSi) [18]. The compound shows FM ordering below $T_C$= 300 K in HTM, while the LTM phase is an AFM below $T_N$=280 K [18]. It is worth mentioning here that the neutron diffraction data of this compound reveal that the Tb moments are aligned along the $c$-axis in both the cases [18]. HoTiGe shows field induced metamagnetic transition and shows magnetic ordering which is slightly lower than mentioned above [157].

## 3.3 $R$Mn$X$ compounds

Among all the $RTX$ compounds, $R$MnX are the compounds, in which Mn contributes considerable moment. Ac and dc magnetic susceptibility show two magnetic transitions in GdMnAl. The authors have reported the second anomaly as due to reorientation of Gd moments [25]. The amorphous phase of GdMnAl shows different magnetic properties [158]. After mechanical grinding the high temperature magnetic transition disappears while the milled sample shows a spin glass transition at 53 K. NdMnAl does not show long range ordering [26]. Neutron diffraction study show AFM ordering in TbMnAl compound and magnetization measurements reveals metamagnetic behavior in this compound [159]. Spin glass state has been observed in $R$MnGa ($R$=Ce, Pr, Nd, Gd, Tb, Dy) compounds [27]. It has also been found that among $R$MnGa compounds, NdMnGa and DyMnGa show simple spin glass state while PrMnGa, GdMnGa and TbMnGa show reentrant spin glass behavior [27]. The spin glass behavior is also observed in $R$MnIn ($R$= Gd, Dy) compounds as any feature indicating FM or AFM ordering is absent in the magnetization data [26]. Moreover, the heat capacity data do not show any sharp peak related to magnetic ordering in these compounds.

The most important aspect in $R$MnSi series is the high magnetic transition temperature of the compounds, which can be correlated with the occurrence of AFM Mn layers [20]. It has been observed that some other compounds with same crystal structure and having antiferromagnetically/ferromagnetically coupled Mn layers still have low magnetic ordering temperature than $R$MnSi compounds, which indicates that the correlation in this series is a result of a particular electronic structure having high Mn moment and a strong AFM coupling [20]. $R$MnSi compounds have magnetic character with two magnetic sublattices, which are $R$ and Mn and have atoms located in layers separated by Si layers alternating in the following order: $R$-Si-Mn$_2$-Si-$R$ [160]. Since Mn is magnetic in these compounds therefore three main types of exchange interactions take place which are; $R$-$R$ interaction, $R$-Mn interaction and Mn-Mn interaction. The Mn moment in these compounds is found to be in the range of 2.8-3.3 $\mu_B$. It has been observed that the nature and strength of magnetic ordering in these compounds essentially depend on the Mn-Mn interatomic distance within the magnetic layers [160]. The in-plane Mn-Mn coupling is found to show FM interaction for $d_{Mn-Mn} < 2.84$ Å and AFM for $d_{Mn-Mn} > 2.89$ Å and coexistence of FM and AFM for intermediate distances [23 and ref. therein].

In $R$MnSi series, compounds with $R$=La, Ce are paramagnetic as shown by magnetization data, but are AFM according to neutron diffraction data [20]. The AFM ordering at high temperatures in these compounds arises due to Mn sublattice. Compounds with $R$=Pr-Sm are antiferromagnets and show multiple magnetic transitions [20]. It has been reported that GdMnSi, which has a $T_C$ of 310 K and an inflection point at 300 K exhibits spontaneous magnetization over whole ordering temperature range [20]. Ovtchenkova et al. [161] reported the the FM ordering in GdMnSi at $T_C$=314 K. Later, it was reported that in SmMnSi negative moment can be induced by cooling at constant magnetic field below its magnetic transition temperature [162]. The unusual magnetic behavior in SmMnSi has two magnetic transitions at $T_t$=130 K and $T_N$= 240 K along with a compensation temperature $T_{Com}$=215 K. Welter et al. [23] reported TiNiSi type structure of TbMnSi, DyMnSi and NdMnGe in high temperature form. All these compounds are AFM below $T_N$ and show collinear magnetic structure with Mn moment aligned along [100] for TbMnSi and DyMnSi and [001] for NdMnGe between $T_N$ to $T_t$ (<150 K). Below $T_t$, these compounds show flat spiral structure characterized by a propagation vector ($k_x$, 0, 0) with Mn and $R$ (=Tb, Dy and Nd) moments lying in the (001) plane [23]. TbMnSi and DyMnSi show additional transition at $T_t$=140 and 55 K, respectively [23]. Welter et al. [163] reported magnetization in 4.2 -300 K range for TbMnSi and found that the compound shows FM nature below 260 K. The neutron diffraction data reported by these authors show magnetic ordering in Mn and Tb sublattices take place at 260 K where Mn and Tb FM (001) layers are coupled anitferromagnetically [163]. A large coercive field ($H_C$=6 kOe at 4.2 K and asymmetry of the hysteresis loop has been observed in TbMnSi [163]. It has been reported that HoMnSi exhibits three different magnetic regimes; a collinear AFM structure from 300 to 55 K, a canted AFM structure due to non-collinear arrangements of Ho moments between 55 and 15 K and incommensurate structure below it, while LuMnSi exhibits flat spiral AFM structure below 255 K [164].

*R*MnGe (*R*=La-Nd) compounds are antiferromagnets below 410 K [165]. Transition temperatures of some of these compounds were not detected by magnetization data, but determined from neutron diffraction data. The authors reported that the magnetic structure of these compounds can be described in such a manner, in which rare earth sublattice orders in FM (001) layers with various ordering schemes along the stacking axis at low temperatures and by a stacking of AFM (001) Mn layers which persists down to 2 K. [165]. PrMnGe exhibits FM ordering followed by FM-AFM transition below 80 K where it shows crystallographic transition (tetragonal to orthorhombic symmetry), while NdMnGe shows a FM ordering of Nd sublattice below 200 K which persists down to 2 K [165].

It has been reported that all the compounds in *R*MnGe (*R*=Dy-Tm, Y) show AFM ordering of Mn sublattice above room temperature, while *R* sublattice orders antiferromagnetically at 85, 70, 60, 30 for *R*=Dy, Ho, Er and Tm, respectively [21]. The authors reported that compounds with *R*=Y, Er and Tm are simple collinear with a spin reorientation transition below 25 K in TmMnGe, while DyMnGe shows cycloid and HoMnGe exhibits a conical magnetic structure [21]. They also reported that inverse susceptibility data in GdMnGe show a small deviation at around $T_N$=490 K, which might be due to the ordering of Mn sublattice [21]. Later, Ivanova et al. [166] reported complex magnetic behavior of GdMnGe. These authors reported spin reorientation transition at $T_1$=100 K and AFM ordering temperature, $T_N$=350 K [166]. The sign of $\theta_p$ in these compounds suggests dominant AFM interaction in YMnGe, ErMnGe and TmMnGe while FM character in GdMnGe, DyMnGe and HoMnGe [21].

Magnetization isotherms show metamagnetic like transitions in ErMnGe and TmMnGe compounds [21]. It is worth noting that in all the compounds, the effective magnetic moment ($\mu_{eff}$) is larger than $R^{3+}$ free ion moment, which confirms the non-zero magnetic moment of Mn atom. Ivanova et al. [167] performed high field magnetization experiment on TbMnGe and DyMnGe compounds and reported that destruction of FM spiral structure takes place in high fields. It has been observed that TbMnGe shows collinear magnetic structure with both Tb and Mn moments aligned in [100] between $T_N$=510 to $T_{t1}$=200 K and a flat spiral structure characterized by propagation vector ($k_x$, 0, 0) with Mn and Tb moments almost lying in the (001) plane and below $T_{t2}$=50 K the $k_x$ becomes 1/4 and the magnetic structure is commensurate [168].

## 3.4 *RFeX* compounds

There are many reports on the magnetic properties of *R*FeX compounds with different *X* elements. Usually, the coupling between the *R* and Fe moments is FM in the case of light rare earths, whereas it is AFM in the case of heavy rare earths. Compounds in this series do not show Fe moment, if exists, is very small. However, Oesterreicher [169] reported that YFeAl and LuFeAl show magnetic ordering. Magnetic ordering in these non-magnetic rare earth compounds indicates localized moment on Fe atom. Further clarification regarding Fe localized moment in these compounds need neutron diffraction experiment results. GdFeAl shows FM ordering at 265 K with a saturation moment of 5.8 $\mu_B$/f.u., which is lower than the expected saturation moment of $Gd^{3+}$ ion [170]. The authors [170] also reported that the decrease in saturation moment arises due to the AFM coupling between Gd and Fe moments. DyFeAl shows FM ordering at $T_C$= 129 K with negligible hysteresis loss in its magnetization isotherms [171]. ErFeAl and TmFeAl have been studied by performing neutron depolarization and diffraction experiments [172]. TmFeAl shows a first order magnetic transition at low temperature where it consists of magnetic nanodomains, while the behavior is different in ErFeAl which shows a magnetic domain size of 0.4 μm [172].

Welter et al. [173] reported that no moment on the Fe atom could be observed in *R*FeSi from neutron diffraction data, which was also seen from the magnetization data. The $\mu_{eff}$ values in these compounds are close to the free rare earth ion ($R^{3+}$) moment values. Furthermore, it can be seen from the susceptibility data of LaFeSi that it does not show any magnetic ordering down to low temperature, confirming the absence of moment on Fe [173]. Among all the compounds, SmFeSi does not follow a Curie-Weiss law, which occurs most of Sm compounds in *RTX* series and TbFeSi shows large coercive fields (9 kOe at 4.2 K), which shows strong uniaxial anisotropy [173]. It has been observed that PrFeSi does not show any magnetic ordering down to 2 K, which is also confirmed by neutron diffraction. Most of compounds in $R$FeSi are FM.

Prasad et al. [174] substituted Ru at the Fe site in PrFeSi compound and observed that the Ru substituted compound orders antiferromagnetically. The $T_N$ of the compound increases with increase in Ru content and becomes between 17 and 27 K for x⩾0.15 in PrFe$_{1-x}$Ru$_x$Si [174]. Later on, Napoletano et al. [175] and Zhang et al. [176] reported the magnetic and magnetocaloric properties of $R$FeSi ($R$=Gd, Tb, Dy) compounds. They found different values of $T_C$, $\theta_p$ and $\mu_{eff}$ for GdFeSi, TbFeSi and DyFeSi as compared to those reported by Welter et al. [173]. Among all the compounds of RFeSi, HoFeSi shows multiple magnetic ordering: one at $T_C$=29 K, where it changes its state from paramagnetic to FM and other at $T_t$ = 20 K, where FM to AFM or ferrimagnetic transition take place [177]. ErFeSi shows FM to PM transition around 22 K and found to show thermomagnetic irreversibility (TMI) at low temperatures below $T_C$ [178]. Generally, the difference in ZFC and FC below ordering temperature arises due to TMI.

TMI was seen in many of $RTX$ compounds, which arises due to many reasons: (i) Magnetic system with large magnetocrystalline anisotropy and low $T_C$ possess narrow domain walls and owing to domain wall pinning effect, show TMI, (ii) TMI also arises due to spin glass behavior and magnetically frustrated systems, (iii) In superparamagnets.

## 3.5 $R$CoX compounds

As in the case of many $R$-Co intermetallics compounds, Co sublattice is almost non-magnetic in $R$CoX series. In $R$CoAl, light rare earth compounds were not found to be stable, unlike the compounds with heavy rare earths. Compounds with $R$=Gd, Tb, Dy, Ho order ferromagnetically and show TMI at low temperatures [179,180]. Generally, it has been seen that the saturation moment in these compounds is smaller than the expected value of corresponding rare earth ion. The reduction in saturation moment generally arises due to coexistence of non-FM state, crystal field effect and magnetic crystalline anisotropy. In addition to this, intrinsic magnetic disorder was also reported by some authors to bring down saturation magnetic moment in $R$CoAl compounds [180 and ref. therein]. Report on CeCoGa compound reveals complex magnetic behavior [181]. It shows moderate heavy fermion behavior, undergoing a complex AFM state below 4.3 K where it coexists with spin-glass like state. It has been observed that TMI in CeCoGa is a characteristic of spin glass state and large value of $\theta_p$ suggests the presence of Kondo type interactions [181]. The compound also shows a field induced metamagnetic transition and shift in the hysteresis loop (exchange bias).

Among the $R$CoSi compounds, LaCoSi shows Pauli paramagnetic behavior, which confirms that there is no moment on Co atom [31]. Neutron diffraction measurement on these compounds also confirmed the absence of localized moment on the Co sites. PrCoSi and CeCoSi do not show any magnetic ordering down to 4.2 K temperature [31]. Later, Chevalier and Matar [182] reported AFM ordering in CeCoSi below 8.8 K. Leciejewicz et al. [183] reported that magnetic moments of Ho atoms in HoCoSi form a conical spiral structure at 1.7 K and transforms to FM structure on increasing temperature with a $T_C$ of 13 K. Later on, we have performed magnetic and magnetocaloric measurements on HoCoSi compound and found slightly lower $T_C$ of 14 K [184]. Neutron diffraction data on $R$CoGe shows no local moment on Co site, as expected [185]. The authors reported that PrCoGe and CeCoGe show paramagnetic behavior down to 1.6 K, while NdCoGe orders antiferromagnetically below 8 K. Neutron diffraction reports [186, 187, 188] on $R$CoSn ($R$=Tb, Dy, Ho, Er) show that rare earth moment in these compounds form different kinds of modulated magnetic structures. In RCoSn (R=Dy, Ho, Er), each magnetic structure can be considered as two modes A and G along the $b$ and $c$ axes, respectively, which results in non-collinear magnetic structures [186]. André et al. [187] reported magnetization data showing three anomalies at temperatures 5.4, 11.6 and 20.5 K in TbCoSn and two, at temperatures 4 and 8 K, in HoCoSn.

### 3.6 $R$Ni$X$ compounds

Generally, there are many similarities between the magnetic properties of $R$Co and $R$Ni compounds because of the nonmagnetic $T$ sublattice in both of them. A similar scenario exists between $R$Co$X$ and $R$Ni$X$ compounds. YbNiAl and LuNiAl do not show any magnetic ordering, which confirms that only rare earth atoms contribute to the moment [189]. Later Schank et al. [190] reported magnetic and transport properties of YbNiAl and YbPtAl, which are found to be antiferromagnetic at 2.9 and 5.9 K, respectively. These authors reported

YbNiAl to be the first aniferromagnetic Yb-based heavy fermion. Neutron diffraction [191] shows only one magnetic phase transition in light rare earth $R$NiAl compounds ($R$=Pr, Nd) with an incommensurate sine wave modulated magnetic structure, while two or three transitions are seen in the heavy rare earth compounds ($R$=Tb, Dy, Ho), which show commensurate magnetic structure. TbNiAl shows geometrically induced frustration, which gives rise to the splitting into two different types of antiferromagnetically ordered Tb moments [192]. TbNiAl and TbPdAl show magnetic frustrations of moments, which arises due to the AFM exchange coupling between two Tb ions and strong magnetocrystalline anisotropy [193].

Single crystal study has also been done by some researchers on TbNiAl and DyNiAl [37, 194]. TbNiAl shows an abrupt transition in the lattice parameters around 110 K (while the space group is conserved), which is accompanied by sudden change in the resistivity [36, 37]. Application of hydrostatic pressure on TbNiAl lowers the structural transition temperature, which vanishes completely for pressures ⩾0.55 GPa [37]. A similar behavior is seen in resistivity data on the application of pressure. The application of hydrostatic pressure in this compound prefers a low $c/a$ phase and stabilizes AFM ground state, while the uniaxial pressure along the $c$-axis leads to a FM order in this compound [37]. Some compounds in RNiAl (R=Dy, Ho) show thermo-magnetic irreversibility in magnetization data at low temperatures [195,196]. Generally the TMI decreases with field and disappears at high fields. However, in the case of Dy and Ho, a small irreversibility is still present even at high fields, which indicates magnetic frustration in these compounds. This arises because of the triangular lattice formation of $R^{3+}$ ion in the basal plane accompanied by AFM ordering. It has been reported that compounds of $R$NiAl series show $\mu_{eff.}$ higher than the theoretical value, which may arise due to polarization of 3d moment in Ni, and termed as magnetic polaronic effect, which is also responsible for some unusual magnetic properties in these compounds [195, 196]. Most of the compounds of $R$NiAl series are metamagnetic in nature [195, 196, 197].

CeNiGa has two phases with different crystal structures and both show intermediate valence state and Kondo effect [38]. However, the Kondo temperatures ($T_K$) are quite different for these two; $T_K$= 300 K for LTP and $T_K$= 95 K for HTP [38]. Neutron diffraction data on $R$NiGa ($R$=Tb-Tm) [198,199, 200, 201, 202] show that only TbNiGa has commensurate magnetic structure, while others (Dy-Tm) show incommensurate structures with a propagation vector along the $a$- axis for Tb, Dy and Er and along $b$ and $c$-axes for Tm and Ho compounds. No moment was observed for Ni atoms in neutron diffraction studies. In $R$NiIn series, CeNiIn shows valence fluctuating behavior [203], no long range order was seen in PrNiIn above 1.5 K [204]. Neutron diffraction and magnetization studies [204, 205, 206] show that RNiIn compounds with $R$=Nd, Ho and Er show FM ordering and $R$=Tb, Dy and Tm show AFM ordering below their respective ordering temperatures. At 1.5 K, Tb and Dy compounds show two coexisting magnetic phases; non-collinear AFM phase and a phase described by propagation vector (0.5, 0, 0.5) [204]. Alignment of moments is parallel to the $a$-axis for Nd, $c$-axis for Ho, Er and Tm. GdNiGa shows FM odering [207]. Zhang et al. [208] reported that compounds with $R$=Tb, Dy, Ho in RNiIn undergo two magnetic transitions on increasing the temperature. $R$=Tb and Dy compounds show field induced metamagnetic transition. The susceptibility data of these compounds show TMI, which is attributed to the domain wall pining effect.

LaNiSi is found to show phonon mediated superconductivity with a superconducting transition temperature ($T_C$) = 1.23 K as revealed by magnetic, heat capacity and resistivity data [209]. In case of CeNiSi no magnetic ordering was seen down to 2.4 K [210]. NdNiSi shows two magnetic transitions; at 6.8 K and 2.8 K, in which first transition is AFM in character, while second results from the reorientation of moments in the AFM state [211]. Neutron diffraction and magnetometric measurements performed on $R$NiSi ($R$=Tb-Er) compounds disclose AFM ordering at low temperatures [212]. At 1.5 K, HoNiSi shows collinear magnetic structure, while TbNiSi and DyNiSi show square modulated magnetic structure, which transforms to sine modulated structure on increasing temperature. ErNiSi shows collinear antiferromganetic structure at 1.6 K, which changes to an incommensurate structure at 2.5 K. The magnetic moments in these compounds are aligned along $b$-axis for $R$=Tb, Dy, Ho, while they are aligned along $a$-axis for the Er compound. Recently, we have studied the magnetic and magnetocaloric properties of HoNiSi and ErNiSi in the temperature range 1.8 K to RT. ErNiSi shows soft ferromagnetic behavior [213]. The magnetization data show only one magnetic transition in these compounds with field induced metamagnetic transition.

Neutron diffraction and magnetization measurements [214, 215, 216] on $R$NiGe ($R$=Gd-Tm) show that these compounds order antiferromagnetically at low temperatures. The magnetic structure in TbNiGe and DyNiGe can be described by square modulated at low temperature, which changes to sine modulated for TbNiGe and cycloidal spiral for DyNiGe compound with moments parallel to $c$-axis [215]. HoNiGe and ErNiGe are AFM and shows collinear magnetic structures at low temperatures, which changes to modulated structure at 2.5 and 2.3 K, respectively [216]. Most of compounds in RNiSn series show AFM ordering at low temperatures [217, 218, 219, 220, 221, 222, 223]. LaNiSn shows superconductivity below $T_C$=0.59 K, while PrNiSn is paramagnetic down to 0.9 K [223, 224]. Single crystal study of TbNiSn and DyNiSn belonging to this series shows multiple magnetic ordering and field induced multi-step metamagnetic transitions [219, 221]. Report on neutron diffraction data shows that NdNiSn, GdNiSn, TbNiSn, DyNiSn and HoNiSn exhibit incommensurate magnetic structure, which become nearly commensurate at 6 K for TbNiSn, while for DyNiSn it still remains incommensurate [220, 225 226, 227, 228 229, 230]. Single crystal neutron diffraction study of DyNiSn shows that the compound has incommensurate modulation along $c$-axis and commensurate modulation along $b$-axis (easy magnetization axis) on the application of an external magnetic field at 1.6 and 2 K, respectively [231]. Later, some authors reported that TbNiSn shows sine wave modulated magnetic structure below 20 K ($T_N$) and then forms an antiphase/square wave type structure below its 7 K transition [232, 233]. Among the other compounds of $R$NiSn series, CeNiSn is found to show some extraordinary properties. Many experiments [234, 235] in CeNiSn suggest that a pseudo-gap opens in the Kondo state at low temperatures. The heat capacity data (C/T vs. T) show a change in slope around 0.5 K, indicating some additional structure inside the pseudo gap of density of sates (DOS) around Fermi energy ($E_F$) [235]. The pseudo gap behavior in this compound is reported to be due to the interplay between spinons and soft crystal field states [236]. YbNiSn shows Kondo behavior with small ferromagnetic moment below 5.5 K [237].

LaNiSb, LuNiSb and YNiSb are Pauli paramagnets in the RNiSb series, which suggests that Ni atoms do not possess magnetic moment in these compounds as well [43, 44]. Mössbauer study of GdNiSb reveals that in cubic phase, compound shows magnetic ordering below 9.5 K and in hexagonal phase it orders at about 3.5 K [45]. YbNiSb shows Curie-Weiss behavior above 10 K in magnetic susceptibility data and temperature independent behavior in the temperature range 2-8 K in zero field heat capacity data [238, 239]. Field dependence of heat capacity data show a broad peak, which shifts to higher temperature with increase in field.

## 3.7 $R$Cu$X$ compounds

Reports on neutron diffraction, magnetization and heat capacity data in $R$CuAl compounds suggest that compounds with $R$=Gd-Er show FM ordering, PrCuAl and NdCuAl are AFM, YCuAl and LuCuAl are Pauli paramagnets [240, 241, 242, 243, 244, 245, 246, 247]. It has also been observed that GdCuAl and DyCuAl have some additional magnetic transitions. PrCuAl shows collinear AFM ordering [242]. NdCuAl shows complex magnetic structure, which can be ascribed to the co-existence of FM and AFM components of Nd moments [243]. An interesting property of these compounds is that substitution of Cu in place of Ni retains the crystal structure, but the magnetic properties are modified from AFM to FM [244, 245, 247].

In the $R$CuIn family, CeCuIn is paramagnetic down to 1.9 K, while the compounds with $R$=Nd, Gd, Tb, Ho, Er show AFM ordering in which the magnetic structure for Nd, Tb and Ho is non-collinear and for Er it is collinear [248, 249, 250]. A number of authors have studied the magnetization and magnetic structure of RCuSi compounds. The compounds with $R$= Pr, Gd, and Tb with AlB$_2$ type crystal structure show FM behavior below their Curie temperatures of  14 K for PrCuSi, 49 K for GdCuSi and 47 K for TbCuSi [251]. Later, it has been reported that PrCuSi and GdCuSi order antiferromagnetically with $T_N$=5.1 and 14 K, respectively [252, 253]. Kido et al. [254] performed magnetization measurements in the temperature range of 77-300 K for RCuSi compounds and found that $\theta_p$ is negative for light rare earth compounds and positive in case of heavy rare earth compounds. Neutron diffraction data showed that TbCuSi has cosinusoidally modulated magnetic structure at 4.2 K, while $R$= Dy and Ho compounds do not show any magnetic ordering down to 4.2 K [255]. Later on, Oleś et al. [50] reported that DyCuSi and HoCuSi are AFM with  $T_N$ = 11 K for DyCuSi and 9 K for HoCuSi and show sine modulated structure with a magnetic ordering stable in the temperature range of 1.4 K to $T_N$. Soon after this report, Papamantellos et al. [256] made investigations based on simultaneous refinements of high resolution and high flux data and found slightly different results for DyCuSi and HoCuSi as compared to the reported ones in Ref. [50]. The magnetic structure in these compounds can be described as collinear

FM for CeCuSi [257], transverse sine wave modulated for ErCuSi [258], and sine wave modulated for TmCuSi [259]. Recently, we have prepared some compounds such as NdCuSi, GdCuSi and TbCuSi in this series and found that all these compounds are AFM below their ordering temperatures [260]. NdCuSi shows field induced metamagnetic transition.

Neutron diffraction and magnetization measurements show that all the compounds of $R$CuGe series with $R$=Pr, Nd, Tb, Dy, Ho and Er along with Gd are AFM at low temperatures [52]. All the compounds except NdCuSi show sine modulated magnetic structures. It has been observed that direction of moments in compounds with $R$=Tb and Dy is perpendicular to $c$-axis while for $R$=Ho and Er it is parallel to $c$-axis [52]. In the case of ErCuGe, an additional magnetic transition was observed at 3.7 K [261]. The compounds are found to change their magnetic state from AFM to FM on the application of an external field [261]. CeCuGe shows FM behavior below $T_C$=10.2 K with two distinct paramagnetic regimes: at high temperatures, $\mu_{eff}$ is larger than that at low temperatures [56, 262]. The decrease in the $\mu_{eff}$ at low temperature is attributed to crystalline electric field effect.

$R$CuSn ($R$=Ce, Pr, Nd, Gd-Er) compounds show AFM nature at low temperatures [263, 264, 265]. Sakurai et al. [265] synthesized CeCuSn by two methods namely arc melting and Czochralski pulling. They found that former compound shows AFM ordering with $T_N$= 8 K with a weak spontaneous magnetization while the later compound shows AFM ordering with $T_N$=10 K with no spontaneous magnetization. Nakotte et al. [266] reported three magnetic transitions in CeCuSn. They reported that two magnetic transitions, AF1 and AF2, are seen in zero field, while the third magnetic phase transition, AF3 is only stable in an applied field. Baran et al. [264] reported that PrCuSn shows sine wave modulated magnetic structure below 3 K while NdCuSn shows collinear magnetic structure below 9 K.

Later, Pacheco et al. [267] reported that the magnetic structure of NdCuSn is described by two magnetic wave vectors, $k_1$=(0, 3/8, 0), which is attributed to a majority phase (>90%) having an ordered Cu/Sn sublattice. This wave vector, $k_1$ originates from transverse sine wave modulated arrangement of spins oriented parallel to $c$-axis. The other wave vector, $k_2$= (1/2, 0, 0) is attributed to the minority phase (<10%) having a disordered Cu/Sn sublatiice and originates from the collinear arrangement along $a$-axis. Hatzl et al. [268] reported the absence of long range magnetic ordering, but a spin glass behavior suspected below 10 K. Later Baran et al. [264] confirmed long range ordering in this compound. They reported that the difference in ZFC and FC magnetization data is attributed to random orientation of crystallites, and anisotropy. EuCuSn does not show any magnetic ordering down to 4.2 K [269]. It has been reported that compounds with Tb, Dy and Ho exhibit collinear AFM structures at 1.5 K which transforms to incommensurate structures at 12.6 and 5.8 K for TbCuSn and HoCuSn, while it remains stable up to $T_N$ for DyCuSn compound [263, 270]. Magnetic structure of ErCuSn has been described by two wave-vectors, $k_1$ and $k_2$. The magnetic reflections with $k_1$ fall to zero at $T_N$ (4.7 K), while the other vanishes at 3 K [263]. Hence, it is clear that some of these compounds show multiple magnetic transitions. EuCuAs orders antiferromagnetically below 14 K [62]. It has been observed that for both H parallel to *ab* and H parallel to *c* magnetization increases sharply with field and shows saturation below the ordering temperature.

## 3.8 *R*Ru*X* compounds

Many compounds of the *RTX* family have been discovered in which the T component is a 4d element such as Ru, Rh, Pd, Ag. In *R*RuSi and *R*RuGe compounds, LaRuSi and LaRuGe show Pauli paramagnetic behavior [64]. CeRuSi and CeRuGe show no magnetic order down to 4.2 K [64]. It has also been observed that silicide and germanide of Sm do not follow Curie –Weiss law. The neutron diffraction data show that NdRuSi can be described by collinear magnetic structure in which FM (001) Nd layers couples antiferromagnetically along *c*-axis [64]. Similar magnetic structures were observed in PrRuSi and PrRuGe compounds [271]. The heavy rare earth compounds of *R*RuGe show FM ordering below their respective $T_C$ [272, 273]. ErRuSi was found to show soft FM nature with a $T_C$ of 8 K [274]. The neutron diffraction measurements performed on TbRuGe, HoRuGe and ErRuGe reveal the non-collinear FM ordering in first two compounds and cone spiral in ErRuGe [272, 275]. Sereni et al. [276] investigated the physical properties of CeRuSi and CeRuGe. They observed that substitution of Si by Ge does not modify the interaction between 4f and conduction electrons. However, a change in local symmetry modifies the electronic structure of the intermediate crystal field levels [276].

As motioned earlier, in CeRuSn, Ce has two site such as intermediate valent Ce1 and trivalent Ce2. To confirm this, Riecken et al. [68] performed magnetic susceptibility measurement in cooling and heating modes. They observed that the compound shows slightly different transition temperature in cooling and heating data along with large thermal hysteresis, which suggests a first order valence transition for Ce. The analysis of inverse susceptibility data was found to confirm two crystallographically different Ce sites [68]. Later, Mydosh et al. [277] made thermodynamic and transport measurements of CeRuSn. They observed large hysteresis in magnetic susceptibility, the heat capacity as well as electronic and transport properties, just below room temperature, suggesting the formation of an incommensurate charge density wave modulation with temperature dependent $q$ vector. The compound shows metamagnetic transition below $T_N$=2.7 K for half of the Ce sites. After this, a number of experiments such as Hall effect, thermal expansion, NMR, [119]Sn Mössbauer, X-ray absorption/photoelectron spectroscopy, synchrotron XRD etc. were carried out, which gave a detailed information on CeRuSn [278, 279, 280, 281].

## 3.9 $R$Rh$X$ compounds

It has been found that YRhAl and LaRhAl in $R$RhAl series are superconducting below 0.9 and 2.4 K, respectively, while PrRhAl, NdRhAl and GdRhAl show FM nature below their Curie temperatures [282, 283]. One can note that YRhAl and LaRhAl adopt different structure types, though both of them show superconducting nature. This suggests that the structure in these compounds do not play a crucial role for the occurrence of superconductivity [283]. CeRhAl shows mixed valent behavior and does not show any ordering down to 1.7 K [282]. The large and negative $\theta_p$ and small $\mu_{eff}$ values estimated from CW law in CeRhAl suggest Kondo type interactions in this compound. Later Ślebarski et al. [284] reported that the heat capacity data in CeRhAl compound show a $\lambda$-type peak at 3.8 K, which suggest the AFM transition below this temperature while the magnetization measurements suggest weak FM phase. In a strong magnetic field, weak orientation of moments in GdRhAl takes place, which suggests the canted FM structure [285].

Magnetic measurements performed on $R$RhGa series, reveal that NdRhGa and GdRhGa are FM, TbRhGa is AFM, while CeRhGa does not show any magnetic ordering down to lowest measured temperature [70]. Goraus et al. [286] reported that the heat capacity at different fields show Fermi liquid state in CeRhGa. The small value of electronic coefficient ($\gamma$) hints at the 4$f$-shell instability of Ce in CeRhGa [286]. Compounds with $R$=Tb, Ho and Er in RRhGa series show collinear AFM structures with wave vector $k$=(0, 1/2, 0), which is stable in the temperature range of 1.5 to $T_N$ of respective compounds, whereas TmRhGa shows an incommensurate sine wave modulated structure at 1.5 K [71, 287].

CeRhIn shows mixed valent behavior as seen from the magnetization data. Adroja et al. [288] reported the magnetic susceptibility of CeRhIn in temperature range 4.2-300 K, with a broad maximum around 150 K, which signals the mixed valent feature of Ce compound. EuRhIn shows FM behavior with diavalent Eu [73]. It has been observed that at 4 K and 55 kOe field, it shows almost parallel spin arrangement. The magnetization measurements show trivalent state of Yb in YbRhIn [74]. The heat capacity in YbRhIn shows a rapid increase below 4 K, which moves towards higher temperature on the application of field, which suggests that some magnetic transition either FM or spin glass occurs at low temperature in this compound.

The magnetization measurements performed by Chevalier et al. [75] show FM behavior below $T_C$=55 K for TbRhSi and $T_C$= 25 K for DyRhSi, AFM ordering in HoRhSi and ErRhSi. LaRhSi shows superconducting behavior below 4.3 K [76]. Bażela et al. [289] reported the magnetic structures of TbRhSi, HoRhSi and ErRhSi. They observed that all these compounds are AFM at low temperatures and show double flat spiral magnetic structures in case of Tb and Er while collinear magnetic structure for Ho compound. Later, Szytuła et al. [290] reported that ErRhSi is an antiferromagnet below $T_N$ = 8.5 K and has two magnetic phases; one is described by collinear magnetic structure with a propagation vector $\boldsymbol{k}$ = (0, 1/2, 0) between1.5 K and $T_t$ = 2.3 K and other by a sine-wave modulated structure between $T_t$ and $T_N$. Later, Gondek et al. [291] performed neutron diffraction and magnetization experiments on TbRhSi, DyRhSi and HoRhSi and found that temperature dependence of magnetization indicates a small FM component, which disappears at 28 K. DyRhSi shows typical AFM behavior with $T_N$= 14.6 K and a transition at $T_t$=5.4 K. HoRhSi shows single transition corresponding to AFM ordering at 8.5 K. the authors reported that the neutron diffraction data reveal that the magnetic structure in TbRhSi and DyRhSi changes below their ordering temperature.

Magnetization and neutron diffraction studies on CeRhGe and NdRhGe show that these compounds are antiferromagnets and show collinear magnetic structure (C mode) for CeRhGe whereas NdRhGe can be described by the propagation wave vector k=(1/2, 0, 1/2) [117, 292]. The moments in these compounds are aligned along $b$-axis. CeRhGe shows linear magnetization, while NdRhGe shows a metamagnetic transition with increase in applied field. The single crystal magnetization and neutron diffraction studies on CeRhGe show incommensurate magnetic structure, large anisotropy in susceptibility and magnetization, with a magnetic easy axis along the *a*-axis [293]. SmRhGe shows FM ordering below 56 K and has second magnetic transition at 11.5 K [294]. GdRhGe shows two magnetic transitions; at, $T_1$=31.8 K and $T_2$= 24 K [295]. The first magnetic transition in GdRhGe is AFM in nature, while the second one seems to be complex, which can be seen from the ac susceptibility data [295]. The magnetization data show linear behavior and field induced metamagnetic transition in GdRhGe. Both transitions were also observed in heat capacity data [295]. The first peaks suppressed on the application of field, while the other (corresponding to $T_2$) smeared out in 50 kOe.

The magnetic structure of TbRhGe was found to be incommensurate, modulated spin-wave type with a propagation vector $k$=(0,0.45, 0.11) at 1.7 K, which transforms into another structure described by the propagation vector $k$=(0, 0.44, 0) [296]. HoRhGe and ErRhGe show collinear magnetic structures at T=1.6 K with a propagation vector k=(1/2, 0, 1/2) and (0,1/2,0), respectively, which changes to incommensurate sine modulated structure at 5 K for ErRhGe [297]. Magnetization measurements reveal that DyRhGe and TmRhGe show AFM ordering with field induced metamagnetic transition [298]. Neutron diffraction study of TmRhGe reveals collinear magnetic structure, where moments are coupled ferromagnetically in chemical unit cell and antiferromagnetically in magnetic cell [298]. YbRhGe is AFM with $T_N$=7 K [77]. Recently we have performed magnetization measurements on $R$RhGe ($R$=Ho-Tm) compounds and found that all these compounds are AFM at low temperature, which is also confirmed by heat capacity data [299, 300]. HoRhGe shows AFM ground state which transforms into FM state on the application of field [300]. The weak antiferromagnetic ordering was also confirmed in this compound by electronic band structure calculations [300]. Compounds with $R$=Tb, Dy and Er show an upturn in magnetic susceptibility below transition temperatures, which arises due to the change in magnetic structures at low temperature. This is also seen in previously reported neutron diffraction data. All these compounds show field induced metamagnetic transitions. There are two metamagnetic transitions in DyRhGe.

Compounds in RRhSn series show interesting magnetic properties. LaRhSn is a superconductor with $T_C$=2 K [301]. CeRhSn shows valence fluctuation and highly anisotropic behavior as shown by single crystal study [302]. A strong enhancement in the heat capacity and magnetic susceptibility data below 7 K in this compound was observed, which was reported to arise due to the presence of unquenched moments in the quasi-Kagome lattice of Ce-Rh plane [302]. Some reports show that CeRhSn exhibits non-Fermi liquid behavior. The results obtained for CeRhSn can be interpreted in terms of Griffiths phase [303, 304]. Results obtained by Higaki et al. [305] for CeRhSn are slightly different for those reported earlier. They found that 4$f$ electrons in CeRhSn have a localized nature. The elastic modulii and specific heat suggest that crystal field effect plays an important role, instead of Kondo effect. Subsequent to this, there have been many reports [306, 307, 308, 309, 310, 311], which give a detailed report on CeRhSn.

Compounds TbRhSn, DyRhSn, ErRhSn and TmRhSn show AFM ordering at low temperatures, while PrRhSn, SmRhSn, HoRhSn and NdRhSn show FM ordering [312, 313, 314, 315]. Mihalik et al. studied single crystals of NdRhSn and reported that at 9.8 K, a transition from paramagnetic to AFM state takes place, which has narrow temperature interval and goes to a FM like state, which has spontaneous magnetization and is about 2/3 of the value of free $Nd^{3+}$ moment [316]. Such behavior in NdRhSn may be due to crystal field effects or complex magnetic structure of ground state [316]. Later, a detailed report on NdRhSn appeared, which shows that NdRhSn has FM ground state ($T_t$=7.6 K), which is followed by an incommensurate AFM state between $T_N$=9.8 and $T_t$ with moments aligned along easy magnetization axis, which is $c$-axis in both the magnetic states [317]. It has been reported that the magnetic phase transitions temperatures mentioned above in NdRhSn are sensitive to hydrostatic and uniaxial pressures. In a recent report, SmRhSn is shown to exhibit two magnetic transitions; one from paramagnetic to FM (at 14.5 K) and other at 7 K [315]. GdRhSn shows non-collinear magnetic ordering as seen from dc and ac magnetization as well as [155]Gd Mössbauer spectroscopy studies [318]. Very recently, we observed that there is a sharp transition in the inverse magnetic susceptibility data at 245 K [80], which is also confirmed by heat capacity data. The low temperature XRD reveals that this transition is iso-structural.

It has been observed that TbRhSn and DyRhSn show an upturn in magnetic susceptibility data, which may be due to reorientation of magnetic moments at low temperatures [313, 319, 320, 321, 322]. TbRhSn and DyRhSn also show metamagnetic transition and is characterized by spin flop transition. Many authors have reported that HoRhSn is FM, however the neutron diffraction data show AFM behavior [323, 324, 325]. ErRhSn does not show any magnetic order down to 1.8 K, but the negative sign of $\theta_p$ suggests AFM interactions in this compound. The neutron diffraction experiment at low temperature is underway for this compound. The compounds $R$=Tb-Tm releases only 10-32 % entropy at their respective ordering temperatures, which indicates a crystal field split ground state multiplet in these compounds [313]. A hump like shape in heat capacity data above ordering temperature, suggests that PrRhSn and ErRhSn show Schottky anomaly above their ordering temperatures [313, 326]. YbRhSn shows AFM ordering below 2 K [132]. The estimation of high $\gamma$ value (1200 mJ/K$^2$ mol) from heat capacity data confirms heavy fermionic state in this compound.

In $R$RhSb compounds, LaRhSb is a BCS superconductor below 2.1 K [81]. Magnetization and heat capacity data in PrRhSb show two peaks, at 18 and 6 K. The authors [81] have reported the lower transition as FM and the upper transition as AFM. CeRhSb in this series is one of the most studied compounds in $RTX$ family. More than 100 reports are available on this compound alone and hence in this review, we only discuss the most important findings on this compound. Like CeNiSn, CeRhSb also shows semiconducting behavior. It has been observed that both these compounds adopt TiNiSi type orthorhombic crystal structure and show Kondo behavior at high temperatures with a feature of gap formation at low temperatures [327]. Though having many similar properties, these compounds show different pressure dependencies of electrical resistivity; under high pressure the gap in CeNiSn (single crystals) collapses whereas in CeRhSb, it increases [328 and ref. therein].

The Hall study in CeRhSb shows that Hall coefficient is positive and shows rapid increase along with Hall mobility below 8 K, which corresponds to gap opening in the single carrier model [328]. It has been observed that with the application of pressure in CeRhSb, the Kondo temperature increases and gap is quenched at 80 kbar [327]. The nuclear spin-lattice relaxation study of CeNiSn and CeRhSb demonstrates that the gapped state in spin excitation possesses the v-shaped structure [329]. The magnetic heat capacity in the form of $C_m$/T exhibits a maximum and substantially decreases at low temperatures, suggesting the energy gap opening in the heavy quasiparticle band [330]. A broad maximum about 113 K seen in the magnetic susceptibility data, which suggests valence fluctuation in Ce of this compound [331]. Malik et al. [332] reported that $\Delta C/T$ ($\Delta C$ =magnetic heat capacity) shows a maximum at 10 K ($T_{max}$) and is linear below it. The linear nature of $\Delta C/T$ below 10 K, which is the gapped state, is attributed to the decrease in electronic DOS with decrease in temperature [332].

In CeRhBi, the analysis of heat capacity data provides large $\gamma$, which is 500 mJ/mol K$^2$ and reveals heavy fermion behavior of this system [333]. Occurance of magnetic ordering and gap formation in CeRhBi compound has been ruled out by electrical resistivity and $^{209}$Bi-NQR measurements [333]. The magnetic susceptibility data obeys Curie Weiss law above 70 K, while the susceptibility is temperature independent up to 250 K in CeRhAs [333]. It has been observed that suppression of gap in CeRhAs take place on the application of external pressure, which may be due to the opposite effect of chemical and the external pressure on the gap size [333]. In addition to this, CeRhAs shows mixed valent semiconductor behavior with an energy gap of 144 K, which is much larger than that of CeRhSb [333].

## 3.10 *R*Pd*X* compounds

CePdAl is found to show heavy fermion behavior. The triangular coordination symmetry of Ce atoms results in geometrical frustration in this compound [334]. The neutron diffraction data show incommensurate AFM ordering and a longitudinal sine wave modulated structure below 2.7 K. The measurements performed at low temperatures give an evidence of second magnetic transition between 0.6 and 1.3 K in CePdAl [334]. Later, Keller et al. [335] reported that in CePdAl, the Ce atoms located on 3f crystallographic site show a coexistence of ordered (2/3 of Ce) and disordered (1/3 of Ce) moments. High field magnetization measurements were carried out on a single crystal CePdAl and on a powdered sample show multiple magnetization steps, which may be attributed by geometrical frustration effects [336]. Hexagonal TbPdAl shows AFM ordering below $T_N$=43 K, while DyPdAl and HoPdAl show FM nature below 24 and 6.5 K, respectively [84]. Magnetic and neutron diffraction measurements show that PrPdAl and NdPdAl exhibit double magnetic transitions [337].

PrPdAl and NdPdAl show two magnetic transitions; at $T_{N1}$=4.2 K and $T_{N2}$=1.5 K and $T_{N1}$=5 K and $T_{N2}$=4 K, respectively. The triangular coordination of $R$ atoms on 3f sites in these compounds leads to geometrical frustration and results in incommensurate AFM structures [337]. Two magnetic transitions at 50 of FM nature and 20 K due to spin reorientation were observed in GdPdAl [85]. The magnetic isotherms at temperature higher than $T_C$ show non-linear behavior, which might be attributed to critical spin fluctuations near the transition temperature or to the 4$d$ band (of Pd) polarization [85].

The magnetic susceptibility shows two magnetic transitions at $T_{N1}$=43 K and $T_{N2}$=22 K, indicating the AFM transition and is insensitive to iso-structural transition in TbPdAl [86]. The magnetic susceptibility data show TMI, arising due to frustration effects [86]. Donni et al. [338] performed the neutron diffraction measurements on TbPdAl samples showing hexagonal and orthorhombic structure modifications, which were synthesized with different heat treatments. They observed that hexagonal TbPdAl which shows geometrically frustration transforms into non-frustrated orthorhombic modification on heat treatment. In orthorhombic form, the magnetic structure below $T_{N1}$ show incommensurate modulation with a propagation vector $k_1$=(0.271, 0.523, 0) which goes to a lock-in transition to a commensurate structure with $k_2$=(0, 1/2, 0) at $T_{N2}$. In hexagonal TbPdAl, the rare earth magnetic moments form a triangular lattice in basal plane, which results in geometrically frustrated structure. At $T_{N1}$, only 2/3 of non-frustrated Tb moments order with commensurate structure with a propagation vector $k_1$= (1/2, 0, 1/2) while 1/3 of frustrated Tb moments transform into commensurate structure [$k_1$= (1/2, 0, 1/2)] below $T_{N1}$ to a complete incommensurate magnetic structure with $k_2$ =(1/2, 0.12, 1/2) below $T_{N2}$ [338]. The magnetization isotherms of TbPdAl show metamagnetic transition, which can destroy the AFM structure and induces a FM like state with field application [339]. DyPdAl shows two magnetic transitions at 25 and 17 K, similar to other compounds in this series [340]. The magnetization measurements performed on single crystals of DyPdAl show strong magnetic anisotropy [341]. The magnetization recorded along $a$-axis shows hysteresis and multiple metamagnetic transitions, while these effects were not observed when it was measured along $c$-axis.

The temperature dependence of magnetization in DyPdAl and HoPdAl suggests AFM character along *a*-axis and FM character along *c*-axis [341, 342]. Xu and Shen [343] studied the magnetic and magnetocaloric properties of both hexagonal and orthorhombic HoPdAl. They reported that these compounds show AFM ordering at $T_N$= 12 and 10 K, respectively. A second transition at $T_t$=4 K was also observed in hexagonal HoPdAl, while there is only a single transition in the orthorhombic form [343]. Both compounds show field induced metamagnetic transition with a critical field value in hexagonal HoPdAl smaller than that of orthorhombic HoPdAl.

CePdGa is a trivalent Kondo antiferromagnet below 2.2 K [344]. The compound follows Curie-Weiss law above 180 K and deviates below, due to crystal electric field effects [345]. Magnetic and neutron diffraction studies reveal that *R*PdGa (*R*=Tb-Er) show complex magnetic structure [346]. At 1.5 K, TbPdGa exhibits a complex magnetic structure, while HoPdGa and ErPdGa show collinear AFM structure, which transforms into sine wave modulated structure at 26 K in TbPdGa, at 6.8 K in HoPdGa and at 5 K in ErPdGa [346]. DyPdGa and GdPdGa show FM order below 20 K and AFM ordering below 5.1 K, respectively. Magnetization measurements reveal that all these compounds show additional anomalies at $T_t$ =24 K, 3.9 K, and 2.1 K for R=Tb, Ho and Er compounds [346].

The C/T data shows a remarkable increase below 10 K, indicating heavy fermion nature of CePdIn and a sharp peak at 1.7 K signaling the AFM ordering [347]. Very large negative $\theta_p$ value, which is much larger than its $T_N$=1.3 K hints at a strong Kondo interaction in this compound [348]. Fujii et al. [349] reportet that CePtIn is an antiferromagnetic material with $T_N$=1.8 K, while the complete replacement of Pd by Pt does not show any magnetic ordering. PrPdIn does not show any magnetic ordering, whereas the large value of temperature independent magnetic susceptibility is of the order of those characteristic of strong spin fluctuators [348]. Nishigori et al. [350] reported that TbPdIn and DyPdIn are spin glasses due to spin frustration of rare earth ions forming a deformed Kagome lattice. Magnetization measurements performed on *R*PdIn (*R*=Gd-Er) reveal that except GdPdIn, which is a ferromagnet, all other compounds show ferrimagnetic ordering with an additional transition below their $T_C$ [87]. The authors [87] claimed that TMI observed in these compounds is not due to spin glass state. Later, Gondek et al. [351] performed neutron diffraction measurements on NdPdIn, HoPdIn and ErPdIn and reported that all these compounds are FM. The moments in HoPdIn and ErPdIn are parallel to *c*-axis, while in NdPdIn the moments change their direction from basal plane to *c*-axis [351]. NdPdIn shows two magnetic transitions, one at $T_C$=30 K, which is FM in nature and occurs due to the ordering of Nd moments, whereas the other is due to the reorientation of the easy axis [351]. Later, Li et al. [352] observed two magnetic transitions in HoPdIn consistent with Ref. [296]. They reported that first transition is para-ferro and the second transition arises due to spin reorientation.

Li et al. [353] carried out magnetic measurements along with high field magnetization and showed that GdPdIn and TmPdIn are FM and AFM, respectively, having single magnetic phase transition. TmPdIn shows no TMI and shows field induced metamagnetic transition. Li et al. [354] reported later that PrPdIn and NdPdIn exhibit FM ordering at $T_C$= 11.2 and 34.3 K, respectively. An additional transition occurs in NdPdIn at 18.3 due to the AFM coupling. The TMI seen in these compounds below their ordering temperatures may be the consequence of the domain wall pinning effect or the magnetic frustration due to non-collinear magnetic structure [354]. SmPdIn is an extremely hard magnetic material [355]. The magnetic entropy estimated from heat capacity data was found to be 80 % of theoretical value (Rln6), which reveals that most of spins take part in magnetization and leads to FM phase in this compound [355]. Cirafici et al. [89] reported the susceptibility data in the temperature range 100-1000 K for EuPdIn. Later Pöttgen [137] reported the susceptibility measurement down to 4.2 K for EuPdIn and estimated magnetization parameters which are different from the earlier report. He reported that the compound orders antiferromagnetically and transforms to FM state on the application of field. Neutron diffraction study on TbPdIn reveal non-collinear magnetic structure below 66 K with several additional small reflections in the ordered state, which can be indexed by multiple propagation vectors [356]. DyPdIn shows two magnetic phases; FM in the temperature between 15 to 31 K with moments along *c*-axis and AFM below 15 K [356]. Later it was reported that TbPdIn and DyPdIn are ferromagnets below $T_C$=74 K and 38 K, respectively with metastable magnetic properties [357]. Different measurements performed on these compounds confirm that the TMI observed in these compounds arises due to domain wall pinning effect [357]. The authors [357] reported that DyPdIn shows another magnetic phase transition at $T_N$=23 K characteristic of AFM coupling. Later, Javorský et al. [358] reported that magnetic structure of TbPdIn can be described by long range magnetic ordering with a wave vector (0, 0, 0). YbPdIn shows anomalous behavior because of valency variation of Yb, which changes from diavalent to trivalent with increase in temperature [89].

The magnetic susceptibility data in CePdSi and PrPdSi show field dependence at low temperatures, suggesting ferro- or ferrimagnetic ordering with transition temperatures 7 and 5 K, respectively [91]. These compounds show negative $\theta_P$, indicating complex magnetic ordering. EuPd(Pt)Si show no clear magnetic ordering from susceptibility data however [151]Eu Mössbauer data shows hyperfine splitting at 4.2 K in EuPtSi, hinting the magnetic phase transition at this temperature [92]. YbPdSi orders ferromagnetically below 8 K [359]. The heat capacity of this compound suggests the role of only the ground state doublet in the magnetic ordering and heavy fermion behavior at low temperatures as evidenced by the enhanced value of $\gamma$ (300 mJ/mol K$^2$).

Kotsanidis et al. [360] studied magnetic and neutron diffraction data of RPdGe (R=Pr, Nd, Tb, Dy, Er), which show AFM nature of these compounds. Others with R=Gd, Ho and Tm do not show any signature of magnetic ordering down to 3.6 K. At 1.8 K, the magnetic structure of NdPdGe is characterized by incommensurate sine wave modulated type with propagation vector $k$=(0, 0, 0.01) [360]. Later, it was reported that CePdGe shows sine modulated magnetic structure below $T_N$=3.4 K and TbPdGe shows magnetic ordering below 35 K with a non-collinear magnetic structure at 2 K, which changes the direction of moments at 20 K [361]. After that, Penc et al. [362] reported that RPdGe (R=Gd, Dy, Er) compounds are antiferromagnets, while HoPdGe is paramagnetic down to 1.6 K. DyPdGe shows non-collinear magnetic structure below ordering temperature, which is stable with temperature, while ErPdGe shows complex magnetic structure which shows variation with temperature [362]. EuPdGe shows AFM ordering below 8.5 K with an effective magnetic moment close to free Eu$^{2+}$ ion [363]. Compound shows metamagnetic transition at 15 kOe and shows almost saturation of magnetization for a field of 55 kOe.

Magnetic properties show that RPdSn (R=La-Nd, Sm, Gd-Ho) are AFM [364]. Adroja and Malik [95] reported AFM ordering in RPdSn (R=Ce, Sm, Eu, Gd-Dy, Er) compounds and did not observe any magnetic ordering in the compounds with R=Pr, Nd, Ho and Tm down to 4.2 K. YbPdSn show Curie –Weiss behavior in the temperature range of 150-300 K and deviates from it below 150 K [95]. Due to crystalline electric field effects in RPdSn compounds, $T_N$s of these compounds do not follow de Gennes scaling [95]. Large negative value of $\theta_p$ in case of CePdSn implies strong hybridization between Ce $4f$ and conduction electrons [95]. Neutron diffraction results show spiral magnetic structure in CePdSn [365]. Later, Zygmunt and Szytuła [366] reported magnetic ordering in PrPdSn, NdPdSn and HoPdSn at 4.3, 2.4 and 3.7 K, respectively. They observed additional anomalies in ErPdSn and DyPdSn. The neutron diffraction data show collinear magnetic structure for PrPdSn and sine modulated for NdPdSn and TbPdSn, which changes below 10 K for TbPdSn in which the magnetic moments form an AFM conical spiral [367].

Neutron diffraction studies of EuPdSn show incommensurate antiferromagnetic structure at 13.2 and 3.6 K, which can be indexed by propagation vectors $k_1$=(0, 0.217, $q_z$), where $q_z$ ≤0.02 and $k_2$=(0,0.276,0), respectively [368]. Magnetic properties of single crystalline TbPdSn show three successive magnetic transitions at $T_N$=23.8, $T_1$=12.2 and $T_2$=2.5 K with multistep metamagnetic transitions in the three ordered phases along $b$-axis, which is the easy axis of magnetization [369]. Magnetization and neutron diffraction of single crystals of HoPdSn show two magnetic transitions at $T_1$=2.5 and $T_N$=3.6 K with two types of magnetic structures, which can be described by wave vectors; $k_1$=(1, 0, $k_z$) and $k_2$=(0.33, 0, 0.93) in which $k_z$ component varies discontinuously from 0.38 to 0.35 at 2.5 K, whereas neutron diffraction in polycrystalline sample shows a magnetic structure characterized by the wave vector k=(1/3, 1/2, 1/3) [370]. Andre et al. [371] reported an AFM transition in ErPdSn at 5.2 K. They found that below $T_1$=2.5 K a square wave modulated structure exists, which changes to sine modulated structure above $T_1$.

Malik and Adroja [99] reported the magnetic properties of $R$PdSb ($R$=La-Tm) in the temperature range of 4.2- 300 K. They observed FM ordering in CePdSb and AFM ordering in NdPdSb, SmPdSb, EuPdSb, and GdPdSb, while others were found to be paramagnetic down to 4.2 K. Later, it has been observed that $R$PdSb ($R$=Ce, Pr, Sm) show FM ordering, while compounds with $R$=Nd, Gd, Tb, Dy and Ho are AFM at low temperatures [366]. A ferromagnetic Kondo state has been reported in CePdSb by many authors. This seems to be unusual because generally Kondo systems show AFM ground state and CePdSb has been reported to be half metallic FM [372]. It has been observed that the application of pressure on susceptibility data shows an increase in $T_C$ of CePdSb [373]. The magnetic heat capacity data shows absence of peak near $T_C$; however, a broad maximum near 9 K has been observed [374]. The magnetic properties of single crystalline CePdSb show anisotropy, which is well explained by CEF model by taking into account the strong exchange interaction along $c$-axis compared to that within the $c$ plane [375]. These results suggest that FM coupling in CePdSb along c-axis plays an important role to realize the one dimension like magnetic order just below $T_C$ [375]. Magnetometric and neutron diffraction experiments reveal no long range magnetic order in PrPdSb at 1.5 K and a sine wave modulated magnetic structure in NdPdSb, which transforms into square modulated structure below 5.8 K [376]. GdPdSb was found to show two magnetic transitions; one AFM at $T_N$=13.1 and other FM at $T_C$=8 K [377]. Magnetic measurements show that ErPdSb is paramagnetic while YPdSb is a weak diamagnet [378]. The studies show that CePdSb, GdPdSb, ErPdSb and YPdSb are half metallic in $R$PdSb series.

In the $R$PdBi series, GdPdBi, DyPdBi and HoPdBi are AFM, YPdBi is diamagnetic, while ErPdBi is paramagnetic throughout the temperature under study [100]. CePdBi shows a magnetic transition at 2 K and the gradual drop in susceptibility data below this can be associated with AFM ordering [379]. At the lowest temperature, ac susceptibility exhibits diamagnetic behavior hinting the superconducting state in this compound. Below 5 K, the ac susceptibility shows frequency dependence and shows an anomaly at 2.5 K, implying a spin-glass like state, which may attributed to atomic disorder along with the clusters formation of Ce and Pd moments within the cubic lattice [379]. The significant atomic disorder observed in this compound by various measurements is the typical characteristic of Heusler alloys. The curvature in the inverse susceptibility data and the fact that experimental $\mu_{eff}$ is lower than the free $Ce^{3+}$ moment suggest the presence of CEF and Kondo effect in this compound [379]. GdPdBi which is known as half Heusler alloy, is an antiferromagnet below $T_N$=13 K, whose $\theta_p$ is 3.7 times larger than $T_N$, which indicates partly frustrated $Gd^{3+}$- $Gd^{3+}$ magnetic interaction [380]. Recently Pan et al. [381] reported superconductivity at $T_C$=1.22 K and magnetic order at $T_N$=1.06 K in half Heusler ErPdBi. Overall, it has been observed that in RPdSb and RPdBi compounds with MgAgAs type structure are also known as half Heusler phase and found to show many intriguing and excellent properties such as half metallic behavior, semiconducting behavior, giant magneto-resistivity and heavy fermion like behavior [100 and ref. therein]. These compounds provide a unique opportunity to study the interplay of AFM order, superconductivity and topological quantum states [381].

### 3.11 *R*Ag*X* compounds

The compounds in *R*AgAl (*R*=Pr, Tb, Er) series show TMI in susceptibility plots and show a maxima in ZFC magnetization below their $T_C$s, which shifts to lower temperature on the application of higher fields [102]. The ac susceptibility in these compounds show frequency dependence which hints of spin glass state and the heat capacity in these compounds show broad peaks around the $T_C$ which reveals the inhomogeneous magnetic state [102]. Thus, these compounds show a complex magnetic state. Similar results were also seen in GdAgAl, which shows a spin glass state at low temperatures and follow the Curie-Weiss law at high temperatures [382]. The $\mu_{eff}$ estimated from Curie Weiss fit (8.3 $\mu_B$/f.u.) was found to be slightly higher than free $Gd^{3+}$ moment, which implies polarization of 3d electrons of Ag [382 and ref. therein]. TMI in magnetization data have been also been seen in RAgAl (*R*=Dy-Er) compounds [103]. CeAgAl is found to show FM ordering [383]. The heat capacity data in this compound suggest a spin glass state while magnetization and resistivity data rule out such a state in this compound [383]. Later, Singh et al. [384] reported that magnetization data in TbAgAl show competing interactions, which leads to the formation of Griffiths-like phase. These authors also reported that the temperature and time dependence of magnetization along with heat capacity data suggest the glassy state in this compound [384]. As can be seen from the Table II, almost all the compounds in *R*AgAl series have $\theta_p$ smaller than the respective ordering temperature, indicating competing FM and AFM interactions, which can also cause the occurrence of TMI [384 and ref. therein].

*R*AgGa compounds with *R*=Gd, Ho and Er show FM nature, TbAgGa is an antiferromagnet, while DyAgGa shows metamagnetic behavior [104]. No compounds in these show saturation, which indicate that magnetization process in these compounds is complex and characteristic of non-collinear spin structure [104]. All compounds have positive $\theta_p$, which indicates that the FM interactions are dominated in these compounds. Adroja et al. [385] reported that CeAgGa is a ferromagnet at about 5.5 K and does not follow Curie-Weiss law below 70 K. Goraus et al. [386] reported the CeAgGa shows spin glass like behavior with at $T_f$=5.1 K and a FM ordering at $T_C$=3.6 K. The Arrott plots shown by these authors do not reveal FM characteristics and suggest inhomogeneous magnetic phase below 5 K. The heat capacity in CeAgGa does not show λ- shaped anomaly while C/T shows a maximum at 3.95 K and a shoulder at about 5 K [386]. The field dependence of heat capacity shows FM/spin glass behavior [386]. Later, Zygmunt et al. [387] reported spin glass characteristic of TbAgGa and FM behavior in DyAgGa and HoAgGa. The neutron diffraction measurements performed by these authors show sine wave modulated magnetic structure in TbAgGa and FM ordering in DyAgGa and HoAgGa compounds. Silva et al. [388] reported slightly different value of $T_C$ that reported by other authors and stated that the compound shows negligible hysteresis therefore is promising as a magnetic refrigerant at low temperatures.

Magnetometric and neutron diffraction data show AFM ordering in $R$AgSi ($R$=Gd-Er) compounds and magnetization isotherms show field induced metamagnetic transitions (except in Er) [389]. TbAgSi was found to show an additional magnetic transition below $T_N$ [389]. The ZFC magnetization shows a broad maximum at 11 K followed by a sharp jump at 44 K, which becomes more pronounced in ac susceptibility data in NdAgSi [390]. Baran et al. [391] reported that TmAgSi is AFM below $T_N$=3.3 K. The magnetic structure in TmAgSi is non-collinear, which can be characterized by three wave vectors such as; $k_1$= (1/2, 0, 0), $k_2$= (-1/2, 1/2, 0), and $k_3$= (0, -1/2, 0).

Magnetization and heat capacity data show a peak at 4.8 K in CeAgGe, which is assumed to be corresponding to an AFM transition [106]. The neutron diffraction confirms the AFM arrangement of moments, which lie within the *ab* plane [392]. Neutron diffraction measurements show sine modulated structure for TbAgGe, square modulated structure for DyAgGe and HoAgGe and collinear AFM structure in ErAgGe [107]. Magnetic structure in TbAgGe is affected by temperature changes [107]. Magnetic measurements performed on GdAgGe show AFM ordering below 15.6 K [107]. TbAgGe and DyAgGe show some additional anomalies in magnetization data at temperature below $T_N$ [107]. Polycrystalline TbAgGe, DyAgGe and HoAgGe show field induced metamagnetic transition [107]. EuAgGe shows magnetic phase transition below 18 K, which can be characterized as a spin glass transition [393].

Morosan et al. [108] performed magnetic and thermodynamic measurements on single crystals of $R$AgGe ($R$=Tb-Lu) compounds. All these compounds show AFM transitions.

TbAgGe, DyAgGe and HoAgGe show multiple magnetic transitions. One or more metamagnetic transitions were observed with the field application along $c$-axis (for TbAgGe) or perpendicular to it (for HoAgGe, ErAgGe and TmAgGe), whereas DyAgGe shows metamagnetic transition in both the directions [108]. It has been observed that TmAgGe shows extreme anisotropy of magnetization [108]. YbAgGe shows enhanced $\gamma$ value and magnetic ordering below 1 K, which hints that this compound is close to a quantum critical point [108]. Later, it was reported that YbAgGe shows heavy fermion and field induced non-Fermi liquid behavior and has two magnetic transitions below 1 K and [394]. Recently, Tokiwa et al. [395] reported quantum bicriticality in heavy fermion YbAgGe compound. The points at which two distinct symmetry-broken phases become simultaneously unstable are called bicritical points and are typical for spin-flop metamagnetism [395]. Baran et al. [396] studied the physical properties of TmAgGe and reported an AFM ordering below $T_N$=4.2 K. The neutron diffraction data show a non-collinear magnetic structure within the basal plane, which can be characterized by following wave vectors: $k_1$= (1/2, 0, 0), $k_2$= (-1/2, 1/2, 0), and $k_3$= (0, 1/2, 0).

Most of compounds in $R$AgSn are AFM at low temperatures [397, 398]. Compounds with $R$=Nd, Tb and Ho show collinear AFM structure with the orthohexagonal magnetic unit cell and multistep metamagnetic transition [398, 399]. Baran et al. [112] carried out magnetization and neutron diffraction measurements on RAgSn (R=Ce-Nd, Gd-Er) and results show sine modulated magnetic structure in PrAgSn (between T=1.6 to $T_N$=3.6 K), while incommensurate sine modulated magnetic structure near 10.2 and 5.6 K in HoAgSn and ErAgSn compounds, respectively [112]. No magnetic ordering has been seen in TmAgSn down to 2 K [400]. The magnetization data reveal AFM ordering ($T_N$=4.3 K) and diavalent Eu in EuAgSb [61]. The compound shows difference between ZFC and FC data at $T_N$, which can be attributed to an unstable AFM ground state.

### 3.12 $R$Ir$X$ compounds

Magnetic measurements performed on $R$IrAl ($R$=Y, La-Nd) reveal mixed valent state with a very high Kondo temperature (1300 K) for CeIrAl, FM state for PrIrAl and NdIrAl and Pauli paramagnetic behavior in YIrAl and LaIrAl compounds [401].

LaIrSi shows a superconducting transition at 2.3 K [76], NdIrSi is a ferromagnet with $T_C$=10 K and does not show saturation up to 20 kOe, which suggest non-collinear magnetic ordering below $T_C$ [76]. AFM ordering has been observed in $R$IrSi ($R$=Tb-Er) compounds at low temperatures as revealed by magnetization and neutron diffraction measurements [120]. TbIrSi and ErIrSi show sine wave modulated magnetic structures at 1.5 K [120]. HoIrSi shows collinear AFM structure at 1.5 K, which co-exist with sine modulated structure near $T_N$=4.6 K. Field dependence of magnetization shows one-step metamagnetic transition in TbIrSi and two-step metamagnetic transitions in HoIrSi and DyIrSi [402]. DyIrSi shows multiple magnetic transitions [402]. The large difference in ZFC and FC curves was observed below and above $T_N$ in $R$IrSi ($R$=Tb, Dy, Ho), indicating short range order above $T_N$ [402].

TbIrGe and ErIrGe show sine modulated magnetic structures below $T_N$ [116]. Below $T_t$=2.9 K, the collinear and sine modulated structures coexist in ErIrGe and no magnetic ordering has been observed in HoIrGe down to 1.5 K [116]. Compounds with $R$=Gd, Tb and Dy in $R$IrGe series show metamagnetic transition on application of field [116]. CeIrGe is an intermediate valent system and is paramagnetic down to 1.5 K [117]. YbIrGe shows AFM ordering below $T_N$=2.4 K. The estimation of magnetic entropy at $T_N$ from heat capacity shows a gain of 85 % of the expected $R$ln 2, which rules out heavy fermion behavior in this compound [119].

CeIrSb shows non-Fermi liquid behavior and a very large negative $\theta_p$ of -1300 K, which is much larger than in CeRhSb. This suggests that $c$-$f$ (conduction band - 4$f$ band) hybridization in CeIrSb is much stronger than that in CeRhSb [118]. YbIrSn shows AFM ordering at $T_N$=3.2 K with Yb ion in trivalent state [403].

### 3.13 $R$Pt$X$ compounds

Magnetic measurements on compounds in $R$PtAl ($R$=Ce-Nd) show large magnetocrystalline anisotropy and reveal that dominant FM contribution, in which moments align along $a$-axis for CePtAl, PrPtAl and $b$-axis for NdPtAl [404]. CePtAl has multiple magnetic transitions at $T_C$=5.9 K, $T_2$=4.3 K, and $T_3$=2.5 K and its magnetic structure can be characterized by two coexisting wave vectors [404]. The complex magnetism in CePtAl may be attributed to competing anisotropic exchange interactions [404]. NdPtAl shows FM ordering below $T_C$=19.2 K. It has been reported that PrPtAl has a non magnetic CEF ground state singlet with its first excited state at 21 K from ground state where long range magnetic ordering take place $T_C$=5.8 K because of strong enough exchange interactions [404, 405]. The high field magnetization measurements on single crystals of PrPtAl reveal a large magnetic anisotropy and show an induced-moment ferromagnet parallel $a$-axis [406]. Drescher et al. [407] have done pressure study on antiferromagnetic Kondo lattice YbPtAl and reported that the compound shows complex magnetic state at ambient pressure and an unusual volume induced change of $T_N$. The authors suggested that the anomalous volume dependence of ordering temperature is attributed to the interplay between frustrated anisotropic exchange interactions and magnetocrystalline anisotropy. The neutron diffraction measurement of YbPtAl shows incommensurate magnetic structure below $T_N$ in this compound [408].

Kotsanidis et al. [409] reported the magnetic properties of $R$PtGa ($R$=La-Nd, Gd-Tm) compounds in the temperature range of 5-300 K. These results show that magnetic ordering is present in $R$=Gd, Tb, Dy and Tm, while it is absent in Ce, Pr, Nd, Ho and Er down to 5 K. From the magnetic susceptibility plots of these compounds, it seems that compounds with $R$=Gd, Tb, Dy and Tm are antiferrromagnets at low temperatures. Penc et al. [410] carried on the magnetic and neutron diffraction measurements on these compounds, which reveal collinear magnetic structure for TbPtGa, DyPtGa, HoPtGa at 1.5 K, which transforms into sine wave modulated structure at 18 K for TbPtGa, at 7 K for DyPtGa and at 2.4 K for HoPtGa, while ErPtGa shows sine wave modulated structure between 1.5 and 3.2 K ($T_N$). Compounds with $R$=Tb-Ho also show metamagnetic transition on the application of field [410]. Single crystalline CePtGa shows anomaly at 3.5 K, which corresponds to AFM ordering [411]. The compound shows metamagnetic transition along $a$ and $c$ axes with higher critical field along $c$ axis. The magnetic anisotropy in susceptibility data of CePtGa suggests the existence of CEF [411]. The susceptibility data in YbPtGa shows a broad peak at 4 K, while the heat capacity data in this compound shows $\lambda$- type anomaly at 3.8 K [412].

Remarkable increase in the C/T data below 10 K indicates heavy fermion nature of CePtIn [347, 413]. The heat capacity data do not give any signature of magnetic ordering in CePtIn down to 50 mK [413, 347]. PrPtIn remains paramagnetic down to 1.7 K, while SmPtIn shows FM ordering below $T_C$=25 K [123]. The magnetization in SmPtIn shows TMI [123]. Magnetic properties of single crystalline $R$PtIn ($R$=Y, Gd-Lu) reveal that $R$=Tb and Tm compounds order antiferromagnetically at $T_N$=46 and 3 K, respectively, while the ones with $R$=Gd, Dy-Er seem to possess at least a FM component of the magnetization along $c$ axis [414]. It has been reported that the magnetic ordering temperature of all the compounds follow the de Gennes scaling, whereas the switching from FM to AFM ordering is correlated with the change in anisotropy [414]. The anisotropic susceptibilities of YPtIn and LuPtIn show temperature independent behavior [414]. The TbPtIn in this series shows extremely anisotropic behavior with moments confined to ab plane [414]. Angular dependent magnetization measurements were performed on single crystals of TbPtIn reveals the easy axis of magnetization, which coincides with high symmetry directions [120] for TbPtIn [415]. The neutron diffraction data reveal that DyPtIn and HoPtIn compounds show FM component parallel to $c$ axis and an AFM one in the $ab$ plane at 2 K [416]. In HoPtIn, the AFM component disappears with increasing temperature and remains FM up to the ferro-paramagnetic transition [416]. The magnetization data show that both compounds show TMI [416]. The neutron diffraction in TbPtIn shows an AFM structure described by wave vector $k$=(1/2, 0, 1/2) [417]. The neutron diffraction shows FM ordering in ErPtIn [418] below 13 K with moments along $c$-axis, while non-collinear triangular AFM structure with a wave vector, $k$=(1/4,1/4,1/2) in TmPtIn [419] with moments confined in basal hexagonal plane.

It has been observed that CePtSi shows heavy fermion, which does not show any magnetic ordering down to 70 mK and has enhanced $\gamma$=800 mJ/mol K$^2$, which also confirms the Kondo behavior of this compound [420]. In $R$PtSi series, compound with $R$=Nd shows AFM ordering below $T_N$=3.8 K, $R$=Sm shows FM ordering below 15 K and $R$=La shows superconductivity below 3.8 K [421]. The heat capacity data in NdPtSi and SmPtSi show large contribution from CEF effect [421]. Isothermal magnetization data show metamagnetic transition at 2 K in NdPtSi. HoPtSi, HoPtGe, ErPtSi and ErPtGe show collinear magnetic structures with propagation vectors $k$=(1/2, 0, 1/2) and (0, 1/2, 0), respectively [422]. With increase in temperature, the magnetic structure in ErPtSi snf ErPtGe transforms to sine wave modulated [422]. Except GdPtSi, which is ferromagnet below 16 K, all other compounds GdPt$X$, TbPt$X$ and DyPt$X$ ($X$=Si, Ge) are AFM at low temperatures [423]. The neutron diffraction data reveal complex magnetic structure below $T_N$=12.5 K in TbPtSi, amplitude modulated structure in TbPtGe below $T_N$=15 K, a collinear AFM structure below in DyPtSi (8.2 K) and DyPtGe (8 K) [423]. All TbPt$X$ and DyPt$X$ compounds show metamagnetic transition as revealed by their magnetization isotherms [423].

Neutron diffraction report on $R$PtSn ($R$=Ce, Tb, Ho) reveals that all these compounds are AFM at $T_N$=8, 13.8 and 9.2 K, respectively and show sine modulated magnetic structures, which changes at $T_t$=10 and 3 K for TbPtSn and HoPtSn, respectively [424]. Magnetization data show that these compounds show further anomaly at $T_t$= 5.2 K for CePtSn, 10 K for TbPtSn and 3 K for HtPtSn and field induced metamagnetic transition in TbPtSn [424]. It has been reported that in CePtSn, magnetic structure can be described by two wave vectors [$k_1$=(0,0.470,0) and $k_2$=(0, 0.428,0)] at 2 K [425]. Subsequently, it was reported that neutron diffraction [425] and μSR [426] show contradictory results, which has been resolved by a spin slip model [427], in which the $k$ vector comes to be incommensurate for both phases because of periodic occurrence of spin slip planes within the commensurate phase with $k$=(0,1/2,0) [428]. Later, Szytuła et al. [429] studied magnetic and neutron diffraction properties in $R$PtSn ($R$=Dy-Er) compounds and found that all these compounds are AFM and has collinear AFM order in DyPtSn and HoPtSn compounds. The HoPtSn shows spiral AFM order below 3 K, while ErPtSn shows an AFM ordering below 3.5 K [429]. Magnetization data show three anomalies in HoPtSn and field induced metamagnetic transtitions in DyPtSn and HoPtSn [429]. CePtSn is found to show very interesting properties in this series. The compound show very interesting and unusual behavior when a magnetic field is applied along $b$-axis, in which the AFM structure moves incommensurably with the crystal lattice and leads to the creation of the spin-slips [430]. The magnetization study of single crystalline CePtSn shows that two magnetic transitions are rather insensitive to application of field along $c$-axis and show changes when the field is applied along $a$ and $b$ axes [431]. PrPtSn is paramagnetic down to 0.4 K, which may be due to the singlet ground state of $Pr^{3+}$ ion resulting in van Vleck behavior below 20 K [432]. NdPtSn shows two AFM transitions at $T_N$=2.4 K (second order) and $T_M$=1.9 K (first order) and a field induced metamagnetic transition along $c$-axis [433]. The high-pressure modification of NdPtSn shows spin glass like state at 4.5 K, which is confirmed by heat capacity data [130]. The heat capacity data suggest antiferromagnetic character of SmPtSn with $T_N$=3.5 K [434]. The heat capacity data in TmPtSn show one sharp peak at 0.9 K and other broad peak at 1.8 K, which might be due to two consecutive magnetic transitions at 1.7 and 1.9 K [435]. YbPtSn orders antiferromagnetically below 3.5 K [132].

*R*PtSb (*R*=Ce-Nd, Sm, Gd, Yb) compounds show magnetic ordering at low temperatures, however the nature of magnetic order was not revealed by the authors [133]. From susceptibility plots it seems to be FM ordering in *R*=Ce, Pr, Nd, Sm compounds. Single crystal studies in CePtSb shows strong anisotropy between H parallel to *a* and H parallel to *c* modes [375]. The magnetization measurements show saturated magnetic moment of 0.91 μ$_B$ for CePtSb, which lie within the *c* plane in the FM state [375]. Matsubayashi et al. [436] reported an AFM ordering at T$_N$=0.34 K in YbPtSb. The authors also reported that the large value of γ and A coefficients estimated from the heat capacity and resistivity data, respectively, suggest Fermi liquid state in this compound.

The compounds of *R*PtBi series have been considered as a potential topological insulator, which opens a new era in the solid state physics [437]. The topological insulator shows a no. of applications including quantum computing [437 and ref. therein]. Liu et al. [437] reported studies on three members of this series viz: GdPtBi, DyPtBi and LuPtBi using angle resolved photoemission spectroscopy and discussed metallic surface electronic structure of these compounds, which show a clear spin orbit splitting of the surface band that crosses the chemical potential and Kramer's degeneracy of spin at the Γ and M points in these half-Heusler compounds.

### 3.14 *R*Au*X* compounds

In *R*AuAl series, LaAuAl shows Pauli parmagnetic behavior, CeAuAl orders antiferromagnetically below 3.8 K and NdAuAl shows FM order with a *T$_C$*=10 K [134]. NdAuAl shows a broad peak around 10 K in ZFC magnetization, which gradually decreases at low temperatures because of large anisotropy of the sample, while the FC magnetization shows behavior of typical FM compound [134]. The large negative θ$_p$, as generally observed in Ce compounds, suggests the presence of Kondo type interaction or CEF effect in CeAuAl compound. *R*AuIn (*R*=Ce, Gd-Er) compounds show AFM ordering below their *T$_N$* [135, 136, 438]. Sill and Hitzman [439] performed magnetization measurements for RAuGa (R=Sm, Gd-Tm) in the temperature range 2.5-300 K and found that only GdAuGa shows AFM below 6 K, while others do not show. The magnetic structure in CeAuIn, TbAuIn and DyAuIn can be characterized by wave vector *k*=(0,0,1/2), while the magnetic structure of HoAuIn is characterized by two wave vectors *k$_1$*=(0,1/2,1/2) and *k$_2$* =(0,0,1/2) [135, 136]. It has been observed that TbAuIn shows spin glass behavior between $T_N$ and $T_{SG}$=58 K and two-step metamagnetic transition [135]. ErAuIn is found to show triangular AFM magnetic structure [136]. Szytuła et al. [440] reported that $R$=Ce, Tb, Dy and Er compounds in RAuIn series can be described by non-collinear magnetic structure and show geometrical frustration. EuAuIn shows AFM ordering below 21 K and metamagnetic transition on the application of field [137].

YbAuGe and LuAuGe show diamagnetic behavior, whereas LaAuGe and PrAuGe show paramagnetic nature. $R$AuGe ($R$=Nd, Gd-Er) compounds show AFM ordering [138, 139, 441]. The neutron diffraction measurements in $R$AuGe ($R$=Pr, Nd, Tb, Ho, Er) compounds reveal that PrAuGe does not show any additional magnetic reflections at 1.6 K, while magnetic structure of NdAuGe and ErAuGe show the magnetic structure, which can be described by two propagation vectors [442]. HoAuGe shows incommensurate magnetic structure below $T_N$, while the magnetic structure in TbAuGe was reported to be complex [442]. Later, Baran et al. [443] reported the neutron diffraction results in HoAuGe and ErAuGe. These results show that at 1.5 K, both HoAuGe and ErAuGe show collinear magnetic structure characterized by wave vector $k$=(1/2, 0, 0), which transforms into transverse sine wave incommensurate magnetic structure at 4.5 and 3 K, respectively. However, in ErAuGe some magnetic reflections show significant broadening attributed to magnetic domain size effect [443]. Among all the compounds of RAuGe, only CeAuGe shows FM ordering below $T_C$=10 K, which is confirmed by neutron diffraction study [392, 441]. Recently, Bashir et al. [444] reported that NdAuGe shows bifurcation at 7.3 K, which is well above $T_N$ (3.7 k), and may arise from an inhomogeneous magnetic ground state, which may result in a spin glass like state.

The magnetometric and neutron diffraction results show AFM ordering in $R$AuSn ($R$=Gd-Ho) with the magnetic structure characterized by propagation vector $k$=(1/2, 0, 0) [141]. The hexagonal $CaIn_2$ - type form of ErAuSn has a complex antiferromagnetic structure with $T_N$=12.1(2) K [140] while the cubic one does not show any long range magnetic order down to 1.6 K [141]. Later, Baran et al. [445] reported neutron diffraction data on polycrystalline ErAuSn, which show an AFM ordering below 1.3 K with a magnetic structure described by wave vector $k$= (1/2, 1/2, 1/2). CeAuSn shows two magnetic anomalies in the ac susceptibility data, which are confirmed as AFM in nature and are termed as $T_{N1}$= 7 K and $T_{N2}$= 4.5 K [142]. Similar to CeAuSn, two magnetic transitions ($T_{N1}$= 3.3 K and $T_{N2}$= 2.7 K) were also seen in PrAuSn with a difference that the second transition is accompanied with non-zero value in imaginary component of ac susceptibility [142]. The dc and ac magnetic susceptibility data show two magnetic transitions at $T_N$=10.5 K (corresponding to AFM to another magnetic transition) and $T_C$= 13.8 K (transition from magnetic to paramagnetic) in NdAuSn [446, 447]. It has been reported that the ac susceptibility measurements in TmAuSn show no magnetic ordering in the temperature range of 4.8-31.2 K. Łątka et al. [448] reported that TbAuSn orders antiferromagnetically, while TmAuSn does not show any magnetic order in whole temperature range under study.

## 4. Magnetocaloric effect (MCE)

One of the reasons behind the renewed interest in *RTX* family at present is the realization that many of these compounds would be potential magnetocaloric materials. This is due to the fact that rare earth component contributes to a large magnetic entropy change in the case of *R-T* intermetallics. Though the temperatures at which the MCE peaks is below 100 K in most of the *RTX* compounds, the magnitudes of MCE and the sign change occurring in some of them have made it an interesting research topic. Usually, Gd and its compounds exhibit large MCE values. This is mostly true in the present series as well. Magnetocaloric parameters of interest are isothermal magnetic entropy change ($\Delta S_m$), adiabatic temperature change ($\Delta T_{ad}$) and the relative cooling power (RC). These parameters reported for different compounds are also listed in Table III. In the following, we present the most important MCE results reported on various *RTX* materials.

Compounds with Fe as *T* elements of *RTX* series are found to show large MCE around 100 K. GdFeSi is one of the best refrigerant materials found in *RTX* series till now. It shows a MCE of 22.3 J/kg K and RC of 1940 J/kg for a field change of 90 kOe around 118 K [175]. RTAl compounds are known to show good MCE. The large MCE in DyNiAl and HoNiAl was reported to arise from the polarization of 3*d* band of Ni [195, 196]. GdNiGa shows very good MCE value around its ordering temperature [449]. RNiIn (R=Gd-Er) compounds also show large MCE [208]. DyNiIn and HoNiIn show two peaks in the MCE data. The peaks show negative and positive MCE for DyNiIn, while it is positive for both the peaks. This results in a large RC in HoNiIn. MCE studies have been reported in ErNiSi from magnetization and heat capacity data [213]. Similar to ErNiSi, HoNiSi also shows large MCE. Large MCE observed in these two compounds is attributed to field induced metamagnetic transition. Compounds of $R$CuSi ($R$=Dy, Ho) show giant MCE accompanied by metamagnetic transition [450, 451]. $R$=Nd and Gd compounds of this series also show considerable MCE [260].

Most of compounds of $R$CuAl series were studied for evaluating their potential for magnetic refrigeration applications [246, 452, 453, 454, 455, 456]. Some authors have synthesized amorphous and crystalline forms of these materials [246, 455, 456]. The amorphous form is found to show spin glass state. Dong et al. [455] reported the influence of crystal grain dimension on the magnetic and MCE properties. They found that reduction in crystal grain size leads to a decrease in $T_C$, MCE and an enhancement in magnetic anisotropy. Amorphous form of NdFeAl ribbon shows hard magnetic behavior at low temperatures. The $\Delta S_M$ value for this compound is found to be 5.65 J/kg K for the field change of 50 kOe [457]. GdNiAl shows large value of RC because of multiple magnetic transitions [25].

GdRhGe shows a sign change in MCE [295]. The compound show positive peak near $T_1$, which reverses its direction and become negative around $T_2$. The change in sign around $T_2$ in MCE as well as in MR in GdRhGe may be due to complex magnetic ordering below $T_2$. The considerable MCE has been observed in $R$RhGe ($R$=Tb-Tm) compounds [299]. HoRhGe among these compound shows highest MCE [300]. At low temperatures, compounds with $R$=Tb, Dy and Er show sign change in MCE attributed to change in magnetic structures as seen in neutron diffraction and magnetization data [299]. All these compounds show positive MCE around their ordering temperature, which suggests that the application of field destroy the antiferromagnetism in these compounds.

The $R$RhSn ($R$=Tb-Tm) shows considerable MCE around their ordering temperatures [313]. The peak is broadened in some compounds, due to non-collinear magnetic structures. Among all the compounds of $R$RhSn series, HoRhSn shows the highest MCE value. TbRhSn and GdRhSn show a sign change of the entropy change at low temperature, which arises due to reorientation of moments [80, 313].

A large MCE and RC have been observed in TbPdAl because of saturation of magnetization due to field induced metamagnetic transition [339]. MCE changes its sign as AFM state changes to field induced FM state. As mentioned above, very small field is required for metamagnetic transition for hexagonal HoPdAl compared to that of orthorhombic HoPdAl. This results in a larger MCE in hexagonal HoPdAl in comparison to the orthorhombic counterpart [343]. Both the compounds show small negative MCE at low temperatures, which disappears on increasing the field. Both PrPdIn and NdPdIn show positive MCE around their ordering temperatures [354]. NdPdIn shows two peaks in MCE; one corresponds to $T_C$ and other broad peak is a consequence of second transition at 18.3 K. Among the compounds of $R$PdIn, HoPdIn shows large MCE along with the large value of RC [352]. The MCE has a broad beak around its ordering temperature in TbAgAl, which also suggests the spin glass state in this compound. Despite the spin glass state, TbAgAl shows considerable MCE around 60 K [384].

Kaštil et al. [194] studied anisotropic magnetocaloric effect on single crystals of DyNiAl. The authors observed two peaks in isothermal magnetic entropy change with H parallel to $a$ axis and single peak when H is parallel to $c$ axis. The latter result is similar to that reported by Singh et al. [195] on polycrystalline samples, but the magnitude of the MCE is considerably enhanced (by 40% at 20 kOe) as that of Ref. [195]. In some of RTiGe compounds, due to field induced metamagnetic transition, the MCE is considerable [458]. Some compounds of $R$CuGe also show good value of MCE at low temperatures [459].

## 5. Magneto-transport properties

As many $RTX$ compounds undergo a metamagnetic transition or multiple magnetic transitions, it is of interest to study the electrical resistivity as a function of temperature and magnetic field. Many of them show large magnetoresistance, positive or negative, in the magnetically ordered regime. Some of them also show good MR in the vicinity of magnetic order-disorder transition as well as in the paramagnetic regime. Complex magnetic structure, crystal field effect and magnetic polaronic effect are seen to influence the transport properties considerably. Generally, the compounds which show large MCE also show large MR. Furthermore, since both these properties are related to the change in the magnetic state brought about by the applied field, the signs of MCE and MR also correlated. In this section, we discuss the resistivity and MR measurements reported by many authors and the values of MR are given in Table IV.

DyTiGe shows an anomaly in resistivity data corresponding due to the onset of magnetic ordering and an abrupt change in MR below $T_N$ at critical field [460]. GdMnSi shows positive MR at low fields which changes the sign with increase in field and shows negative MR, which increases with field [461]. The single crystal study of CeCoGe shows negative as well as positive MR [462]. The compound shows negative MR near $T_N$ and the sign of MR change from negative to positive, attributed to magnon scattering and spin fluctuations [462]. HoNiAl shows large negative MR at low temperatures and small positive MR in the paramagnetic regime [196]. The effect of magnetic polarons was also seen in the MR data. It has been reported that the polaronic effect increases the resistivity of compound due to strong confinement of conduction electrons. On the application of magnetic field, size of polarons grow, coalesce, which causes the delocalization of conduction electrons and results in large MR [196]. YbNiAl shows logarithmic increase of resistivity by 25 % in the temperature between 100 to 10 K, which indicates heavy fermion behavior in this compound [190]. As mentioned earlier, HoNiSi and ErNiSi show large MCE accompanied by field induced metamagnetic transition. The field induced metamagnetic transition along with suppression of spin scattering results in large negative MR in these compounds [213].

The MR data in single crystal TbNiSn show large changes at liquid helium temperature, which is in the response to the multi step magnetization process [463]. The resistivity of TbNiSn was found to be affected by hydrostatic pressure applied along the *b* axis [464]. The application of pressure in YbNiSn shows a decrease in room temperature resistance while an increase in the ordering temperature [465]. Karla et al. [44] reported that the resistivity of *R*NiSb (*R*=La-Nd) increases with increase in temperature and hence shows metallic nature, while compounds with *R*=Tb-Ho show semiconducting behavior with small energy gap at the Fermi level. The resistivity in heavy rare earth compounds decreases in the range between 150 and 300 K. YNiSb shows metallic behavior. Since the value of resistivity is very large, it was termed as a semimetal. The light rare earth compounds show negative as well as positive MR, while heavy rare earth compounds in this series show large negative MR. LuNiSb shows small positive MR [44]. The large MR in these semiconductors may arise either from suppression of spin disorder scattering because of orientation of spins along field direction under the application of field or from the splitting of the up and down spin sub-bands, which reduces the gap [44].

NdCuSi shows large MR near its ordering temperature [260], which arises due to suppression of critical spin fluctuation following the metamagnetic transition. GdCuSi shows both positive as well as negative MR [260]. At 1.5 K, it shows positive MR with a magnitude of 25 % and near ordering temperature (at 15 K) it shows large negative MR of 27 %. Baran et al. [263] reported MR in some $R$CuSn compounds. They observed positive MR below and above $T_N$ for GdCuSn and DyCuSn compounds. HoCuSn shows very large MR which falls rapidly below 6.8 K ($T_N$).

The electrical resistivity in $R$RhAl ($R$=Y, Ce, Pr, Nd, Gd) shows metallic nature [282, 283]. The resistivity in YRhAl and LaRhAl drop sharply and becomes nearly zero at 0.7 K and 2.4 K, respectively [283, 282]. Application of even a small field suppresses the transition temperature in YRhAl, which becomes below 0.37 K in a field of 500 Oe. The broad maximum due to magnetic scattering in the resistivity data in CeRhIn reveals the mixed valent nature of Ce in this compound [288]. Later, Higaki et al. [305] performed the resistivity measurements on single crystalline sample and observed that the anisotropy is not so large in the whole temperature range and it shows quadratic temperature dependence below 50 K. The resistivity of polycrystalline LaRhIn sample decreases with temperature and drops suddenly to zero at the superconducting transition temperature 2 K [305, 466]. YbRhIn shows logarithmic temperature dependence of resistivity with decrease in temperature and shows a saturation of resistivity below 20 K [74].

CeRhGe shows that the electrical resistivity is flat in the temperature range 150- 270 K and falls slowly below 150 K and shows a shoulder around 15-30 K and then falls rapidly below 15 K [117]. A careful observation suggests that double peak behavior of resistivity is due to Kondo lattice behavior, maximum near 150 K is due to crystalline electric field effect and the shoulder around 15-30 K signals the onset of coherency in this compound [117]. Later, Asai et al. [467] studied the pressure dependence of resistivity in CeRhGe compounds. It has been observed that CeRhGe shows extremely huge resistivity peak at 10 K and 1.2 GPa, which may be associated with the formation of spin density waves. With further increase in pressure, the peak in resistivity was found to be diminished at a critical pressure of 1.9 PGa. From the relation, $\rho=\rho_0+AT^2$, where A is constant, it is found that the value of A increases with increase in pressure and has maximum value at critical pressure, which is of the order of the value for heavy fermion superconductor $CeCu_2Si_2$ [467]. As it is mentioned earlier that GdRhGe shows two magnetic transitions; $T_1$ (AFM) and $T_2$ (complex in nature). Fit to electrical resistivity below $T_2$ in the GdRhGe gives n= 2, which is generally observed in case of FM materials. This suggests that the transition corresponding to $T_2$ in GdRhGe is not simple AFM but is complex one [295]. The high temperature resistivity fit reveals that in this regime resistivity is determined by *s-d* scattering [295]. Both temperature and field dependence of MR in GdRhGe yield slight negative value around $T_1$, changes its sign below $T_1$, and shows large positive MR below $T_2$ and may attributed to the peculiar magnetic state of the compound [295]. All compounds in *R*RhGe (*R*=Tb-Er) show metallic nature with negative MR around their magnetic ordering temperatures and positive MR at low temperature which again changes its sign above the critical field [299, 300]. The positive MR at low temperature in these compounds may arise due to the AFM ordering or complex magnetic structure.

The resistivity in CeRhSn shows high anisotropy, which indicates the valence fluctuation in this compound [302]. At high temperature, the compounds of *R*RhSn (*R*=Tb-Tm) show non-linear temperature dependence with negative curvature, attributed to *s-d* scattering of conduction electrons and is a characteristic of itinerant systems [313]. When the resistivity data at low temperatures is fitted by the equation, $\rho=\rho_0+AT^n$, where n is ab exponent, it has been observed that the AFM compounds (TbRhSn and DyRhSn) give n= 3 and the FM (HoRhSn) one gives n=2 [313]. These observations are in general agreement with theory. The fit at high temperatures confirms *s-d* scattering. The temperature and field dependencies of MR are also studied for TbRhSn, DyRhSn and HoRhSn [313]. The MR data shows negative values near their respective ordering temperatures and changes to positive at lower temperatures. The change in MR in these compounds may be due to different contributions [313]. One reason might be the large magnetic anisotropy with low transition temperatures, which results in narrow domain walls. At low temperatures, the probability of domain walls getting pinned is larger, which limits the suppression of domain wall contribution to negative MR and results in saturation of MR with positive values. The second reason may be the Lorentz contribution from conduction electrons and fluctuating Rh moments. The non-collinear magnetic structure may also play an important role in increasing the resistivity with field. The resistivity data in GdRhSn shows two anomalies, one corresponding to AFM transition at 16 K and other attributed to iso-structural transition at about 245 K [80].

MR in single crystals of LaRhSb has been studied in transverse and longitudinal modes [468]. For transverse mode, MR increases as $B^n$ with n=1.3 for B>20 kOe and n becomes 0.2-0.3 at 95 kOe. Non-saturating behavior of MR in transverse mode suggests compensated metal behavior of LaRhSb, without open orbits [468]. In longitudinal mode, MR shows different behaviors: for B parallel to I and *a* and B parallel to I and *b*, the MR saturates in high fields, while it increases above 95 kOe for B parallel to I and *c* [468]. The electrical resistivity in CeRhSb shows a broad peak about 113 K (consistent with susceptibility data) and a sharp rise below 21 K, which reveals formation of gap of about 4 K in the electronic DOS [331]. The pressure dependence of resistivity in CeRhSb shows characteristic temperature for valence fluctuation and gap energy increases with increase in pressure at low temperatures [469]. Usually the application of field gives a negative MR in Kondo insulator systems, however CeRhSb shows positive MR on the application of field [469]. It has been seen from MR data that at 4.2 K and in 4 T, the MR in CeRhSb shows an increase with pressure from 5 % at 1 bar to 34 % at 17 kbar and then a decrease to 29 % at 23 kbar [469]. As it was reported that [470] CeRhSb shows strongly anisotropic gap and stateted that physical properties are expected to be very sensitive to impurities. Keeping this in mind, Takabatake et al. [471] studied high quality single crystals of CeNiSn and CeRhSb. They observed that the increase in resistivity at low temperatures previously reported for these compounds is suppressed in single crystalline samples [471]. The Hall coefficient values show an increase on decreasing temperature [471]. The authors reported that the metallic character seen along *a*-axis resistivity is due to the significant increase of relaxation time of carriers and suggested that the energy gap was closed along *a*-axis [471]. These authors also concluded that CeNiSn and probably CeRhSb are semimetals and not semiconductors [471].

CeRhBi shows double peak in resistivity data which corresponds to Kondo effect while low temperature peak signals the onset of coherent Kondo scattering [333]. The resistivity of CeRhAs shows shallow maximum on decreasing temperature and then shows an rapid increase below 150 K [333]. The energy gap in CeRhAs estimated from resistivity data is 144 K [333].

CePdAl shows an increase in resistivity with pressure (0.6 GPa) after that it shows rapid decrease at 1 GPa [472]. The $T^{1.5}$ dependence of resistivity on pressure above 0.6 GPa shows non-Fermi liquid behavior [472]. The application of field suppresses the instability and stabilized the Fermi liquid state for field of 7 T and at 1.2 GPa pressure [472]. Electrical resistivity in GdPdAl shows thermal hysteresis on cooling and heating [85]. A sharp jump at 180 K and transitions at 48 and 24 K were observed in the resistivity data. The sharp jump at 180 K arises due to first order isostrutural transition, the ones at 48 and 24 K are due to the ordering of Gd moments and and reorientation of spins, respectively [85]. Similar to GdPdAl, the resistivity in TbPdAl shows a sharp jump around iso-structural transition and two anomalies corresponding to two AFM transitions $T_{N1}$ and $T_{N2}$ [86]. The resistivity in DyPdAl shows two anomalies in agreement with magnetization data. The resistivity shows an increase at $T_N$, which is attributed to magnetic superzone gaps at Fermi level expected for AFM materials [341]. At high temperatures, both GdPdIn and TmPdIn show non-linear behavior of resistivity, which arises due to the s-d interband scattering of conduction electrons [353]. The resistivity at 5 K in TmRhSn shows an upturn due to the creation of superzone boundary gap as generally seen in AFM compounds [353]. YbPtGe and YbPdGe show breaks in electrical resistivity at 4.7 and 11.4, respectively, which are the characteristic of FM Curie temperatures and magnetic resistivity of these compounds shows Kondo behavior [473].

The resistivity for light rare earth compounds in RPdSn series is higher than those for compounds with heavy rare earths [364]. The resistivity data of CePdSn shows two linear portions, typical of Kondo resistivity [364]. The application of pressure on the resistivity of CePdSn shows a drastic change [474]. With increase in pressure, the resistivity at low temperatures increases drastically compared to that at room temperature. The peak corresponding to the AFM transition gets broadened with increase in pressure and disappears at a pressure higher than 60 kbar, which indicates that the pressure transformed the magnetic state into a non magnetic state [474]. In SmPdSn and DyPdSn, the resistivity shows two bends; first is due to the onset of AFM ordering and second may correspond to change in magnetic structure [364].

The electrical resistivity in CePdSb was found to show a broad peak about 150 K and an ln (T) dependence at high temperatures, which hints the presence of crystal field and Kondo effect [99]. Two anomalies are observed in GdPdSb resistivity data, which corresponds to magnetic transitions [377]. The electrical resistivity shows a metallic nature at low temperatures and activated behavior at high temperature for ErPdSb, while it shows semiconducting behavior in the whole temperature range under study for YPdSb [378]. Gofryk et al. [100] reported that temperature dependence of electrical resistivity in $R$PdSb ($R$=Ho, Er, Y) and $R$PdBi ($R$=Y, Nd, Dy, Ho, Er) shows semimetal or narrow-gap semiconductor behavior. Large value of electrical resistivity with its shape signals strong atomic disorder in CePdBi [379]. At 5 K, the field dependence of resistivity shows a drop attributed to AFM fluctuations above 2 K and below 1.5 K, the resistivity shows a clear superconducting transition. It has been observed that $T_C$ depends on field application and decreases with increase in field, which confirms its superconducting nature in CePdBi [379].

The resistivity in CeAgAl shows a rapid drop at T=2.8 K, corresponding to the FM ordering and shallow minimum at 16 K followed by a weak peak near 9.5 K, which are characteristic of dense Kondo systems. [383]. The resistivity in NdAgSi shows two transitions as seen in magnetization data [390]. Below 40 K, the resistivity follows $T^2$ behavior, which is expected for FM materials [390]. In LuAgGe, the transverse MR shows $H^2$ behavior, which is expected for normal metals [108]. Field dependence of resistivity in TbAgGe shows metamagnetic transitions as also seen in magnetization data.

Compounds with $R$=Gd, Tb, and Dy in $R$AgSn series show a rapid decrease in resistivity below $T_N$, which arises due to reduction in the spin disorder scattering [397]. Compounds with $R$=Nd, Sm, Ho and Er exhibit a prominent peak of critical divergent nature at $T_N$ superposed on a background which resembles that of TbAgSn. Authors reported that the divergent behavior resulted from the existence of conduction electrons associated with long wavelengths, which are strongly scattered by the large amplitude of spin fluctuations at the ordering temperature [397]. In CeAgSn, the resistivity increases below $T_N$, possibly due to a decrease of carriers due to AFM zone boundaries or a subtle combination of lattice and impurity effects [397]. HoAgSn exhibits slightly negative MR above $T_N$, which becomes positive below $T_N$. A similar behavior of MR has been seen in DyAgSn and ErAgSn as well [397]. The positive MR below $T_N$ may be attributed to the reorientation of AFM structure in presence of field [397]. SmAgSn, NdAgSn and TbAgSn show positive MR below and above $T_N$ [397].

CeIrAl and CeRhSb crystallize in the same crystal structure and show mixed valent state. However, there is no rise in low temperature resistivity data, typical of gap forming systems, which was seen in CeRhSb [401]. The resistivity does not show any Kondo type behavior in the temperature between 2 to 300 K, consistent with its high Kondo temperature. It has been observed that below 50 K, CeIrGe shows $T^2$ dependence of resistivity and at high temperatures, it shows a negative curvature [117]. The resistivity of CeIrSb is larger than that of CeRhSb and CeRhBi. It has been observed that compounds which show gradual decrease in resistivity with a positive curvature seem to be valence fluctuating system [118].

Single crystal of CePtAl shows negative MR in the paramagnetic state (T>5.9 K) and exhibits a minimum in ordered state (T<5.9 K) along *a*-axis, while it shows positive MR below 5.9 K, which is almost field independent along *b* and *c*-axes [475]. The pressure effect on resistivity for CePtAl has also been studied and the results show remarkable change from heavy fermion state with magnetic ordering at pressures lower than 5 GPa to valence fluctuating state at pressures higher than 6 GPa [476]. Electrical resistivity in CePtGa has been measured with pressure and field [477]. The results show that $T_N$ decreases with increasing pressure and disappears above 1 GPa and a negative MR at 4.2 K with a $H^2$ dependence. The negative MR at 4.2 K is a characteristic of incoherent Kondo systems and is suppressed by the pressure, keeping the sign of MR unchanged [477]. It has been observed that $H^2$ coefficient decreases linearly with increasing pressure and changes its sign above 3 GPa, indicating the crossover from incoherent Kondo state [477]. The magnetic scattering contribution to the resistivity in YbPtGa shows a behavior consistent with Kondo-type interactions [412]. The resistivity data in PrPtIn and SmPtIn show metallic character with non-linear behavior at high temperatures, which reveals that in addition to the electron-phonon scattering, the interband scattering process may also be considered [123]. The hump in resistivity data of PrPtIn in the temperature range of 4.2-50 K ascribed to crystal electric field effect [123].

The magnetic resistivity in CePtSi shows a characteristic of dense Kondo systems [420]. NdPtSi shows the slope in the resistivity data corresponding to AFM ordering, while LaPtSi shows a jump at 3.8 K, which corresponds to the superconducting transition temperature [421]. The low temperature resistivity data in LaPtSi, NdPtSi and SmPtSi were fitted by power law relation [421]. The resistivity in CePtGe does not show any change at high temperatures on the application of pressure, however at low temperatures the shoulder accentuates on increasing pressure upto the highest pressure of 17.4 kbar [478]. It has been observed that $T_N$ is not very sensitive to pressure in this compound [478]. CePtGe shows an increase in resistivity at ambient _pressure on cooling from room temperature and shows a broad peak around 150 K, which deceases steeply down to 15 K and resumes a step drop at 10 K [478]. Such a double peak structure of resistivity suggests the Kondo behavior in this compound as seen in many compounds mentioned earlier.

The resistivity in CePtSn exhibits a rapid fall at 8 K, which corresponds to the onset of AFM ordering [479]. The magnetic resistivity shows ln(T) dependence in two temperature regimes suggesting Kondo type behavior [479]. MR measurements performed on single crystals of CePtSn at 1.8 K and in the field applied along *a* and *b*-axes show sharp anomalies at 12.5 T for field along a-axis and 4 and 11 T for field along *b*-axis, which are due to the metamagnetic transitions in this compound [431]. Later, it has been observed that CePtSn shows an irreversible MR effect around 3 T field along *b*-axis, [428]. Mišek et al. [430] studied the effect of pressure on MR and found gradual increase in the critical field up to 1.2 GPa, where only one transition is observed at 11.5 T. On further increase of pressure above 1.2 GPa, two transitions were observed and after that the critical field decreases with increase in pressure and reaches 7.5 T at 2.5 GPa [430].

It was reported that *R*PtSb (*R*=La-Sm) are metallic and show anomaly in resistivity plots at 4.5, 8, 15.5, and 24 for R=Ce, Pr, Nd and Sm, respectively [133]. GdPtSb and YbPtSb show semimetal characteristic in resistivity data [133]. It has been reported that *R*PtSb has low carrier concentration, which shows a transition from semimetallic to metallic behavior as *R* goes from Gd to Yb [480]. It has been observed that in the single crystal study of CePtSb, it shows strong anisotropy in ρ, $\rho_c > \rho_a$ [375]. The compound shows large positive MR for H parallel to *a-axis*, which may arise due to the anomalous spin dynamics in the FM state of this compound [375].

All *R*AuAl (*R*=La, Ce and Nd) compounds show metallic character at high temperatures in resistivity data [134]. The resistivity of CeAuAl shows shallow minimum near 20 K followed by a maximum at about 5 K, below which there is a sharp drop in resistivity [134]. The magnetic resistivity shows a decrease with increase in ln (T). These characteristics show a Kondo behavior in CeAuAl [134]. At 5 and 10 K, CeAuAl shows negative MR and at 2 K it shows a peak in MR after which it crosses to positive value [134]. The change in the sign of MR in CeAuAl is characteristic feature of Kondo lattice and is attributed to the coherence effect in Kondo system [134]. The negative MR at all temperatures down to 10 K in NdAuAl, confirms the FM ordering in this compound [134]. GdAuIn shows negative MR, which changes sign at low temperatures [481]. Complex behavior of MR arises from the interplay between the AFM coupling and the strong magnetic field, both of which suppresses magnetic fluctuations but also act against each other [481]. NdAuGe shows tendency of saturation in its resistivity data above room temperature and a sharp drop around 3.8 K [444].

## 6. Hydrogenation of *RTX* compounds

Hydrides of rare earth intermetallics have been considered as an interesting topic because of possibility of change in different physical properties resulting from the formation of *R*-H bond. Moreover, this study has led to the discovery of many potential hydrogen storage materials, many of which can store hydrogen with densities much more than the liquid hydrogen density. Interstitial modification by hydrogen causes changes in the unit cell dimensions, unit cell volume and consequently the electronic structure. The changes are as a result of chemical effect involving the charge transfer between hydrogen and the host compound as well as the magnetovolume effect caused by lattice expansion. A large number of RTX compounds have been subjected to hydrogenation. In the following some of the most important reports on the effect of hydrogen on these compounds are discussed. A few studies deal with the effect of deuterium insertion on the magnetic and related properties. For a clear view the change in crystal structure and magnetic nature of the compounds on hydrogenation is shown in Table V.

Hydrogen insertion in both forms of GdTiGe (CeFeSi and CeScSi-type) shows interesting change in their structure and magnetic properties [16]. Hydrogenation of both forms of GdTiGe forms same hydride GdTiGeH, which crystallizes in CeScSi-type structure. The authors reported that GdTiGe which was AFM ($T_N$=412 K) and FM ($T_C$=376 K) before hydrogenation for CeFeSi and CeScSi-type forms, respectively become paramagnetic in the temperature range 4-300 K after hydrogenation [16]. The hydrogenated compound GdTiGeH shows paramagnetic nature in the temperature range of 4-300 K. The hydrogenation of NdMnSi shows a significant change in structure and magnetic properties from its parent compound [482]. The hydride adopts ZrCuSi type structure, while its parent compound shows CeFeSi type structure [482]. The hydride NdMnSi shows an increase in unit cell volume by 3.3 % with an increase in $T_{N1}$, and a decrease in $T_{N2}$. It has been observed that the $T_{N1}$ of NdMnSiH increases from 280 K to 565 K, while the $T_{N2}$ decreases from 185 to 103 K [482]. The parent compound CeMnGe shows two magnetic anomalies at $T_{N1}$= 313 K and $T_{N2}$ = 41 K, the first one is corresponding to AFM ordering of Mn moments and the second is due to Ce moments [483]. CeMnGeH shows an anomaly at $T_C$=316 K characteristic of FM, ferrimagnetic or canted systems and shows saturation below $T_C$, which reveals that no anomaly ascribed due to the Ce moment ordering [483]. Thus, hydrogen insertion in CeMnGe suppresses the magnetic ordering of Ce moments and shows canted magnetic ordering of the Mn sublattice [483].

Hydrogenated CeCoGe has same structure as CeCoGe but the hydrogenation of this compound causes an increase of the unit cell, in which the lattice parameter $a$ decreases by -3.1 % whereas the c parameter shows a drastic increase by 12.7% [484]. In addition to this, the hydrogenated compound shows spin fluctuation behavior unlike CeCoGe, which orders antiferromagnetically. An expansion of unit cell volume together with change of AFM states to FM states have been observed in NdCoSi and NdCoGe on the process of hydrogenation [485].

Upon hydrogen insertion, CeNiAl changes its crystal structure from ZrNiAl to AlB$_2$ type hexagonal type. In addition to this, hydrogenation also leads to a valence change from +4 to +3 for Ce [486]. The crystal structure retains in case of SmNiAl after hydrogenation [487]. SmNiAl shows two peaks in ZFC M-T data at 60 and 10 K, indicating the AFM transitions in this compound. The hydrogenated SmNiAl retains both these peak positions in ZFC mode. However, the FC data show a positive peak and negative peak at 52 and 12 K, respectively. The positive peak corresponds to AFM transition similar to that seen in the ZFC curve (but with a slight lowering from 60 K to 52 K), while the second (positive) peak as seen in ZFC curve disappears in the FC mode [487]. The negative peak suggests an anomalous diamagnetic response, which is attributed to hydrogenation. Later on, He [488] reported that hydrogenated SmNiAl is a zero magnetization ferromagnet, which has a $T_C$ of 75 K and a compensation temperature of 38 K where the orbital contribution to moment in the compound is balanced by the spin moment.

Hydrogen insertion in HoNiAl is found to change its crystal structure from hexagonal to orthorhombic [489]. Hydrogenated HoNiAl shows very different properties compared to those of the parent compound. There are two magnetic phase transitions in HoNiAl, as reported by different authors [189, 191, 196, 489]. But the hydrogenated HoNiAl shows only one transition at 6 K, at which the magnetic state changes to AFM from an amplitude modulated FM. CeNiGa is found to show two-crystal structures (LTP-hexagonal, HTP-orthorhombic) for different temperatures. However, hydrogenation of the compound shows only one crystal phase (hexagonal AlB$_2$ type) [38]. Along with the change in the crystal structure, hydrogenation also changes the valence from the intermediate valence for the parent compound to a purely trivalent state for the hydrogenated compound.

There are many reports on the hydrogenation of $R$NiIn compounds. These reports [490, 491, 492, 493, 494, 495] show that the crystal structure of compounds is unchanged on hydrogenation, while there is an expansion of unit cell on hydrogenation. It was seen that the hydrogenated CeNiIn shows FM behavior below $T_C$= 6.8 K with a change in Ce valence from intermediate to the trivalent state [492]. Marcos et al. reported the MCE of hydrogenated CeNiIn compound. They found that the compound shows isothermal entropy change of 0.9 J/mol K and adiabatic temperature change of 3.4 K at the FM transition for a field change of 90 kOe. Later, Shashikala et al. [494] studied CeNiIn with different hydrogen compositions (CeNiInH$_y$ y=0.3-0.6 and 1.6) and reported that it shows two distinct hydride phases. It was shown that higher hydride phase (y=1.6) is relatively unstable and after short period of time it goes into lower hydride phase. The magnetization measurements show change in Ce valence from mixed valent to trivalent passing through a heavy fermion-like state as a function of hydrogen concentration [494]. The hydrogenated GdNiIn shows two modifications; (i) decrease in moment per formula unit and (ii) decrease in $T_C$ [495]. These two effects can be interpreted as the hydrogen insertion in GdNiIn induces a moment in Ni atoms, which is antiferromagnetically coupled with the Gd moments and hence reduces the magnetic moment per formula unit. On the other hand, hydrogen insertion increases the interatomic distance which decreases the strength of the Gd-Gd coupling and results in the decrease in $T_C$ [495].

Hydrogenation of $R$NiSi ($R$=La, Ce, Nd) has been reported in Ref. [210, 211]. Hydrogenation in these compounds does not affect the crystal structure except slight change in unit cell parameters. The hydrogenation in CeNiSi shows anomalous increase in strength Kondo effect, which is linked with the unusual site occupancy by H-atoms [211]. Hydrogenation in NdNiSi shows a change in the magnetization data from parent compound [211]. The hydrogenated NdNiSi is paramagnetic down to 2 K, unlike the parent compound. Many authors [496, 497, 498, 499, 500] studied the effect of hydrogenation in CeNiSn. CeNiSn shows multiple magnetic transitions such as; from Kondo insulator (CeNiSn) to an antiferromagnet (CeNiSnH$_{1.0}$) to a ferromagnet (CeNiSnH$_{1.8}$). At the same time, it also changes its crystal structure CeNiSn (orthorhombic, TiNiSi) to CeNiSnH$_{1.8}$ (hexagonal, ZrBeSi). Macros et al. [499] also reported the MCE in hydrogenated CeNiSn, which showed $\Delta S_M$=-1.7 J/mol K and $\Delta T_{ad}$=4.2 K for a field change of 90 kOe. Deuteride TbNiSn shows a change in magnetic structure as well as a reduction in Tb magnetic moment [501]. The compound shows a canted magnetic structure below 10 K. The hydrogenation in HoNiSn causes a crystal structure change from orthorhombic (TiNiSi type) to hexagonal (ZrNiAl type) [502].

CeCuSi shows a significant effect of hydrogenation [503]. The parent compound shows FM ordering below 15.2 K, while the hydrogenated compound shows a complex magnetic phase diagram with three magnetic transitions at 13.7, 10.6 and 8.4 K. The hydrogenated compound shows superstructure deriving from the ZrBeSi type structure. Hydrogen insertion in CeCuGe does not change its crystal structure, but the magnetic state of compound changes [504]. It has been observed that parent compound has FM ordering, while its hydride is paramagnetic down to 1.8 K [504].

Hydrogenated CeRuSi also shows a change in its physical properties, as compared to the parent compound [505]. The compound does not show any change in its space group upon hydrogen insertion; however hydrogenation results in expansion of the unit cell. CeRuSi does not show any magnetic ordering, whereas its hydride shows two AFM transitions at $T_{N1}$=7.5 K and $T_{N2}$=3.1 K [64, 505]. The neutron diffraction experiments performed on CeRuSi hydride show that between these transition temperatures, Ce moments order in a collinear AFM sinusoidal structure, which becomes square wave modulated at $T_{N2}$ [506]. Fernandez et al. [507] studied the pressure effect of CeRuSi hydride. They found that the Néel temperatures ($T_{N1}$ and $T_{N2}$) and the critical fields ($H_c$) for metamagnetic transitions increase with pressure, while there was a decrease in the net magnetization. Similar to CeRuSi, hydrogen insertion destroys the non-magnetic heavy fermion behavior in CeRuGe [508]. Different measurements suggest an AFM ordering below 4 K in CeRuGeH. The space group is same for both CeRuGe and its hydride, though the type of crystal structure changes from CeFeSi (parent) to ZrCuSiAs (hydride) type tetragonal structure [508].

It has been observed that CeRhIn has two different hydride phases such as CeRhInH$_{1.55}$ and CeRhInH$_{0.55}$ [509]. The hydride phase with higher hydrogen concentration is found to be unstable and transforms to low hydrogen concentrated phase, which retains hexagonal crystal structure of parent compound with a slight expansion of the unit cell. The second point observed in lower hydride phase is that the Ce shows its mixed valent state, though the reduction in Kondo temperature hints its trivalent state [509]. Hydrogen insertion in CeRhGe modifies its crystal structure from orthorhombic to hexagonal and shows a transformation from an AFM state to an intermediate valence, which was evidenced by magnetization, resistivity and thermoelectric power measurements [510].

The hydrogenated CeRhSn and CeIrSn adopt the same crystal structure as the parent compound with an increased unit cell volume [511]. Magnetization, resistivity and thermo-power measurements confirm a change in ground state from intermediate valence to nearly trivalent upon hydrogenation, which arise due to an expansion of unit cell volume resulting a decrease in hybridization between $4f$ and conduction electrons [511]. Hydrogenated CeRhSb adopts the same crystal structure as its parent compound except a slight increase in its cell volume [512]. The magnetic measurements show that the hydrogen insertion in CeRhSb changes its state from intermediate valence to trivalent with an AFM ordering below 3.6 K [512]. Hydrogen insertion in CePdIn and CePdSn retains their crystal symmetry of the parent compounds but their unit cells show an expansion [513]. Magnetization measurements show AFM ordering below $T_N$= 3 and 5 K for CePdInH and CPdSnH, respectively [513].

Hydrogen insertion in CeIrGa shows interesting effect on the physical properties of parent compound. After hydrogenation only single phase CeIrGaH$_{1.7}$ is stabilized and adopts the hexagonal crystal structure, which is different from its parent compound, which crystallizes in orthorhombic crystal structure [514]. Upon hydrogenation, the mixed valent Ce transforms into $Ce^{3+}$ state [514]. Similar to CeIrGa, the structural transition also take place in CeIrGe upon hydrogen insertion [510]. The hydrogenated compound shows ZrBeSi type hexagonal crystal structure [510]. In addition to this, the strength of Kondo effect has decreased in CeIrGe by hydrogen insertion. There is no change in the crystal structure of CeIrSb after hydrogenation but shows an expansion of unit cell by 4% and an anisotropic unit cell expansion by 4.3 % in $c$ direction [515]. The increase in volume of CeIrSb hydride is greater than those detected in CeRhSb, CePdSn or CeNiSn hydrides and also exhibits higher transition temperature compared to those of the above mentioned hydrides.

Hydrogenated CePtAl shows different crystal structure which is different from its parent compound, but was not known by authors [516]. The hydrogen also influenced its magnetic properties by increasing of its FM character, i.e. $T_C$ of $CePtAlH_{1.1}$ is 11.6 K, while it is 5.6 K for CePtAl [516]. The structural and magnetic measurements were performed on $R$PtIn ($R$=Tb, Er, Tm) deuterides [517]. The results show that there is no change in crystal structures except for some changes in lattice constants (contraction in $a$ and increase in $c$ parameter) upon introducing deuterium [517]. Magnetic properties of these compounds after deuteridization show significant changes. It has been reported that the magnetic transition temperature is enhanced in case of $R$=Tb and Er compounds, whereas diminished in the case of Tm compound [517]. $TbPtInD_{1.3}$ shows two metmagnetic transitions [517].

## 7. Some other probes/ properties

## 7.1 Electronic structure calculations and electron spectroscopy

The optical and electronic structure studies of GdTiSi and GdTiGe compounds show a strong hybridization between the $d$ states of Gd and Ti [518]. Apart from this, a significant width of the $f$ band was also observed in the calculations. These calculations also indicate a significant polarization of the Ti $3d$ electrons, which may be related to the indirect exchange interaction with Gd moments. The electronic structure of GdMnGe has been studied by X-ray photoemission spectroscopy (XPS) and calculated using the tight-binding linearized muffin tin orbitals (TB-LMTO) [519]. The spin polarized calculations for GdMnGe show that Mn and Gd $d$ bands are strongly polarized and reveal a moment of 3.22 $\mu_B$ on Mn, which is directed in opposite direction with that of Gd [519]. The XPS studies in the case of GdCoAl,

GdNiAl and GdCuAl reveal that d electrons in GdCoAl and GdNiAl contribute to the density of states (DOS) at the Fermi energy, while in GdCuAl the main contribution is related to $s$ and $p$ electrons of Al and to the $5d$ electrons of Gd [35]. The contribution of DOS at Fermi level decreases when one goes from Ni to Cu [35].

Band structure calculations using local spin density approximation (LSDA) reveal magnetic ground state in CeCoGa while LSDA+U approach shows a stable AFM ground state [181]. It has been reported that the magnetic ordering is present not only in the Ce sublattice but also in the Co sublattice and a small moment at Ga sites when calculations were performed for moderate U parameter [181]. Due to the AFM ordering with different values of local moments for these sublattices, the compound shows an overall FM behavior [181]. XPS studies in $R$CuIn ($R$=La-Nd, Lu) indicate that the valence bands are mainly determined by the Cu $3d$ band [248]. The $4f$ states in Pr and Nd hybridize with the Cu $3d$ state and the appearance of $3d^9 4f^0$ component in CeCuIn gives a clear evidence of intermediate valence behavior of Ce [248]. Electronic structure calculations for LaCuSn show relatively low density of states at Fermi energy. The mixed valent nature of Ce in CeRhAl shown in the x-ray diffraction data was confirmed by X-ray absorption spectra performed at $L_3(2p$-$5d)$ and $M_{5,4}$ ($3d$-$4f$) edges of Ce as well. These studies also show the trivalent nature of La compound, which is a superconductor at low temperatures [520]. The band structure calculation in CeRhGa shows a nonmagnetic ground state, which is consistent with valence band XPS analysis [286]. The electronic structure calculations in CeRhGe show strong contribution of the Ce 4f states at the Fermi level [521]. XPS measurements in CeRhGa confirmed a mixed valent behavior of Ce, as indicated by the magnetization data [286]. The measurements also reveal the strong onsite hybridization between 4f and conduction electron states in the band [286]. *ab initio* calculations for HoRhGe with the LSDA+$U$ method show an antiferromagnetic ordering of the Ho magnetic moments as the ground state while rhodium ions were found almost nonmagnetic [300]. Electronic structure calculations for other compounds of RRhGe series are under way and will be reported elsewhere. The electronic band structure calculations for YbRhSn shows that the exchange interaction between $f$-electrons and conduction electrons plays a key role in the heavy fermion character of them [522]. Szytula et al. [523] reported electronic band structures for some Yb$TX$ ($T$=Au,

Pd, Rh, Pt; $X$=Sn, Bi) compounds and showed that valence band in these compounds are mainly formed by 4f orbitals of Yb and $4d(5d)$ orbitals of the $T$ element.

Kumigashira et al. [524] used high resolution photoemission spectroscopy to investigate Kondo insulator pseudogap behavior in CeRhSb and CeRhAs. These authors have reported that the size of pseudogap is independent of temperature [being scaled with the Kondo temperature ($T_K$)], whereas the temperature evolution of this is dominated by $T_{coh}$ ($<T_K$), which is another characteristic temperature. Kumigashira et al. [525] also reported that in these compounds 4f derived density of states shows a pseudogap at Fermi level in contrast to metallic Kondo materials. They observed that the size of the pseudogap is smaller than that of conduction pseudogap, though both scale well with $T_K$. These results suggest that for Kondo gap in Ce compounds, the hybridization of 4f and conduction electrons is essential near Fermi level [525]. XPS study on GdPdAl reveals a dominant contribution of $d$ electrons at the Fermi level [85]. Unlike in GdPdAl, Pd does not contribute to the moment in DyPdAl and HoPdAl [341].

Theoretical electronic structure calculations show metallic ground state in CePdSn and CePtSn. Spin polarized calculations suggest stable magnetic state in CePdSn, while it is observed in CePtSn only when lattice is expanded [526]. Electronic structures of $R$PdSb ($R$=Pd, Nd) were studied using XPS [527]. The XPS results show that valence bands in these compounds are mainly determined by Pd 4d band and show hybridization of 4f orbitals with conduction band, which is large in the case of PrPdSb [527]. Interestingly, the electronic structure calculations show half metallic nature in CePdSb, i.e., metallic for majority spins and semiconducting for minority spin bands [528]. Similar calculations in GdPdSb predict a pseudogap at the Fermi energy in the AFM ground state, while for the FM case it predicts itinerant states in the minority band and weakly localized states in majority band [377]. It has also been observed that AFM state is more stable than ferromagnetic configuration in GdPdSb [377]. Band structure calculations reveal that GdPdSb is a half metallic ferromagnet [377]. Both ErPdSb and YPdSb show semimatellic or narrow gap semiconducting behavior [378]. Electronic band structures show formation of narrow gaps in ErPdSb and YPdSb compounds. It shows inverted band order and thus may support topological quantum states in ErPdBi [381].

The XPS measurements in CeAgAl reveal stable configuration of $4f$ shell and rather a weak intra-site hybridization effect, which is also supported by *ab initio* calculations [383]. Both calculated and measured valence band spectra show that Ag and Al atoms have a tendency to randomly occupy the same crystallographic site, which changes the local charge density on Ce atoms and results in local surrounding dependent exchange interactions between Ce atoms [383]. Band structure calculations performed by Goraus et al. [386] for CeAgGa, give the Ce moment, which is insensitive to Ga/Ag off stoichiometry in the 8h position. XPS valence bands for $R$AgSn ($R$=Ce, Pr, Nd, Dy) are compared with those calculated using spin polarized tight binding linear muffin tin orbital method. This reveals that the valence bands are mainly determined by Ag $4d$ band [529]. The electronic structure studied by XPS and *ab initio* calculations shows that valence bands in LaPtIn and CePtIn are dominated by the Pt $5d$ and In $5s$ and $5p$ states [530]. It has been observed that the Ce $3d$ and Ce $4d$ spectra do not give any evidence of intermediate valence of $Ce^{3+}$ in CePtIn [530]. X-ray magnetic circular dichroism (XMCD) and X-ray absorption near edge structure (XANES) spectra at Ce $L_{2,3}$-edges confirm the trivalent state of Ce and the XMCD signal at Pt $L_{2,3}$-edges suggests that the Pt $5d$ states are also involved in the magnetic interaction in CePtSn [531].

The electronic structure study of Ce$T$In ($T$=Ni, Cu, Pd, Au) compounds by XPS show main contribution of $T$nd band in valence bands, while few percent share of $R$ $4f$ states [532]. The electronic structure studied by XPS measurements and calculated ones were found in agreement in $R$AuSn ($R$=Sc, Ce, Gd, Er, Au) [533]. The results shows that R $4f$ and Au $5d$ well defined identified spectral features far away from the Fermi level [533]. It has been observed that spectral weight in the vicinity of the Fermi level is mainly contributed by Au/Sn $sp$ and $R$ $spd$ bands [533]. Band structure calculations performed for GdAuSn show a pseudogap at the Fermi level [534].

Mishra and Dhar [535] studied the magnetic and electric hyperfine interactions for the [111]Cd probe nucleus in NdScGe using the time differential perturbed angular correlation (TDPAC) technique. They found that Cd probe occupying the Sc site experiences a large magnetic hyperfine field with saturation value $B_{hf}$ (0) = −8.5 T. They also indicated considerable spin polarization of the conduction electrons in NdScGe. Theoretical calculations done for GdScGe shows that the exchange interaction increases with decreasing unit cell volume, which results in a pressure induced enhancement of the Curie temperature in this compound [536].

## 7.2 Inelastic neutron scattering

Crystalline electric field in $R$NiSb ($R$=Ce-Nd, Tb-Er) has been investigated by inelastic neutron scattering measurements [537]. It has been observed that moments are directed along a fourfold crystal axis in DyNiSb and ErNiSb and along a twofold one in TbNiSb and HoNiSb with an additional moment rotation in TbNiSb below $T_N$ [537]. CeCuSn has been probed by a number of experiments. Inelastic neutron spectra show two well-defined crystal field excitations [57]. The presence of quasielastic line whose width has dependency on wave vector and temperature has been observed in low energy neutron scattering [57]. Inelastic neutron scattering spectra show no clear evidence of a gap type excitation for low energy while the high energy spectra show a broad excitation centered around 35 meV in CeRhSb, which is the characteristic of a valence fluctuating compound [538]. In case of PrRhSb, the inelastic spectra show the presence of crystal field excitation with crystal field splitting of 25 meV [538]. The low energy inelastic neutron scattering studies at 2 K show no gap-type excitation in CeRhSb as expected for gap-forming Kondo semimetallic compounds while the high energy neutron scattering spectra reveal a broad magnetic response centered at 35 meV, which is consistent with mix-valent behavior of Ce ions in CeRhSb [539].

Inelastic neutron scattering performed on TbNiAl and TbPdAl reveals crystal field excitation below 15 meV and show that the scattering intensity shifts gradually towards lower energy when temperature is increased from 4 to 140 K. This is due to gradual population of the excited crystal field states [540]. There is no significant deviation around structural transition, which indicates that crystal field potential experienced by Tb ion has similar strength and local point symmetry on both sides of the structural transition [540]. Inelastic neutron scatting spectra in CePdGa show two crystal field transitions at 18.9 meV and 33.8 meV [345]. The same study for $R$PdIn ($R$=Ce, Pr, Nd) shows that there is no excitation related to crystal field in CePdIn (due to the delocalization of 4f states) and seven crystal field levels in non magnetic PrPdIn (in which the level distribution does not show significant temperature dependence). In the case of NdPdIn, crystal field distribution is temperature dependent which occurs due to the existence of molecular field [541]. The inelastic neutron spectra in CeAgSn shows two well-defined crystal field excitations and presence of quasielastic line whose width has dependency on wave vector and temperature [57].

The inelastic neutron scattering data show two very broad crystal field transitions in CeAgGa compound [385]. The broadening of CEF transitions detected by inelastic neutron scattering may arise due to the crystallographic disorder between Ag and Ga atoms [385].The inelastic neutron spectra reveal two inelastic lines corresponding to crystal field transition from ground state to first (6.3 meV) and second (17.8meV) excited state in CePtSi [542]. The Kondo temperature inferred from the quasielastic line width in CePtSi was found to be 16 K [542]. Inelastic neutron scattering in CePtSn shows two well-defined crystal field doublets with the excitation energy of 24 and 34.9 meV [479]. The inelastic neutron scattering spectra show two well-defined peaks corresponding to transition between the ground state doublet to two excited state doublets with excitation energies of 25.2 and 29.1 meV [543].

## 7.3 µSR, NMR and ESR

Muon spin relaxation (µSR) experiments performed on CeCuSn indicate an unusual evolution of the magnetic ordering in this compound. It shows short range correlations near 11 K and leads to spin freezing state around 8.6 K and then attains a coexistence of long-range ordered and spin frozen phase below 7.5 K [544]. µSR experiment performed on YbPdSb shows unusual features at low temperatures, which can be explained by the formation of a spin-liquid phase. Such a phase can be stablised by the Kondo effect in the materials close to Doniach's magnetic instability [545]. ESR measurements performed on GdRhAl reveal a small contribution of the Rh $4d$ electrons to the conduction band [285]. The NMR and NQR measurements were performed to study Kondo semiconductors CeNiSn and CeRhSb [546]. The temperature dependence of nuclear spin lattice relaxation rate, $1/T_1$ reveals pseudo type with a V shaped structure of energy gap estimated the band width (D) and pseudogap ($\Delta$) such as D= 140 and 210 K and $\Delta$=14 and 24 K for CeNiSn and CeRhSb, respectively [546]. Ekino et al. [470] studied CeNiSn and CeRhSb by tunneling spectroscopy and reported that the differential conductance at 2-4 K show energy gaps of 8-10 and 20-27 meV for CeNiSn and CeRhSb, respectively, which are comparable to Kondo temperatures of these compounds. They found that with increasing temperature, the zero bias conductance displays a crossover from well- developed gap state to partial gap state and then to a Kondo metallic state [470].

[75]As NQR/NMR studies performed on CeRhAs show that an activation type T dependence of $1/TT_1$, which suggests a gap opening over the entire Fermi surface (which is different in shape compared to those in CeNiSn and CeRhSb) [547]. Results show a gap of 272 K and the bandwidth of about 4000. This is one order of magnitude larger than those in CeNiSn and CeRhSb [547]. Fay et al. [548] reported that the Knight shift, the line width and the spin-lattice ($T_1^{-1}$) and spin-spin ($T_2^{-1}$) relaxation rates of [27]Al for YbNiAl, which show heavy fermion behavior. The authors also reported that in the paramagnetic regime the Knight shift and the line width show linear dependence on the bulk susceptibility, indicating that all contributions to the anisotropic shift separate into a local susceptibility. It is also reported that the susceptibility shows presence or absence of temperature dependence, depending on the orientation of the CEF [548]. NMR measurements on CePdAl below 1 K show heavy fermion state. The second magnetic transition observed previously in this compound is a modification of magnetic fluctuations along $c$-axis, attributed to the change in incommensurate structure along the $c$-axis [549]. Later, Oyamada et al. [550] performed [27]Al NMR measurements on CePdAl and ruled out the second magnetic phase transition down to 30 mK in CePdAl. They reported that paramagnetic Ce moments in the partially ordered state are in a heavy fermion state and is stable down to 0 K. It has also been observed that antiferromagnetic correlations in this compound develop from much higher temperatures above $T_N$ [550]. A strong molecular field produced by Gd polarizes Pd $d$ band, which is confirmed by ESR measurements as well [85]. Lord et al. [551] studied zero field Sb NMR as a function of temperature and pressure in CePdSb. They reported that effective magnetic field at Sb nucleus increases under pressure, comparable to the rate of Curie point. The results show that the pressure dependence of electric field gradient at nucleus is 28 times greater with an opposite sign than that is expected from a simple point charge model of elastic isotropy [551]. ESR measurements on single crystalline GdPdBi shows AFM ordering below $T_N$=13 K and a single $Gd^{3+}$ Dysonian ESR line and no Korringa relaxation in high temperature regime [380]. A strong line broadening in ESR was also seen at low

temperatures, which indicates that spin–spin $Gd^{3+}$ interaction is a dominant relaxation mechanism. ErPdBi shows the combination superconductivity and AFM ordering, which is unusual [381].

Later Kawasaki et al. [552] performed Sb NMR and NQR measurements on CeIrSb, which reveal formation of pseudogap at Fermi level in the density of states and also supported by temperature dependence of Knight shift measured in high fields. They observed that $1/T_1T$ (where $1/T_1$ is the nuclear spin-lattice relaxation rate) has a maximum around 300 K and decreases significantly 1 at low temperatures. These results suggest V-shaped energy gap with a residual density of states at Fermi level [552]. The size of energy gap estimated for CeIrSb is 350 K, which is larger than that of in CeRhSb and CeNiSn by one order of magnitude. Koyama et al. [553] performed $^{121}$Sb and $^{195}$Pt NMR experiments on YbPtSb to probe the local magnetim in this compound and they reported the hyperfine coupling constants of -3.8 and -4.6 kOe/$\mu_B$ at Sb and Pt sites respectively, estimated from the slopes of Knight shifts and susceptibility plots.

In addition to these NMR studies, $^{119}$Sn NMR, $^{45}$Sc NMR [554, 555, 556, 557, 558] and $^{89}$Y NMR [559, 560, 561] experiments were also carried out for certain *RTX* compounds. $^{119}$Sn NMR study in ScAuSn, YAuSn and LuAuSn reveals only one Sn site in all the these compounds, which is in agreement with the crystal structure [554]. From the NMR measurements the authors suggested unusually low s-elctron densities at the Sn nuclei in these compounds. $^{45}$Sc NMR spectra show sharp line at 57 ppm relative to aqueous scandium nitrate solution, which reflect the cubic site symmetry of Sc position. $^{45}$Sc NMR experiment performed on ScAgSn gives an unambiguous proof of the superstructure [555].

## 7.4 Mössbauer studies

$^{155}$Gd Mössbauer spectra obtained at 4.2 K in GdMnSi, GdFeSi and GdCoSi show moderate values of electric field gradients at the Gd site and fits are consistent with the easy magnetization direction observed from neutron diffraction studies of these compounds [562]. $^{57}$Fe Mössbauer spectra for NdFeAl show hypefine splitting at 240 K, while room temperature spectrum is typical for paramagnetic material [457]. Mulders et al. [563] performed $^{169}$Tm and $^{57}$Fe Mössbauer experiments for TmFeAl and reported that below 10 K the $^{169}$Tm Mössbauer spectrum shows sextuplet, characteristic of magnetically ordered Tm

sublattice. The hyperfine field at Tm nuclei estimated from the energy difference between two outer Mössbauer absorption peaks is of 710 T, which is close to the value expected from free Tm ion moment of 7 $\mu_B$ (720 T).   Above 60 K, [169]Tm Mössbauer spectra show asymmetric doublet, characteristic of paramagnetic materials. [57]Fe Mössbauer spectra show distribution of Fe moment with an average $H_{hf}$ of 7.5 T [563].

The [119]Sn Mössbauer spectra for $R$CoSn ($R$=Tb-Tm) show a quadrupole doublet at 295 K, indicating the occupancy of one site of non-cubic symmetry by Sn atoms [564]. Study of low temperature (liquid nitrogen temperature) and high temperature (295 K) reveals that the anisotropic quadrupole pattern at 295 K in these compounds does not arise from an anisotropic Debye-Waller factor and may arise due to presence of impurity phases in these compounds [564]. These compounds show a decrease in isomer shift when going from Tb to Lu is consistent with the lanthanide contraction [564]. Görlich et al. [565] reported that non-collinear complex magnetic structure of $R$CoSn ($R$=Tb-Er) result in a multi-component appearance of the Mössbauer spectra recorded below their ordering temperatures. In addition, the Mössbauer spectra reveal hyperfine split above $T_N$ due to non-vanishing transferred fields at Sn sites in paramagnetic regime in DyCoSn and a change in magnetic ordering at 11.6 K in TbCoSn. Łątka et al. have performed [119]Sn Mössbauer experiments on many Sn containing compounds such as HoAuSn [144], Sm$T$Sn ($T$=Ag, Au) [566], and $R$AuSn [567]. [170]Yb Mössbauer spectra at 1.48 K shows the hyperfine splitting with a hyperfine field, $H_{hf}$=870 kOe at Yb nuclei for YbNiSn [568]. [119]Sn Mössbauer studies in $R$CuSn ($R$=Gd-Er) compounds reveal that small transferred hyperfine fields are detected at the Sn nuclii at 4.2 K which give a hint about the magnetic transitions in these compounds at low temperatures.

[151]Eu Mössbauer studies confirm divalent Eu for EuRhIn [73]. The quadrupole splitting of 10 mm/s at 78 K and hyperfine field of 22.6 T at 4.2 K have been observed from this study. [151]Eu Mössbauer studies show that the isomer shift for isostrutural compounds EuPdIn, EuPtIn and EuAuIn is linearly correlated with the shortest Eu-Eu distance in the structure [126]. Similar to isomershift, the hyperfine field at Eu nuclei at 4.2 K can also be almost linearly correlated with shortest Eu-Eu distance [126]. [170]Yb Mössbauer spectra in YbPdSb show hyperfine splitting below 1 K [569]. To get the information about the strength and anisotropy of the RKKY interaction, the authors performed in field Mössbauer

measurements in the ordered regime for YbPdSb compound. The authors observed that RKKY exchange interaction is anisotropic and $T_{RKKY}$ is of the same order as $T_K$.

Mössbauer studies [312, 314, 570, 571, 572, 573] on $R$RhSn compounds show that there is no magnetic hyperfine field splitting in [119]Sn spectra for LaRhSn and CeRhSn which indicates absence of long range magnetic order. The analysis of [119]Sn Mössbauer spectra reveals that the directions of moments in Pr and Nd compounds are close to $c$-axis [312, 573]. Mössbauer studies suggest non-collinear magnetic structures in NdRhSn [573], GdRhSn [318], DyRhSn [321, 323, 574], HoRhSn [323, 324] and SmRhSn [315, 575] compounds. It was confirmed both by magnetic and Mössbauer studies that in the case of TbRhSn [323, 574, 575, 576], the change in magnetic ordering at $T_{SR}$ = 10.3 K takes place while at the lowest temperature its [119]Sn Mössbauer spectra can be fitted well with a single hyperfine component in accordance with the proposed simple triangular-like magnetic arrangement of Tb moments lying in the hexagonal plane. The [119]Sn Mössbauer studies of CeRhSn and CeIrSn and their hydrides at different temperatures show only one Sn site and larger isomer shift for hydrogenated compounds, which suggests a slightly higher $s$ electron density at the Sn nuclei in hydrogenated compounds [511].

[151]Eu Mössbauer spectra of EuPdSi and EuPtSi show temperature independent behavior of isomer shift which is the characteristic of the divalent Eu ion [92]. The hyperfine splitting has been observed in the case of EuPtSi at 4.2 K, indicating the onset of magnetic ordering. It is worth to mention here that magnetization data of EuPd(Pt)Si do not show the signature of magnetic ordering down to 4.2 K. [151]Eu Mössbauer spectra of EuPdGe show onset of magnetic ordering at 10 K and full hyperfine splitting at 4.2 K with a hyperfine field of 20.7 T at the Eu nuclei [363]. The isomer shift in [151]Eu Mössbauer spectra of Eu$T$Sn ($T$=Cu, Ag, Pt, Pd) indicates no valence fluctuation and divalent nature of Eu while isomer shift in case of [119]Sn is consistent with what is measured in many intermetallic compounds [269]. It has been observed that the value of hypefine field is different and shows a large variation for these compounds below their ordering temperatures. [170]Yb Mössbauer measurements at 25 mK in YbPtAl reveal the modulated magnetic structure in this compound [408]. [119]Sn Mössbauer spectra show broadening below $T_N$ due to magnetic hyperfine field in CePdSn and TbPdSn compounds [364]. [119]Sn Mössbauer studies show non-magnetic

character in LaAgSn and onset of magnetic ordering for the CeAgSn [577] and PrAgSn [577, 578], SmAgSn [566], GdAgSn [579, 580] and ErAgSn [578, 581].

The occurrence of hyperfine field distribution seen in [119]Sn Mössbauer spectra suggest that DyAgSn crystallizes in the disordered CaIn$_2$ type structure [582]. A comprehensive report on the hyperfine interactions as measured by [119]Sn Mössbauer in RAuSn series can be found in ref. [567] and references therein. The experimental results point to a correlation between crystallographic structure and magnetic ordering in the investigated compounds.

Mishra et al. [61] reported [121]Sb and [151]Eu Mössbauer results for Eu$T$Sb ($T$=Cu, Ag, Pt, Au) compounds. Fits to [121]Sb Mössbauer spectra show moderate quadrupole splitting of about -0.29 to 0.24 mm/s. The isomer shift in this case is comparable to ones measured for other intermetallic compounds containing Sb [61 and ref. therein]. No quadrupole splitting was reported in the compounds with ZrBeSi type structure in case of [151]Eu Mössbauer spectra. The hyperfine field of 29.6 and 24.9 T were measured at 4.2 K for EuPtSb and EuCuSb compounds, respectively.

An exhaustive overview of [151]Eu and [155]Gd Mössbauer studies has been published by Pöttgen et al. [7, 583] , which gives all information on their hyperfine interaction parameters and magnetic structures.

## 7.5 Thermopower, thermal expansion and Hall measurements

The thermoelectric power (TEP) was found to be positive in entire temperature range under study in NdCuGe, which indicates that hole conduction dominates in this compound [262]. Unlike NdCuGe, the TEP in CeCuGe shows almost linear T behavior at low temperatures and changes its sign from positive to negative at higher temperatures, which may attributed in part to phonon drag phenomena [584]. Hall measurement for YbNiSn shows an increase in Hall constant ($R_H$) with decrease in temperature and has a sharp peak around ordering temperature [585]. In addition, CeCuGe shows a peak at low temperatures, which is not seen in NdCuGe. The broad maxima seen in TEP of CeRhSn along the $a$ and $c$ axes at 160 K are typical of a valence-fluctuating Ce compounds [466]. Strydom et al. [586] studied thermopower, thermal conductivity and Hall effect of CeRhIn. They found that both thermopower and thermal conductivity show extrema near the $Ce^{3+}$- $Ce^{4+}$ excitation temperature (133 K) [586]. They also found positive and small Hall coefficients, which suggest metal-like carrier concentration [586].

The thermal expansion measurement was performed on CeRhSb and the phonon contribution was subtracted using LaRhSb analog [587]. The results show two distinct features, a broad maximum around 125 K, a large shoulder below 40 K and a temperature where the energy gap opens, $T_g$=10 K [587]. The pressure dependence of thermal expansion in CeRhSb shows that the temperature at which it shows maximum increases with increase in pressure, which is also in agreement with the pressure dependent resistivity data [588]. The Hall coefficient for LaRhSb is found to be -3.7x$10^{-3}$ $cm^3$/C between 0.35-5 K, which corresponds to a carrier density of 89% per f.u. and is 1000 times larger than in the gapped state of CeRhSb [468]. CeRhSb shows a large peak in TEP at 20 K, signaling the onset of Kondo insulator state [589]. TEP in CePdSn shows a broad peak at 110 K and a deep narrow peak at 22 K, which is similar to heavy fermion Ce compound with antiferromagnetic ordering [364]. TEP shows a weak peak around 20 K for other RPdSn (R=Pr, Nd, Sm, Gd-Ho) compounds, which is attributed to the phonon drag [364]. At higher temperatures, the slope of the curve is slightly negative for light rare earth compounds, while slightly positive for heavy rare earth compounds. This difference as well as in resistivity in these compounds hints at the difference in band structures of these two groups [364]. TEP measurements were performed for CeTGe (T=Ni, Pd, and Pt) down to 0.1 K. The TEP shows single and positive peak for CeNiGe and second positive peak was observed in CePdGe and CePtGe [590]. The second positive peak in TEP in CePdGe and CePtGe suggests the existence of Kondo interactions in these compounds and reveal that Kondo interaction may coexist with RKKY interaction even below $T_N$ [590].

Paschen et al. [591] performed TEP and thermal conductivity measurements on single crystal CeNiSn in the temperature range 100 mK-7 K and for field up to 8 T. It has been observed that both TEP and thermal conductivity are highly anisotropic. Below 10 K, the compound shows characteristic features which may be attributed to the opening of a pseudogap in the charge carrier DOS at the Fermi energy for all crystallographic directions which are suppressed by the application of field of 8 T [591]. Thermoelectric and thermo-physical properties of LaPdSb (ZrBeSi-type) and GdPdSb (LiGaGe-type) having hexagonal crystal structures were studied [592]. Both the compounds were found to show relatively high thermoelectric power with very low electrical resistivity, which is caused by the large mobility shown by Hall effect [592]. These materials are promising for new high performance p-type thermoelectric materials. The thermal expansion coefficient of CePdSb shows no feature at $T_C$ (17 K) but shows a broad peak near 10 K, while no unusual behavior has been shown by zero field Sb NMR [593]. The Seebeck coefficient was found to be positive for $R$PdSb (Y, Er), which shows holes are dominant carriers in these compounds [378]. The Hall effect measurements show semimetallic character of the electrical conductivity with holes as majority carriers, which may be due to the presence of some acceptor levels just above the valence band created by atomic disorder [378]. The large TEP together with optimal carrier concentration make $R$PdSb ($R$=Y, Er) compounds promising for thermoelectric applications [378]. TEP of RPdSb and RPdBi is large and positive [100]. The positive TEP shows that the dominant charge carriers in these compounds are holes. The Hall measurements carried out on ErPdSb and ErPdBi show positive Hall coefficient, $R_H$ [100]. Unlike in metals, $R_H$ in these compounds is strongly temperature dependent, which hints at a complex electronic structure with electron and hole band containing carriers with different temperature dependent mobilities [100]. It has been observed that results obtained in $R$PdBi compounds seem to be fully compatible with the theoretical predictions for topological insulators [100 and ref. therein]. Gofryk et al. [100] reported that a non-trivial zero gap semiconducting state has been postulated for YPdBi with relatively small topological band inversion strength. TEP in LaPdBi and GdPdBi is positive and decreases with increasing temperature [594]. The differences in TEP and electrical resistivity between LaPdBi and GdPdBi arise due to the difference in band gap [594]. The thermal conductivity of both the compounds increases with increase in temperature, signalling that the electronic thermal conductivity is predominant [594].

CeIrSn, CePdSb and CeRhSn were found to show large TEP measured in temperature range 1.5-500 K [595]. The large TEP was attributed to the strong scattering of conduction electrons by the 4$f$ electrons in the vicinity of Fermi level [595]. The temperature dependence of thermo power for valence fluctuating Ce has been explained by two-band conductor model assuming a single Lorentzian 4f band [595]. Generally, a large peak appears in TEP when a pseudogap opens in electronic density of states at low temperatures. The TEP for CeIrSb do not show any such type of peak and hence rule out the formation of pseudo gap [118]. Thermal expansion and magnetostriction measurement performed on the CePtSn show that two phase transitions are present where the dl/l vs. T curve shows change in slope. With increase in field (along b-axis), there is no change in the nature of these transition temperatures, rather a change in position of lower transition temperature, which shifts to lower temperatures, occurs [596].

## 7.6 High pressure study

The high-pressure study on CeRhSn shows that features associated with quantum coherence move to higher temperatures [597]. The effect of pressure on $T_N$ (using heat capacity data) of CePdAl is insensitive in the low pressure but show a decrease at higher pressures [598]. Similar effects were also observed in the susceptibility data on the application of pressure. The susceptibility and $T_N$ decrease with pressure and AFM order disappears above 1.1 GPa, which suggest the occurrence of quantum phase transition at this pressure [336]. The pressure dependence on magnetic behavior of CePdSb has been explained by the Doniach phase diagram where the pressure shows an increase in $JN(E_F)$ and shows asymmetry in $T_C(P)$, which was first time observed in the Ce compound by the application of high pressure alone [599]. Umeo et al. [600] studied the pressure effect on YbRhSb compound. The authors observed that with increase in pressure, the ordering temperature increases from 2.7 K to 3.3 K at 2.5 GPa and for pressure $0.9 < P < 1.5$ GPa, another magnetic phase transition occurs. The authors reported that complex magnetic behavior in this compound on the application of pressure may be due to the competition between single ion magneto-crystalline anisotropy with easy magnetization direction along *a*-axis and inter-site exchange interaction with easy direction along *c*-axis. Further, the same group [601] studied this sytem with higher pressures and found some interesting results.

Huppertz et al. [602] reported high temperature and high pressure studies of EuPdIn and EuPtIn compounds. The authors observed that under high pressure and high temperature these compounds decompose into $EuPd_{0.72}In_{1.28}$ and $EuPt_{0.56}In_{1.44}$. These compounds crystallize in $MgZn_2$ type structure. Riecken et al. [603] reported hexagonal high pressure modifications of RPtSn (R=La, Pr, Sm) from the orthorhombic normal pressure modifications. Later, it was observed that RNiSn (R=Ce, Pr, Nd, Sm) and RPdSn (R=La, Pr, Nd), which crystallize in TiNiSi type orthorhombic structure at normal pressure were transformed into corresponding ZrNiAl type hexagonal structure under multi-anvil high pressure (7.5-11.5 GPa) high temperature (1100-1200 ˚C) conditions [604]. High pressure/high temperature phases were also reported for CePdSn [605], RPtSn (R=Y, Gd, Tb) [606], LaPdSn and ErAgSn (a first principles study) [607].

Ab-initio calculation performed by Mihalik et al. shows that Rh has a small moment (0.1 $\mu_B$) due to the polarization of Rh 4$d$ electron state from Nd 5$d$ electron state [316]. *ab initio* calculations using DFT were performed on CeIrGe, which reveal strong contribution of the Ce 4$f$ states at the Fermi level [521].

# 8 Conclusions

Based on the above, it is quite clear that *RTX* compounds show a variety of crystal structures and physical properties. This is probably the most important intermetallic series in which one can look for materials with properties ranging from superconducting to topological insulator behavior. In this sense, this is like the CMR oxides family, which has also been reviewed well. *RTX* series has materials with all possible magnetic interactions and a range of magnetic ordering temperatures. As is clear from this review, a large number of experimental probes, both bulk and local, have been used to investigate this series. An interesting observation from these studies is that except Mn, T atoms generally do not possess any magnetic moment in the entire RTX series. Negligible contribution of moment from the T sublattice is attributed to the hybridization between p states and d states of X and T element, respectively. Because of this, RKKY interaction plays a major role in determining the magnetic interactions and the magnetic structure. Peculiar and tunable electronic band structure is another positive aspect of this series. Study of these compounds gives a deep insight into the fundamental physics of not only this series, but many other systems containing rare earths and nonmagnetic elements. As can be seen, of late, the interest in RTX compounds is to exploit their magnetic and related properties for potential applications and in the design of novel materials. This has been possible because of the extensive work carried out by a large number of groups on this series, which this review has tried to present in a condensed form. There are still many intriguing phenomena unexplored in this series and we hope that an even better understanding in the near future will lead to many more novel and exotic materials and properties from this series.

## Acknowledgment

Sachin Gupta Thanks CSIR, Government of India for providing senior research fellowship.

Table I. List of crystal structures and their space groups of *RTX* compounds

| Compound | Structure type | Crystal structure | Space group | Reference |
|---|---|---|---|---|
| *R*ScSi (*R*=La-Nd, Sm, Gd) | CeScSi | tetragonal | *I4/mmm* | [1, 2,3 4] and ref. |
| *R*ScSi (*R*=Tb-Tm) | Ti$_5$Ga$_4$ | hexagonal | *P6$_3$/mcm* | therein |
| *R*ScGe (*R*= Dy-Tm) | Ti$_5$Ga$_4$ | hexagonal | *P6$_3$/mcm* | [1, 2, 3, 4, 6] and |
| *R*ScGe (*R*=La-Nd, Sm, Eu Gd,Tb) | CeScSi | tetragonal | *I4/mmm* | ref. therein |
| *R*TiSi (*R*=Y, Gd-Tm, Lu) | CeFeSi | tetragonal | *P4/nmm* | [8, 9] |
| *R*TiGe (*R*=Y, La-Nd, Sm, Gd-Tm, Lu) | CeFeSi | tetragonal | *P4/nmm* | [10, 11, 17] |
| CeTiGe (HTM), GdTiGe, TbTiGe | CeScSi | tetragonal | *I4/mmm* | [2, 14, 15, 17, 18] |
| GdTiSb | CeFeSi | tetragonal | *P4/nmm* | [19] |
| *R*MnSi (*R*=La-Sm, Gd) | CeFeSi | tetragonal | *P4/nmm* | [20] |
| *R*MnSi (*R*=Tb-Er) | Co$_2$Si | orthorhombic | *Pnma* | [65] |
| *R*MnSi (*R*=Tb-Ho) | TiNiSi | orthorhombic | *Pnma* | [23, 164] |
| RMnGe(La-Nd) | CeFeSi | tetragonal | *P4/nmm* | [165] |
| NdMnGe | TiNiSi | orthorhombic | *Pnma* | [23] |
| RMnGe (Gd-Er, Y) | TiNiSi | orthorhombic | *Pnma* | [21] and ref. therein |
| TmMnGe** | TiNiSi | orthorhombic | *Pnma* | [21] |
| TmMnGe*** | ZrNiAl | hexagonal | $P\bar{6}2m$ | [21] |
| RMnAl (R=Ce, Nd, Gd) | MgCu₂ | cubic | $Fd\bar{3}m$ | [24, 25, 26] |
| RMnGa (R=Ce, Pr, Nd) Gd, Tb, Dy) | MgCu₂ | cubic | $Fd\bar{3}m$ | [27, 28] |
| RMnGa (R=Gd, Tb, Dy, Ho) | Fe₂P | hexagonal | $P\bar{6}2m$ | [28] |
| RMnIn (R=Gd, Dy, Er, Y) | MgZn₂ | hexagonal | $P6_3/mmc$ | [26] |
| RFeAl (Ce-Nd, Sm) | MgCu₂+* | cubic | | [29] |
| RFeAl (Gd-Lu) | MgZn₂ | hexagonal | $P6_3/mmc$ | [29, 170] |
| RFeSi( R=La-Sm, Gd-Er) | CeFeSi | tetragonal | $P4/nmm$ | [173, 177, 178] |
| RCoAl (R=Gd-Lu) | MgZn₂ | hexagonal | $P6_3/mmc$ | [30] |
| CeCoGa | CeCoAl | monoclinic | $C2/m$ | [181] |
| RCoSi (R=La-Sm, Gd, Tb) | CeFeSi | tetragonal | $P4/nmm$ | [31] |
| TbCoSi | TiNiSi | orthorhombic | $Pnma$ | [32] |
| RCoSi (R=Ho-Lu) | TiNiSi | orthorhombic | $Pnma$ | [564] |
| RCoGe (R=La-Nd) | CeFeSi | tetragonal | $P4/nmm$ | [185] |
| RCoGe (R=Gd-Lu, Y) | TiNiSi | orthorhombic | $Pnma$ | [564] |
| RCoSn (R=Tb-Lu, Y) | TiNiSi | orthorhombic | $Pnma$ | [564] |
| RNiAl (R=Ce-Nd,Sm, Gd-Lu) | ZrNiAl | hexagonal | $P\bar{6}2m$ | [33] |
| CeNiGa (LTP) | ZrNiAl | hexagonal | $P\bar{6}2m$ | [38] |
| CeNiGa (HTP) | TiNiSi | orthorhombic | $Pnma$ | |
| RNiGa (R=Gd-Tm) | CeCu₂ | orthorhombic | $Imma$ | [39] |
| YbNiGa | TiNiSi | orthorhombic | $Pnma$ | [122] |
| RNiIn (R=La-Sm, Gd-Tm) | ZrNiAl | hexagonal | $P\bar{6}2m$ | [204] |
| RNiSi (R=La, Ce, Nd) | LaPtSi | tetragonal | $I4_1md$ | [40, 41, 209] |
| RNiSi (R=Gd-Lu) | TiNiSi | orthorhombic | $Pnma$ | [42] |
| RNiGe (R=Ce, Gd-Tm, Y) | TiNiSi | orthorhombic | $Pnma$ | [214, 504] |
| RNiSn (R=La-Sm, Gd-Lu, Yb) | TiNiSi | orthorhombic | $Pnma$ | [94] |
| RNiSb (R=La-Nd, Sm) | ZrBeSi | hexagonal | $P6_3/mmc$ | [43] |
| RNiSb (R=Gd-Lu) | MgAgAs | cubic | $F\bar{4}3m$ | [43] |
| RCuAl (R=Ce-Nd,Sm, Gd-Lu) | ZrNiAl | hexagonal | $P\bar{6}2m$ | [33] |
| CeCuGa | CeCu$_2$ | orthorhombic | *Imma* | [46] |
| EuCuGa | CeCu$_2$ | orthorhombic | *Imma* | [60] |
| RCuIn (R=La-Nd, Gd, Tb, Ho, Er, Lu) | ZrNiAl | hexagonal | $P\bar{6}2m$ | [248] |
| RCuSi (R=La-Nd, Sm, Gd-Lu, Yb, Y) | Ni$_2$In** | hexagonal | *P6$_3$/mmc* | [48, 49, 50] |
| | AlB$_2$*** | hexagonal | *P6/mmm* | |
| RCuGe (R=La-Lu) | AlB$_2$$^{a*}$ | hexagonal | *P6/mmm* | [51, 52, 53] |
| (R=La-Nd, Sm, Gd) | AlB$_2$$^{a**}$ | hexagonal | *P6/mmm* | |
| (R=Tb-Lu, Y) | CaIn$_2$$^{a**}$ | hexagonal | *P6$_3$/mmc* | |
| RCuSn (R=Ce- Nd) | CaIn$_2$ | hexagonal | *P6$_3$/mmc* | [264, 265] |
| EuCuSn | CeCu$_2$ | orthorhombic | *Imma* | [59] |
| RCuSn (Gd-Er) | LiGaGe | hexagonal | *P6$_3$mc* | [263] |
| EuCuAs | Ni$_2$In | hexagonal | *P6$_3$/mmc* | [62] |
| EuCuSb | ZrBeSi | hexagonal | *P6$_3$/mmc* | [61] |
| YbCuBi | LiGaGe | hexagonal | *P6$_3$mc* | [63] |
| CeRuAl | LaNiAl | orthorhombic | *Pnma* | [47] |
| CeRhAl | LaNiAl | orthorhombic | *Pnma* | [47] |
| RRuSi (R=La-Sm,Gd) | CeFeSi | tetragonal | *P4/nmm* | [64] |
| RRuSi (R=Y, Tb-Tm) | Co$_2$Si | orthorhombic | *Pnma* | [65] |
| RRuGe (R=La-Sm) | CeFeSi | tetragonal | *P4/nmm* | [64] |
| RRuGe (R=Sm, Gd-Tm) | TiNiSi | orthorhombic | *Pnma* | [66] |
| CeRuSn | CeCoAl | monoclinic | *C2/m* | [68] |
| RRhAl (R=Y, Pr, Nd, Gd, Ho, Tm) | TiNiSi | orthorhombic | *Pnma* | [69, 283] |
| RRhAl (R=La, Ce) | Pd$_2$(Mn,Pd)Ge$_2$ | orthorhombic | *Pnma* | [69] |
| RRhGa (R=La-Nd, Gd, Tb, Ho-Lu, Yb, Y] | TiNiSi | orthorhombic | *Pnma* | [67, 70, 71] |
| RRhIn (R=La-Nd, Sm, Gd-Tm) | ZrNiAl | hexagonal | $P\bar{6}2m$ | [72] |





| | | | | |
|---|---|---|---|---|
| RRhIn (R=Eu, Yb, Lu) | TiNiSi | orthorhombic | *Pnma* | [72, 73] |
| RRhSi (R=Y, Gd-Er) | TiNiSi | orthorhombic | *Pnma* | [75] |
| LaRhSi | ZrOS | cubic | *P2$_1$3* | [76] |
| RRhGe (R=Ce-Nd, Sm, Gd-Yb, Y) | TiNiSi | orthorhombic | *Pnma* | [67, 77] |
| RRhSn (R=La-Nd, Sm, Gd-Lu) | ZrNiAl | hexagonal | *P$\bar{6}$2m* | [79] |
| RRhSb (R=La, Pr, Sm, Gd-Tm) | TiNiSi | orthorhombic | *Pnma* | [81, 82] |
| CeRhSb | CeCu$_2$ | orthorhombic | *Imma* | [331] |
| CeRhBi | TiNiSi | orthorhombic | *Pnma* | [333] |
| CeRhAs | TiNiSi | orthorhombic | *Pnma* | [333] |
| RPdAl (R=Sm, Gd-Tm, Y) | TiNiSi | orthorhombic | *Pnma* | [83] |
| RPdAl (R= Ce-Nd, Sm, Gd-Tm, Lu, Y)$^{a***}$ | ZrNiAl | hexagonal | *P$\bar{6}$2m* | [84] |
| RPdGa (R=La-Nd, Sm, Eu, Gd-Tm, Lu, Y) | TiNiSi | orthorhombic | *Pnma* | [60,67] |
| RPdIn (R=La-Sm, Gd-Lu, Y) | ZrNiAl | hexagonal | *P$\bar{6}$2m* | [[87, 88] |
| EuPdIn | TiNiSi | orthorhombic | *Pnma* | [89] |
| RPdSi (R=La-Nd, Sm-Lu, Y) | TiNiSi | orthorhombic | *Pnma* | [90] |
| RPdGe (R= La-Nd, Sm, Gd-Tm, Y) | CeCu$_2$ | orthorhombic | *Imma* | [67] |
| EuPdGe | EuNiGe | monoclinic | *P2$_1$/n* | [93] |
| RPdSn (R=La, Ce-Nd, Sm-Yb) | TiNiSi | orthorhombic | *Pnma* | [95, 96] |
| RPdSn (R=Er, Tm) | Fe$_2$P | hexagonal | *P$\bar{6}$2m* | [95] |
| EuPdAs | Ni$_2$In | hexagonal | *P6$_3$/mmc* | [97] |
| RPdSb (R=Ce-Nd, Sm, Gd-Dy) | CaIn$_2$ | hexagonal | *P6$_3$/mmc* | [366] |
| EuPdSb | TiNiSi | orthorhombic | *Pnma* | [61] |
| RPdSb (R=Ho, Er, Tm) | MgAgAs | cubic | *F$\bar{4}$3m* | [366, 99] |





| | | | | |
|---|---|---|---|---|
| RPdBi (R=La-Nd, Sm Gd-Lu) | MgAgAs | cubic | $F\bar{4}3m$ | [82 and ref therein] |
| RAgAl (R=La-Nd, Sm, Gd-Er, Y) | CeCu$_2$ | orthorhombic | Imma | [101, 102, 103] |
| RAgGa (R=La- Nd, Eu, Gd-Lu, Y) | CeCu$_2$ | orthorhombic | Imma | [60,104, 105 and ref. therein] |
| RAgSi (R=Sm, Gd- Tm, Yb, Lu, Y) | ZrNiAl | hexagonal | $P\bar{6}2m$ | [109, 389] |
| RAgGe (R=La, Ce) | CaIn$_2$ | hexagonal | P6$_3$/mmc | [106] |
| EuAgGe | CeCu$_2$ | orthorhombic | Imma | [111] |
| RAgGe (R=Gd-Lu) | ZrNiAl | hexagonal | $P\bar{6}2m$ | [107, 108] |
| RAgSn (R=La, Ce, Sm, Gd-Er, Yb) | CaIn$_2$ | hexagonal | P6$_3$/mmc | [110] |
| EuAgSn | CeCu$_2$ | orthorhombic | Imma | [59] |
| RAgSn (R=Ce-Nd, Gd-Er) | LiGaGe | hexagonal | P6$_3$mc | [112] |
| EuAgSb | ZrBeSi | hexagonal | P6$_3$/mmc | [61] |
| YbAgBi | LiGaGe | hexagonal | P6$_3$mc | [63] |
| CeOsSi | CeFeSi | tetragonal | P4/nmm | [64] |
| RIrAl (R=Ce-Nd, Sm, Gd-Tm, Lu, Y) | TiNiSi | orthorhombic | Pnma | [115] |
| RIrGa (R=La-Nd, Gd-Tm, Y) | TiNiSi | orthorhombic | Pnma | [67, 70] |
| RIrSi (R=La, Nd) | ZrOS | cubic | P2$_1$3 | [76] |
| RIrSi (R=Tb-Er) | TiNiSi | orthorhombic | Pnma | [120] |
| RIrGe (R=Ce, Gd-Er, Yb) | TiNiSi | orthorhombic | Pnma | [116, 117, 119] |
| RIrSn (R=La-Nd, Sm, Gd, Tb, Ho-Lu) | Fe$_2$P | hexagonal | $P\bar{6}2m$ | [121] |
| RIrSb (Ce, Yb) | TiNiSi | orthorhombic | Pnma | [61,118] |
| RPtAl (R=Sm, Gd-Tm, Lu, Y) | TiNiSi | orthorhombic | Pnma | [83] |
| RPtGa (R=La-Nd, Sm-Lu, Y) | TiNiSi | orthorhombic | Pnma | [60,67] |
| RPtIn (R=La-Nd, Sm, Gd- Lu, Y) | ZrNiAl | hexagonal | $P\bar{6}2m$ | [123, 124, 125] and ref. therein |





| EuPtIn | TiNiSi | orthorhombic | *Pnma* | [126] |
|---|---|---|---|---|
| RPtSi (R=La-Nd, Sm, Gd) | LaPtSi | tetragonal | *I4₁md* | [67] |
| RPtSi (R=Tb-Tm, Lu, Y) | TiNiSi | orthorhombic | *Pnma* | [67] |
| LaPtGe | LaPtSi | tetragonal | *I4₁md* | [67] |
| RPtGe (R=Ce-Nd) | CeCu₂ | orthorhombic | *Imma* | [67] |
| EuPtGe | LaIrSi | cubic | *P2₁3* | [127] |
| RPtGe (R=Sm, Gd-Tm,Y) | TiNiSi | orthorhombic | *Pnma* | [67] |
| RPtSn (R=La-Eu) | TiNiSi | orthorhombic | *Pnma* | [429, 433] |
| RPtSn (R=Gd-Lu) | Fe₂P | hexagonal | $P\bar{6}2m$ | [128] |
| RPtSb (R=La-Sm) | CaIn₂ | hexagonal | *P6₃/mmc* | [133] |
| EuPtSb | TiNiSi | orthorhombic | *Pnma* | [133] |
| RPtSb (R=Gd-Lu) | MgAgAs | cubic | $F\bar{4}3m$ | [133] |
| RAuAl (La, Ce, Nd) | TiNiSi | orthorhombic | *Pnma* | [134] |
| RAuGa (R=La-Nd, Sm-Lu, Y) | CeCu₂ | orthorhombic | *Imma* | [105] |
| RAuIn (R=Ce-Sm, Gd-Tm, Lu] | ZrNiAl | hexagonal | $P\bar{6}2m$ | [135, 136] |
| EuAuIn | TiNiSi | orthorhombic | *Pnma* | [137] |
| RAuGe (R=La-Nd, Sm, Gd-Er, Y) | LiGeGa | hexagonal | *P6₃mc* | [138] |
| EuAuGe | CeCu₂ | orthorhombic | *Imma* | [143] |
| RAuSn ( R= La-Nd) | CaIn₂ | hexagonal | *P6₃/mmc* | [142, 446] |
| RAuSn ( R= Gd-Ho) | LiGeGa | hexagonal | *P6₃mc* | [141] |
| ErAuSn[a*] | MgAgAs | cubic | $F\bar{4}3m$ | [140,144] |
| ErAuSn[a***] | CaIn₂ | hexagonal | *P6₃/m* | [140,144] |
| ErAuSn | MgAgAs | cubic | $F\bar{4}3m$ | [141] |
| YbAuSn | TiNiSi | orthorhombic | *Pnma* | [63] |
| EuAuSb | ZrBeSi | hexagonal | *P6₃/mmc* | [61] |
| YbAuBi | MgAgAs | cubic | $F\bar{4}3m$ | [63] |





*=unidentified phase, **=low temperature phase, ***=high temperature phase, a*= with out annealing, a**= after annealing, a***=rapidly cooled

Table II. Nature of magnetic ordering, ordering temperature, paramagnetic Curie temperature and effective magnetic moment of RTX compounds. In cases where the magnetization data is not available, the ordering temperatures are taken from neutron diffraction data.

| Compound | Type of ordering | Magnetic phase transition(s) | $\theta_p$ (K) | $p_{eff}$ ($\mu_B$/f.u.) | Reference |
|---|---|---|---|---|---|
| PrScSi | AFM | 145 | 99 | 2.65 | [3] |
| NdScSi | FM | 175 | 178 | 2.96 | [3] |
| SmScSi | FM | 270 | | | [3] |
| GdScSi | FM | 318 | 338 | 7.9 | [145] |
| | FM | 352 | | | [5] |
| TbScSi | AFM | 165 | 80 | 10.1 | [145] |
| DyScSi | | | 34 | 11.4 | [145] |
| HoScSi | | | 13 | 11.4 | [145] |
| ErScSi | | | 5 | 10.3 | [145] |
| TmScSi | | | -11 | 8.4 | [145] |
| PrScGe | FM, AFM | 80, 88, 140 | 125 | 2.88 | [3] |
| | AFM, FIM, FIM | 140, 82, 62 | | | [149] |
| NdScGe | FM | 200 | 195 | 2.65 | [3] |





| | | | | | |
|---|---|---|---|---|---|
| SmScGe | FM | 270 | | | [3] |
| GdScGe | FM | 320 | 332 | 7.8 | [145] |
| | FM | 348 | | | [5] |
| TbScGe | FM | 216 | 215 | 9.9 | [145] |
| | FM | 250 | 244 | 9.7 | [150] |
| DyScGe | | | 37 | 11.3 | [145] |
| HoScGe | | | 10 | 11.2 | [145] |
| ErScGe | | | -5 | 10.6 | [145] |
| TmScGe | | | 10 | 8.2 | [145] |
| YTiSi | PPM | | | | [8] |
| GdTiSi | AFM | 400 | 369 | 7.3 | [8] |
| TbTiSi | AFM | 286 | 186 | 10.14 | [8] |
| DyTiSi | AFM | 170 | 110 | 10.43 | [8] |
| HoTiSi | AFM | 95 | 50 | 11 | [8] |
| ErTiSi | AFM | 50 | 25 | 9.7 | [8] |
| TmTiSi | AFM | 20 | 11 | 7 | [8] |
| LuTiSi | PPM | | | | [8] |
| YTiGe | PPM | | | | [13] |
| LaTiGe | PPM | | | | [11] |
| CeTiGe | | | -20 | 2.7 | [11] |
| CeTiGe[HTM] | | | -59 | 2.6 | [14] |





| | | | | | |
|---|---|---|---|---|---|
| CeTiGe | | | -27 | 2.55 | [153] |
| PrTiGe | AFM | 70, 55 | 26 | 3.5 | [11] |
| NdTiGe | AFM | 150, 90 | 93 | 3.6 | [11] |
| SmTiGe | AFM | 260, 215 | NCW | NCW | [11] |
| GdTiGe# | AFM | 412 | 317 | 8.3 | [13] |
| GdTiGe## | FM | 400 | 460 | 7.8 | [11] |
| GdTiGe## | FM | 376 | 416 | 8.39 | [16] |
| TbTiGe# | AFM | 270 | 263 | 9.9 | [11] |
| TbTiGe# | AFM | 288 | 248 | 9.4 | [13] |
| TbTiGe## | FM | 300 | 285 | 10.41 | [18] |
| DyTiGe | AFM | 170 | 155 | 10.7 | [11] |
| | AFM | 175 | 120 | 10.7 | [13] |
| | AFM | 179 | 150 | 10.8 | [156] |
| HoTiGe | AFM | 115, 85 | 103 | 10.7 | [11] |
| | AFM | 115 | 101 | 10.7 | [13] |
| | AFM | 90 | 74.2 | 10.3 | [157] |
| ErTiGe | AFM | 41 | 34 | 10 | [11] |
| TmTiGe | AFM | 15 | 1 | 7.6 | [11] |
| LuTiGe | PPM | | | | [11] |
| GdTiSb | FM | 268 | 350 | 7.5 | [19] |
| GdMnAl | FM | 274 | | | [25] |





| | | | | | |
|---|---|---|---|---|---|
| TbMnAl | AFM | 34 | | | [159] |
| CeMnGa | SG | | -144 | 3.1 | [27] |
| PrMnGa | RSG | 90, 50[r] | -17 | 3.7 | [27] |
| NdMnGa | SG | 10[g] | -11 | 3.4 | [27] |
| GdMnGa | RSG | 220, 80[r] | - | - | [27] |
| TbMnGa | RSG | 120, 85[r] | -6 | 2.7 | [27] |
| DyMnGa | SG | 40[g] | 5 | 3.5 | [27] |
| LaMnSi | AFM | 310[**] | -350 | 3.07 | [20] |
| CeMnSi | AFM | 240[**] | -50 | 3.80 | [20] |
| PrMnSi | AFM | 130, 80 | 75 | 4.69 | [20] |
| NdMnSi | AFM | 175, 80 | 110 | 4.6 | [20] |
| SmMnSi | AFM | 235, 220 , 110 | NCW | | [20] |
| GdMnSi | FM | 310 | 265 | 8.52 | [20] |
| | FM | 314 | | | [161] |
| TbMnSi[###] | AFM | 410, 140 | 44 | 10.2 | [23] |
| | FM | 260 | 164 | 10.37 | [163] |
| DyMnSi | AFM | 400, 55 | 22 | 11.1 | [23] |
| HoMnSi | AFM | 15 | -58 | 12.35 | [164] |
| LuMnSi | AFM | 255 | -201 | 3.17 | [164] |
| LaMnGe | AFM | 420[**] | | | [165] |
| CeMnGe | AFM | 415 [**] | | | [165] |





|  |  | 313, 41 |  |  | [483] |
|---|---|---|---|---|---|
| PrMnGe | AFM | 415 [**] |  |  | [165] |
|  | FM | 150 |  |  |  |
| NdMnGe[###] | AFM | 430 | -31 | 5.3 | [23] |
|  |  | 410 [**] |  |  | [165] |
|  | FM | 190 |  |  | [165] |
| GdMnGe | AFM | 490, 200, 60 | 64 | 8.65 | [21] |
|  | AFM | 350 | 80 | 8.6 | [166] |
| TbMnGe | AFM | 510, 186, 70 | 39 | 10.6 | [168] |
| DyMnGe |  | 70 | 24 | 11.2 | [21] |
| HoMnGe |  | 12 | 11 | 11.2 | [21] |
| ErMnGe |  |  | -34 | 10.39 | [21] |
| TmMnGe |  | 17 | -36 | 8.7 | [21] |
| YMnGe |  |  | -78 | 3.6 | [21] |
| YFeAl | FM | 38 |  |  | [169] |
| GdFeAl | FM | 265 |  |  | [170] |
| TbFeAl | FM | 195 |  |  | [169] |
| DyFeAl | FM | 128 | 143 | 10.76 | [171] |
| HoFeAl | FM | 92 |  |  | [169] |
| ErFeAl | FM | 55 |  |  | [172] |
| TmFeAl | FM | 38 |  |  | [169] |





| | | | | | |
|---|---|---|---|---|---|
| LuFeAl | FM | 39 | | | [169] |
| LaFeSi | PPM | | | | [173] |
| CeFeSi | IV | | | | [173] |
| PrFeSi | | | 35 | 3.62 | [173] |
| NdFeSi | FM | 25 | 20 | 3.90 | [173] |
| SmFeSi | FM | 40 | | | [173] |
| GdFeSi | FM | 135 | 165 | 8.09 | [173] |
| | FM | 118 | 122 | 8.68 | [175] |
| TbFeSi | FM | 125 | 110 | 9.62 | [173] |
| | FM | 110 | 90 | 10.28 | [176] |
| DyFeSi | FM | 110 | 75 | 10.57 | [173] |
| | FM | 70 | 69 | 11.41 | [176] |
| HoFeSi | FM, AFM/FIM | 29, 20 | 24.8 | 11.25 | [177] |
| ErFeSi | FM | 22 | 19.7 | 9.84 | [178] |
| GdCoAl | FM | 100 | | | [179] |
| TbCoAl | FM | 70 | | | [179] |
| DyCoAl | FM | 37 | 44 | 10.73 | [179, 180] |
| HoCoAl | FM | 10 | | | [179] |
| CeCoGa | AFM | 4.3 | -82 | 1.8 | [181] |
| LaCoSi | PPM | | | | [31] |
| CeCoSi | | | -53 | 2.8 | [31] |
| | AFM | 8.8 | -55 | 2.71 | [182] |





| | | | | | |
|---|---|---|---|---|---|
| PrCoSi | | | -10 | 3.9 | [31] |
| NdCoSi | AFM | 7 | 30 | 3.9 | [31] |
| SmCoSi | AFM | 15 | NCW | | [31] |
| GdCoSi | AFM | 175 | 70 | 8.6 | [31] |
| TbCoSi | AFM | 140 | 70 | 9.6 | [31] |
| HoCoSi | FM | 15 | 10 | 10.9 | [183] |
| | FM | 14 | 6 | 10.9 | [184] |
| LaCoGe | PPM | | | | [185] |
| CeCoGe | | | 15 | 2.6 | [185] |
| | AFM | 5 | -39 | 2.6 | [484] |
| PrCoGe | | | -7 | 3.9 | [185] |
| NdCoGe | AFM | 8 | -6 | 4 | [185] |
| TbCoGe | FM | 16 | 17.5 | 9.48 | [183] |
| TbCoSn | AFM | 20.5 | | | [187] |
| DyCoSn | AFM | 10 | 9 | 10.5 | [186] |
| HoCoSn | AFM | 7.8 | 6.5 | 11 | [186] |
| ErCoSn | AFM | 5 | 0 | 9.8 | [186] |
| RNiAl (R=Yb, Lu) | PPM | | | | [189] |
| PrNiAl | | | -10 | 3.73 | [189] and ref. therein |
| | TSW | 6.9 | -23 | 3.7 | [191] |





| | | | | | |
|---|---|---|---|---|---|
| NdNiAl | FM | 15 | 5 | 3.84 | [189, and ref. therein] |
| | TSW | 2.7 | -7.5 | 3.8 | [191] |
| GdNiAl | FM | 66 | 56 | 8.5 | [189] |
| | FM, AFM | 57, 31 | | | [25] |
| TbNiAl | FM | 57 | 45 | 10.1 | [189] |
| | AFM | 47, 23 | 30 | 10.3 | [191] and ref. therein |
| DyNiAl | FM | 39 | 30 | 11.1 | [189] |
| | FM, AFM | 31, 15 | 17 | 10.9 | [191] |
| HoNiAl | FM | 25 | 11 | 10.8 | [189] |
| | FM, AFM | 14.5, 12.5, 5.5 | 7.3 | 10.7 | [191] |
| ErNiAl | FM | 15 | -1 | 9.8 | [189] |
| TmNiAl | FM, AFM | 12, 4.2 | -11 | 7.8 | [189] |
| YbNiAl | AFM | 2.9 | | 4.4 | [190] |
| GdNiGa | FM | 30.5 | 32 | 7.85 | [207] |
| TbNiGa | AFM | 23 | | | [198] |
| HoNiGa | AFM | 12 | | | [200] |
| ErNiGa | AFM | 8 | | | [201] |
| TmNiGa | AFM | 5.5 | -49 | 7.79 | [202] |
| YbNiGa | AFM | 1.9, 1.7 | -30 | 4.4 | [122] |
| GdNiIn | FM | 98 | 101 | 7.96 | [208] |
| | FM | 93.5 | 91 | 7.7 | [205] |





| | | | | | |
|---|---|---|---|---|---|
| TbNiIn | FM, AFM | 71, 12 | 47 | 9.8 | [208] |
| | AFM | 68 | | | [204] |
| DyNiIn | FM, AFM | 30, 14 | 31 | 10.68 | [208] |
| | AFM | 32 | | | [204] |
| HoNiIn | FM | 20, 7 | 18 | 10.61 | [208] |
| ErNiIn | FM | 9 | 8 | 9.64 | [208] |
| TmNiIn | AFM | 2.5 | 0.4 | 7.5 | [206] |
| NdNiSi | AFM | 6.8, 2.8 | | | [211] |
| TbNiSi | AFM | 16 | -6.1 | 9.7 | [40] |
| DyNiSi | AFM | 8.8 | -7.5 | 10.5 | [40] |
| HoNiSi | AFM | 4.1 | -5 | 10.3 | [40] |
| | AFM | 4.2 | -10.1 | 10.9 | [260] |
| ErNiSi | AFM | 3.3 | 0 | 8.9 | [40] |
| | AFM | 3.2 | 0 | 9.57 | [213] |
| GdNiGe | AFM | 11 | -10 | 7.85 | [214] |
| TbNiGe | AFM | 7 | -9 | 9.45 | [214] |
| | AFM | 18.5 | 4.1 | 8.97 | [215] |
| DyNiGe | AFM | 9 | -6 | 10.40 | [214] |
| | AFM | 4.7 | -8.2 | 10.42 | [215] |
| HoNiGe | AFM | | -3 | 10.43 | [214] |
| | AFM | 5 | -2.6 | 11.2 | [216] |





| | | | | | |
|---|---|---|---|---|---|
| ErNiGe | AFM | 6 | -3 | 9.28 | [214] |
| | AFM | 2.9 | -1.4 | 9.19 | [216] |
| TmNiGe | AFM | | -4 | 7.23 | [214] |
| YbNiGe | | | 3 | 4.31 | [22] and ref. therein |
| NdNiSn | AFM | 2.8 | | | [223] |
| SmNiSn | AFM | 9 | -40 | 1.37 | [218] |
| GdNiSn | AFM | 10.5 | -3 | 8.8 | [217] |
| TbNiSn | AFM | 7.2 | 6 | 11.27 | [217] |
| | | 18.5 | -26* | 9.65 | [219] |
| DyNiSn | AFM | 8.2 | -1 | 11.06 | [217] |
| HoNiSn | AFM | - | -2 | 10.75 | [217] |
| | AFM | 4.2 | | | [227] |
| ErNiSn | AFM | - | 6 | 9.85 | [217] |
| TmNiSn | AFM | - | -4 | 7.65 | [217] |
| YbNiSn | FM | 5.5 | -43 | 4.3 | [237] |
| YbNiSb | AFM | 0.8 | -13 | 4.6 | [238] |
| RNiSb (R=Y, La) | PPM | | | | [44] |
| CeNiSb | FM+AFM | 3.5 | -27 | 2.9 | [44] |
| PrNiSb | PM | | -0.7 | 3.8 | [44] |
| NdNiSb | FM | 23 | 13 | 3.7 | [44] |





| | | | | | |
|---|---|---|---|---|---|
| SmNiSb | VVP | | | 1.58 | [43] |
| GdNiSb<sup>cubic</sup> | | | -15 | 8.1 | [43] |
| GdNiSb<sup>cubic</sup> | AFM | 9.5 | | | [45] and ref. therein |
| GdNiSb<sup>hex</sup> | AFM | 3.5 | | | [45] and ref. therein |
| TbNiSb | AFM | 5.5 | -17 | 9.7 | [44] |
| DyNiSb | AFM | 3.5 | -9.8 | 10.9 | [44] |
| HoNiSb | AFM | 2 | -10.8 | 10.7 | [44] |
| CeCuAl | AFM | 5.2 | -37 | 2.53 | [241] |
| PrCuAl | AFM | 7.9 | -0.7 | 3.54 | [240] |
| NdCuAl | AFM | 18 | 20.3 | 3.29 | [240] |
| SmCuAl | | 47 | 29.2 | 0.4 | [240] |
| GdCuAl | FM | 82 | 77.2 | 8.21 | [240] |
| TbCuAl | FM | 49 | 47.3 | 9.97 | [240] |
| DyCuAl | FM | 28 | 25.9 | 10.65 | [240] |
| HoCuAl | FM | 12.1 | 11.5 | 10.6 | [240] |
| ErCuAl | FM | 6.8 | 4.2 | 9.55 | [240] |
| TmCuAl | FM | 2.8 | 9 | 7.45 | [452] |
| YbCuAl | IV | | 33 | 4.33 | [22] and ref. therein |
| CeCuGa | | | -35 | 2.55 | [46] |
| EuCuGa | AFM | 12 | | | [7] and ref. therein |
| CeCuIn | | | -15 | 2.4 | [248] |





| | | | | | |
|---|---|---|---|---|---|
| PrCuIn | | | -3 | 3.65 | [248] |
| NdCuIn | AFM | 4.9 | -7.2 | 2.96 | [249] |
| GdCuIn | AFM | 20 | 20 | 7.9 | [250] |
| TbCuIn | AFM | 14.5 | -5.4 | 9.44 | [249] |
| HoCuIn | AFM | 5 | -10 | 10.51 | [249] |
| ErCuIn | AFM | 3.1 | -12.3 | 9.93 | [249] |
| CeCuSi | FM | 15.5 | -2 | 2.54 | [257] |
| PrCuSi | FM | 14 | 8 | 3.39 | [251] |
| | AFM | 5.1 | -11 | 3.88 | [252] |
| NdCuSi | AFM | 3.1 | -11 | 3.87 | [260] |
| GdCuSi | FM | 49 | 58 | 8.32 | [251] |
| | AFM | 14 | 16.2 | 8.08 | [253] |
| TbCuSi | FM | 47 | 52 | 9.62 | [251] |
| | AFM | 16** | | | [255] |
| DyCuSi | AFM | 11 | 20 | 10.4 | [50] |
| HoCuSi | AFM | 9 | 15 | 10.7 | [50] |
| | AFM | 7 | 8 | 10.62 | [451] |
| ErCuSi | AFM | 6.8 | | | [258] |
| TmCuSi | AFM | 6.5 | | | [259] |
| CeCuGe | FM | 10 | 2 | 2.56 | [504] |
| PrCuGe | AFM | 1.8 | 1 | 3.56 | [52] |





| | | | | | |
|---|---|---|---|---|---|
| NdCuGe | AFM | 3.5 | -7.5 | 3.62 | [52] |
| | AFM | 3.1 | -12 | 3.62 | [262] |
| GdCuGe | AFM | 17 | -0.3 | 7.89 | [52] |
| TbCuGe | AFM | 11.6 | -21 | 9.8 | [52] |
| DyCuGe | AFM | 6 | -12.9 | 10.73 | [52] |
| HoCuGe | AFM | 6.1 | -11.6 | 10.58 | [52] |
| ErCuGe | AFM | 4.3 | -14 | 9.75 | [52] |
| PrCuSn | AFM | 3 | -3.4 | 3.84 | [264] |
| NdCuSn | AFM | 10 | -19.5 | 3.84 | [264] |
| EuCuSn | | | -16 | 7.75 | [269] |
| GdCuSn | AFM | 26 | -36 | 7.8 | [263] |
| TbCuSn | AFM | 17.8 | -18 | 9.8 | [263] |
| | AFM | 15.3 | -38 | 9.82 | [270] |
| DyCuSn | AFM | 5 | -15 | 10.3 | [263] |
| HoCuSn | AFM | 7.8 | -3 | 10.3 | [263] |
| ErCuSn | AFM | 4.8** | 0 | 9.2 | [263] |
| YbCuSn | | | -4.4 | 0.2 | [22] and ref. therein |
| EuCuAs | AFM | 18 | 28 | 7.67 | [7] and ref. therein |
| EuCuSb | AFM | 10 | 3 | 7.84 | [61] |
| YbCuSb | | | -35.8 | 0.5 | [22] and ref. therein |
| EuCuBi | AFM | 18 | -13 | 7.65 | [7] and ref. therein |





| | | | | | |
|---|---|---|---|---|---|
| YbCuBi | | | -1.5 | | [22] and ref. therein |
| CeRuSi | | | -52 | 2.56 | [64] |
| | | | -81 | 2.6 | [505] |
| PrRuSi | AFM | 73 | 7 | 3.52 | [64] |
| NdRuSi | AFM | 74 | 29 | 3.46 | [64] |
| SmRuSi | FM | 65 | NCW | | [64] |
| GdRuSi | FM | 85 | 78 | 8.58 | [64] |
| ErRuSi | FM | 8 | 8.1 | 9.48 | [274] |
| CeRuGe | | | -73 | 2.55 | [64] |
| PrRuGe | AFM | 62 | 0 | 3.82 | [64] |
| NdRuGe | AFM | 65 | 4 | 3.91 | [64] |
| SmRuGe | FM | 45 | NCW | | [64] |
| GdRuGe | FM | 72 | 67 | 8.19 | [272] |
| TbRuGe | FM | 70 | 25 | 9.92 | [272] |
| DyRuGe | FM | 26 | 24 | 10.82 | [272] |
| HoRuGe | FM | 18 | 24 | 10.82 | [272] |
| ErRuGe | FM | 8 | 7 | 9.82 | [272] |
| YRhAl | SC | 0.9 | | 0.1 | [283] |
| LaRhAl | SC | 2.4 | | | [282] |
| CeRhAl | | | -24.8 | 1.17 | [282] |
| | AFM | 3.8 | -11.3 | 1.57 | [284] |





| | | | | | |
|---|---|---|---|---|---|
| PrRhAl | FM | 4.7 | -5.4 | 3.49 | [282] |
| NdRhAl | FM | 10.5 | 6.2 | 3.37 | [282] |
| GdRhAl | FM | 29.8 | 32.9 | 7.82 | [282] |
| CeRhGa | | | 8 | 1.5 | [286] |
| TbRhGa | AFM | 22 | 23 | 10.1 | [71] |
| HoRhGa | AFM | 5.4 | 2 | 10.6 | [71] |
| ErRhGa | AFM | 4.8 | 9 | 9.37 | [71] |
| TmRhGa | AFM | 3.9 | 5.6 | 7.67 | [287] |
| EuRhIn | FM | 22 | 34 | 7.9 | [73] |
| YbRhIn | | | -15 (100) | 4.56 | [74] |
| | | | -18.4 (111) | 4.73 | |
| LaRhSi | SC | 4.3 | | | [76] |
| SmRhSi | AFM, FM | 73, 33 | NCW | | [294] |
| GdRhSi | FM | 100 | 94 | 7.95 | [75] |
| TbRhSi | FM | 55 | 48 | 9.92 | [75] |
| DyRhSi | FM | 25 | 11.5 | 10.31 | [75] |
| HoRhSi | AFM | 8 | 10.5 | 10.71 | [75] |
| ErRhSi | AFM | 7.5 | -3 | 9.54 | [75] |
| CeRhGe | AFM | 10.5 | -56 | 2.3 | [117] |
| NdRhGe | AFM | 14 | -10 | 3.73 | [292] |
| SmRhGe | FM | 56 | NCW | | [294] |





| | | | | | |
|---|---|---|---|---|---|
| GdRhGe | AFM | 31.8 | -4.9 | 8.5 | [295] |
| TbRhGe | AFM | 23 | -3 | 9.43 | [296] |
| | AFM | 24.3 | -18.9 | 10.2 | [299] |
| DyRhGe | AFM | 20 | 1 | 10.3 | [298] |
| | AFM | 20.3 | -3 | 11.5 | [299] |
| HoRhGe | AFM | 5.5 | -1.9 | 10.6 | [300] |
| ErRhGe | AFM | 10.2 | -9.6 | 9.7 | [299] |
| TmRhGe | AFM | 6.2 | -1 | 7.5 | [298] |
| | | 6.8 | -4.4 | 7.6 | [299] |
| YbRhGe | AFM | 7 | -16.4 | 4.48 | [77] |
| CeRhSn | | | -70 | 1.30 | [314] |
| PrRhSn | FM | 3 | 2.9 | 3.64 | [314] |
| NdRhSn | FM | 10.3 | -6.7 | 3.50 | [314] |
| SmRhSn | FM | 14.5 | NCW | | [315] |
| GdRhSn | AFM | 16 | 19 | 7.58 | [314] |
| | AFM | 16.2 | 13.9 | 7.91 | [80] |
| TbRhSn | AFM | 18.3 | -13.2 | 9.8 | [313] |
| DyRhSn | AFM | 7.2 | -19.9 | 10.63 | [313] |
| HoRhSn | FM | 6.2 | 4 | 10.65 | [313] |
| ErRhSn | AFM | | -6.6 | 9.58 | [313] |
| TmRhSn | AFM | 3.1 | -47.1 | 7.01 | [313] |





| | | | | | |
|---|---|---|---|---|---|
| YbRhSn | | | -7 | 4.51 | [63] |
| | AFM | 1.85, 1.4 | -20 | 4.3 | [122] |
| LaRhSb | SC | 2.1 | | | [81] |
| PrRhSb | FM, AFM | 6,18 | -1.3 | 3.5 | [81] |
| CeRhBi | | | -107.5 | 2.87 | [333] |
| GdPdAl | FM | 50 | 49*** | 8.35 | [85] |
| | | | 67**** | 7.94 | |
| TbPdAl | AFM | 43 | 38 | 9.66 | [339] |
| DyPdAl | AFM | 25 | 49.3 | 10.6 | [341] |
| HoPdAl | AFM | 16 | 28 | 10.5 | [342] |
| CePdGa | AFM | 2.2 | | | [344] |
| EuPdGa | FM | 38 | 17 | 7.86 | [7] and ref. therein |
| GdPdGa | AFM | 5.1 | | | [346] |
| TbPdGa | AFM | 34 | | 8.14 | [346] |
| DyPdGa | FM | 20 | | | [346] |
| HoPdGa | AFM | 6.8 | | 7.15 | [346] |
| ErPdGa | AFM | 5 | | 9.2 | [346] |
| CePdIn | AFM | <1.7 | -52.5 | 2.58 | [348] |
| | | 1.8 | -15 | 2.56 | [349] |
| EuPdIn | | | 40 | 7.99 | [89] |
| | AFM | 13 | 13 | 7.6 | [137] |





| | | | | | |
|---|---|---|---|---|---|
| PrPdIn | | | -8.8 | 3.57 | [348] |
| | FM | 11.2 | | | [354] |
| NdPdIn | FM | 30** | 2 | 3.59 | [351] |
| | FM | 34.3 | | | [354] |
| SmPdIn | FM | 54 | NCW | | [355] |
| GdPdIn | FM | 102 | 96.5 | 12 | [87] |
| | FM | 101.5 | 88.3 | 7.99 | [353] |
| TbPdIn | FIM | 70 | 6 | 10.4 | [87] |
| | FM | 74 | 60 | 9.73 | [357] |
| DyPdIn | FIM | 34 | 5.2 | 11 | [87] |
| | FM | 35 | 29 | 10.5 | [357] |
| HoPdIn | FIM | 25 | 7 | 10.8 | [87] |
| | FM | 22** | | | [351] |
| ErPdIn | FIM | 12.3 | 1.6 | 9.7 | [87] |
| | FM | 11** | | | [351] |
| TmPdIn | AFM | 2.7 | -8.1 | 7.47 | [353] |
| CePdSi | FM | 7 | -34 | 2.58 | [91] |
| PrPdSi | FM | 5 | -7 | 3.65 | [91] |
| EuPdSi | | | 9 | 7.45 | [92] |
| YbPdSi | FM | 8 | -41.2 | 4.7 | [359] |
| CePdGe | | | -25 | 2.4 | [360] |





| | AFM | 3.4 | -17 | 2.23 | [361] |
|---|---|---|---|---|---|
| PrPdGe | AFM | 8 | -8 | 3.48 | [360] |
| | | | -12 | 3.7 | [361] |
| NdPdGe | AFM | 4 | -4 | 3.61 | [360] |
| EuPdGe | AFM | 8.5 | 12 | 8 | [363] |
| GdPdGe | | | -32 | 7.92 | [360] |
| | AFM | 17 | -36 | 7.46 | [362] |
| TbPdGe | AFM | 18 | -22 | 9.7 | [360] |
| | AFM | 32 | -24 | 9.69 | [361] |
| DyPdGe | AFM | 6 | -10 | 10.68 | [360] |
| | AFM | 8.4 | -10 | 10.4 | [362] |
| HoPdGe | | | -8 | 10.56 | [360] |
| ErPdGe | AFM | 4 | -6 | 9.3 | [360] |
| TmPdGe | | | -4 | 7.38 | [360] |
| YbPdGe | FM | 11.4 | | | [22] and ref. therein |
| CePdSn | AFM | 7.5 | -68 | 2.67 | [95] |
| | AFM | 6 | -63 | 2.7 | [366] |
| PrPdSn | | | -2 | 3.6 | [95] |
| | AFM | 4.3 | -5.5 | 3.51 | [366] |
| NdPdSn | | | -8 | 4.93 | [95] |
| | AFM | 2.4 | -11 | 3.68 | [366] |





| SmPdSn | AFM | 11 | NCW | | [95] |
|--------|-----|-----|------|------|------|
| EuPdSn | AFM | 15.5, 6 | 13 | 7.78 | [269] |
| EuPdSn | AFM | 13 | 5 | 8.27 | [95] |
| GdPdSn | AFM | 14.5 | -27 | 8.16 | [95] |
| TbPdSn | AFM | 23.5 | -16 | 10.17 | [95] |
| | AFM | 19 | -11 | 10.1 | [366] |
| DyPdSn | AFM | 11.4 | -2 | 11.1 | [95] |
| | AFM | 10 | -7 | 10.5 | [366] |
| HoPdSn | | | -7 | 11.07 | [95] |
| | AFM | 3.7 | -7.5 | 10.7 | [366] |
| ErPdSn | AFM | 5.6 | -0.3 | 9.51 | [95] |
| | AFM | 5.2 | 3 | 9.62 | [366] |
| TmPdSn (hex) | | | -0.1 | 7.98 | [95] |
| YbPdSn | | | -5 | 1.45 | [95] |
| EuPdAs | | | 0 | 7.04 | [7] and ref. therein |
| CePdSb | FM | 16.5 | 11 | 2.6 | [366] |
| PrPdSb | FM | 9 | 3 | 3.6 | [366] |
| NdPdSb | | | 8 | 3.6 | [366] |
| | AFM | 10 | 9 | 2.95 | [376] |
| EuPdSb | AFM | 13 | -35 | 8.19 | [7] and ref. therein |





| | | | | | |
|---|---|---|---|---|---|
| SmPdSb | FM | 3 | | | [366] |
| GdPdSb | AFM | 16.5 | -21 | 8.13 | [366] |
| | AFM | 13.1 | -16 | 7.81 | [377] |
| TbPdSb | AFM | 2.2 | -10 | 9.8 | [366] |
| DyPdSb | AFM | 4.6 | -6 | 10.6 | [366] |
| HoPdSb | AFM | 2.2 | -4 | 10.6 | [366] |
| YbPdSb | | | -9 | 4.39 | [22] and ref. therein |
| CePdBi | SC,SG | 1.4, 2.5 | 0.5 | 2.23 | [379] |
| GdPdBi | AFM | 13.5 | -36.5 | 8 | [100] |
| | AFM | 13 | -48 | 7.9 | [380] |
| DyPdBi | AFM | 3.5 | -11.9 | 10.7 | [100] |
| HoPdBi | AFM | 2.2 | -6.1 | 10.59 | [100] |
| ErPdBi | SC, AFM | 1.22, 1.06 | | | [381] |
| YbPdBi | | | -3.8 | 4.11 | [63] |
| CeAgAl | SG | | -9.6 | 2.6 | [102] |
| | FM | 2.9 | -18 | 2.65 | [383] |
| PrAgAl | SG | 10 | -0.9 | 3.5 | [102] |
| NdAgAl | SG, FM | 16 | 6.7 | 3.4 | [102] |
| GdAgAl | SG | 40 | 53 | 8.3 | [382] |
| TbAgAl | SG, FM | 64 | 45 | 9.9 | [102] |
| | FIM | 59 | 43 | 9.7 | [384] |





| | | | | | |
|---|---|---|---|---|---|
| ErAgAl | SG, FM | 10 | 1.9 | 9.4 | [102] |
| DyAgAl | SG, FM | 43.8, 34.7[a*] | 37 | 10.69 | [103] |
| HoAgAl | SG, FM | 23.7, 18.2 [a*] | 16.4 | 10.79 | [103] |
| ErAgAl | SG, FM | 15.3, 12.5 [a*] | 8.1 | 9.9 | [103] |
| CeAgGa | FM | 5.5 | | 2.49 | [385] |
| | SG, FM | 5.1, 3.6 | -43 | 2.5 | [386] |
| PrAgGa | | | 31 | 3.18 | [104] |
| NdAgGa | | | 4 | 3.65 | [104] |
| GdAgGa | FM | 27 | 52 | 7.95 | [104] |
| TbAgGa | AFM | 18 | 20 | 10.03 | [104] |
| | AFM | 34 | 39 | 9.1 | [398] |
| | SG | 50 | 27 | 9.6 | [387] |
| DyAgGa | | | 17 | 10.6 | [104] |
| | FM | | 2.6 | 10.9 | [387] |
| HoAgGa | FM | 4.7 | 14 | 10.43 | [104] |
| | FM | 16 | 3 | 10.6 | [387] |
| | FM | 7.2 | | 10.9 | [388] |
| ErAgGa | FM | 3 | 12 | 9.43 | [104] |
| TmAgGa | | | 9 | 7.38 | [104] |
| NdAgSi | FM | 44 | | 2.2 | [390] |
| GdAgSi | AFM | 13.4 | -13 | 8.26 | [389] |





| | | | | | |
|---|---|---|---|---|---|
| TbAgSi | AFM | 21.5, 16 | -23.4 | 9.8 | [389] |
| DyAgSi | AFM | 11 | -10.3 | 10.9 | [389] |
| HoAgSi | AFM | 10 | -8 | 10.7 | [389] |
| ErAgSi | AFM | 2.5 | -1.5 | 9.8 | [389] |
| TmAgSi | AFM | 3.3 | -0.8 | 7.45 | [391] |
| CeAgGe | AFM | 4.8 | -9.3 | 2.61 | [106] |
| EuAgGe | SG | 18 | -2 | 7.7 | [393] |
| GdAgGe | AFM | 15.6 | -31.4 | 7.88 | [107] |
| TbAgGe | AFM | 25, 20 | -14.6 | 9.97 | [107] |
| DyAgGe | AFM | 14.5, 11 | 0 | 10.86 | [107] |
| HoAgGe | AFM | 10.3 | 0 | 10.73 | [107] |
| ErAgGe | AFM | 3.6** | 6.4 | 9.54 | [107] |
| TmAgGe | AFM | 4.1 | -14.4 | 7.9 | [108] |
| | AFM | 4.2 | 12 | 7.3 | [396] |
| CeAgSn | AFM | 3.6 | -17 | 2.46 | [112] |
| PrAgSn | AFM | 3.8 | -7.5 | 3.56 | [112] |
| NdAgSn | AFM | 11.5 | -20 | 3.95 | [112] |
| EuAgSn | AFM | 6 | -31 | 7.96 | [269] |
| GdAgSn | AFM | 34 | -26 | 7.65 | [112] |
| TbAgSn | AFM | 33 | -42 | 9.33 | [112] |
| DyAgSn | AFM | 10.6 | -13 | 10.5 | [112] |





| | | | | | |
|---|---|---|---|---|---|
| HoAgSn | AFM | 10.8 | -11 | 10.7 | [112] |
| ErAgSn | AFM | 6.2 | -2 | 9 | [112] |
| TmAgSn | | | 17 | 7.2 | [400] |
| YbAgSn | | | -4.5 | 0.2 | [22] and ref. therein |
| EuAgAs | AFM | 11 | 19 | 7.45 | [7] and ref. therein |
| EuAgSb | AFM | 4.3 | 1.9 | 7.6 | [61] |
| YbAgSb | | | -4.4 | 0.1 | [22] and ref. therein |
| EuAgBi | AFM | 10 | -4 | 7.39 | [7] and ref. therein |
| YbAgBi | | | -1.1 | | [22] and ref. therein |
| CeOsSi | | | -29 | 2.03 | [64] |
| CeIrAl | NCW | $T_K$=1300 | | | [401] |
| PrIrAl | FM | 8 | 2 | 3.51 | [115] |
| | FM | 9 | 3 | 3.41 | [401] |
| NdIrAl | FM | 14 | 0 | 3.69 | [115] |
| | FM | 13 | 3 | 3.46 | [401] |
| GdIrAl | FM | 67 | 40 | 7.97 | [115] |
| TbIrAl | AFM | 36 | 14 | 9.9 | [115] |
| DyIrAl | FM | 8.5 | 8 | 10.5 | [115] |
| HoIrAl | AFM | 7 | 2 | 10.5 | [115] |
| ErIrAl | AFM | 7 | -1 | 9.5 | [115] |
| TmIrAl | AFM | 3.8 | -3 | 7.47 | [115] |





| | | | | | |
|---|---|---|---|---|---|
| LaIrSi | SC | 2.3 | | | [76] |
| NdIrSi | FM | 10 | 12 | 3.62 | [76] |
| TbIrSi | AFM | 32 | | 8.96 | [120] |
| | AFM | 43.5, 32 | -10.3 | 10 | [402] |
| DyIrSi | AFM | 7 | | | [120] |
| | AFM | 7 | 0.2 | 10.77 | [402] |
| HoIrSi | AFM | 4.8 | | 9.9 | [120] |
| | AFM | 4.8 | 3.8 | 10.9 | [402] |
| ErIrSi | AFM | 3.8 | | 6.54 | [120] |
| CeIrGe | | | -10 | 0.27 | [117] |
| GdIrGe | AFM | 24 | -17.8 | 8.2 | [116] |
| TbIrGe | AFM | 23.5 | -11.2 | 9.9 | [116] |
| DyIrGe | AFM | 15 | -13.8 | 11.5 | [116] |
| HoIrGe | | | -3.9 | 10.2 | [116] |
| ErIrGe | AFM | 6.5 | -2.6 | 9.1 | [116] |
| YbIrGe | AFM | 2.4 | -6.6 (H∥$a$) | 4.69 | [119] |
| YbIrSn | AFM | 3.2 | -17.7 | 4.3 | [403] |
| YbIrSb | AFM | 3.9 | -43.6 | 4.13 | [61] |
| CePtAl | FM | 5.8 | -37 | 2.51 | [516] |
| PrPtAl | FM | 5.8 | -6.5 | 3.59 | [405] |
| NdPtAl | FM | 19.2 | | | [404] |





| | | | | | |
|---|---|---|---|---|---|
| TbPtAl | FM | 43 | | | [404] and ref. therein |
| YbPtAl | AFM | 5.9 | | 4.5 | [190] |
| CePtGa | | | -68 | 2.36 | [409] |
| PrPtGa | | | -18 | 3.5 | [409] |
| NdPtGa | | | -15 | 3.55 | [409] |
| EuPtGa | FM | 36 | | | [7] and ref. therein |
| GdPtGa | AFM | 25 | 34 | 7.9 | [409] |
| | AFM | 23 | 19 | 8.14 | [410] |
| TbPtGa | AFM | 20 | 20 | 9.7 | [409] |
| | AFM | 27 | 3 | 9.67 | [410] |
| DyPtGa | AFM | 15 | 10 | 10.68 | [409] |
| | AFM | 12 | 1 | 10.66 | [410] |
| HoPtGa | | | 6 | 10.55 | [409] |
| | AFM | 5.6 | -2 | 10.37 | [410] |
| ErPtGa | | | 5 | 9.55 | [409] |
| | AFM | 3.2 | -6.5 | 9.95 | [410] |
| TmPtGa | AFM | 8 | -5 | 7.4 | [409] |
| YbPtGa | AFM | 3.8 | -23 | 3.95 | [412] |
| CePtIn | | | -73 | 2.58 | [349] |
| PrPtIn | | | -29 | 3.53 | [123] |
| SmPtIn | FM | 25 | NCW | | [123] |





| | | | | | |
|---|---|---|---|---|---|
| EuPtIn | AFM | 16 | 20 | 8 | [126] |
| GdPtIn | FM | 67.5 | -61.6 | 7.6 | [414] |
| TbPtIn | AFM | 46 | -34.7 | 9.7 | [414] |
| DyPtIn | FM | 26.5 | -9.1 | 10.7 | [414] |
| | FM | 37.1 | 32.8 | 10.7 | [416] |
| HoPtIn | FM | 23.5 | -7.7 | 10.5 | [414] |
| | FM | 23.1 | 25 | 10.6 | [416] |
| ErPtIn | FM | 8.5 | 13.2 | 10.1 | [414] |
| | FM | 15 | 13 | 9.3 | [418] |
| TmPtIn | AFM | 3 | 2.5 | 7.7 | [414] |
| | AFM | 3.4 | 6.5 | 7.65 | [419] |
| YbPtIn | AFM | 2.1 | 32.5 | 4.3 | [414] |
| LaPtSi | SC | 3.8 | | | [421] |
| CePtSi | | | -47 | 2.56 | [420] |
| NdPtSi | AFM | 3.8 | -2.89 | 3.68 | [421] |
| SmPtSi | FM | 15 | NCW | | [421] |
| EuPtSi | | | 5 | 7.5 | [92] |
| GdPtSi | FM | 16 | -15.5 | 8.07 | [423] |
| TbPtSi | AFM | 12.5 | -16.9 | 9.78 | [423] |
| DyPtSi | AFM | 8.2 | -9.6 | 10.4 | [423] |
| HoPtSi | AFM | 3.1 | -10.1 | 10.46 | [422] |





| | | | | | |
|---|---|---|---|---|---|
| ErPtSi | AFM | 4.1 | -6.2 | 9.59 | [422] |
| CePtGe | AFM | 3.4 | -82 | 2.54 | [117, 478] |
| EuPtGe | | | 20 | 7.8 | [7] and ref. therein |
| GdPtGe | AFM | 12.5 | -25 | 7.9 | [423] |
| TbPtGe | AFM | 15 | -10.7 | 9.6 | [423] |
| DyPtGe | AFM | 8 | -6 | 10.3 | [423] |
| HoPtGe | | | -6.7 | 10.32 | [422] |
| ErPtGe | AFM | 6 | -11.5 | 9.5 | [422] |
| YbPtGe | FM | 4.7 | | | [22] and ref. therein |
| EuPtSn | AFM | 28.5, 18 | 19 | 8.1 | [269] |
| DyPtSn | AFM | 7.3 | 20 | 10.4 | [429] |
| HoPtSn | AFM | 8.6 | 4.2 | 9.8 | [429] |
| ErPtSn | AFM | 3.2 | 2.3 | 9.06 | [429] |
| YbPtSn | AFM | 3.5 | 9.2 | 4.27 | [63] |
| YbPtAs | VF | | -45 | 3.9 | [22] and ref. therein |
| CePtSb | | 4.5[+] | -34.4 | 2.62 | [133] |
| PrPtSb | | 8[+] | -12.4 | 3.83 | [133] |
| NdPtSb | | 15.5[+] | -3.4 | 3.82 | [133] |
| SmPtSb | | 6.7[+] | | | [133] |
| EuPtSb | AFM | 15.2 | -25.6 | 7.62 | [61] |
| GdPtSb | | | -23 | 8.06 | [133] |





| | | | | | |
|---|---|---|---|---|---|
| YbPtSb | | | -5.3 | 4.4 | [133] |
| YbPtBi | AFM | 0.4 | 2 | 4.2 | [22] and ref. therein |
| CeAuAl | AFM | 3.8 | -35 | 2.51 | [134] |
| NdAuAl | FM | 10 | 2.5 | 3.41 | [134] |
| EuAuAl | AFM | 50 | 52 | 7.6 | [7] and ref. therein |
| EuAuGa | FM | 16 | 16 | 7.72 | [7] and ref. therein |
| GdAuGa | AFM | 6 | -8.5 | 8.06 | [439] |
| TbAuGa | | | -10 | 9.7 | [439] |
| DyAuGa | | | -4.5 | 10.64 | [439] |
| HoAuGa | | | 3.5 | 10.58 | [439] |
| ErAuGa | | | 1.5 | 9.6 | [439] |
| TmAuGa | | | -2 | 7.59 | [439] |
| CeAuIn | AFM | 6 | | | [136] |
| EuAuIn | AFM | 21 | 22 | 7.5 | [137] |
| GdAuIn | AFM | 13 | | | [481] |
| | AFM | 12.5 | -17.1 | 8.2 | [438] |
| TbAuIn | AFM | 58.5, 35 | -17 | 9.8 | [135] |
| DyAuIn | AFM | 11 | -6 | 10.3 | [135] |
| HoAuIn | AFM | 4.8 | -9 | 10.1 | [135] |
| ErAuIn | AFM | 3 | | | [136] |
| CeAuGe | FM | 10 | -5 | 2.55 | [441] |





| | | | | | |
|---|---|---|---|---|---|
| PrAuGe | | | -12.4 | 4 | [139] |
| NdAuGe | AFM | 8.8 | -3 | 3.95 | [139] |
| | AFM | 3.7 | -6.1 | 3.54 | [444] |
| EuAuGe | FM | 32.9 | 33 | 7.4 | [393] |
| GdAuGe | AFM | 16.9 | -2 | 7.40 | [441] |
| TbAuGe | AFM | 7.6 | -16.2 | 9.77 | [139] |
| DyAuGe | AFM | 6.1 | -6.1 | 10.6 | [139] |
| HoAuGe | AFM | 7.6 | -5.1 | 10.1 | [139] |
| ErAuGe | AFM | 5.7 | -5.4 | 9.2 | [139] |
| CeAuSn | AFM | 7, 4.5 | 1.65 | 2.52 | [142] |
| PrAuSn | AFM | 3.3, 2.7 | 2.83 | 3.37 | [142] |
| NdAuSn | AFM | 10.5 | -15.9 | 3.6 | [446,447] |
| SmAuSn | AFM | 36.5 | | | [566, 567] |
| EuAuSn | AFM | 8.5 | -8 | 7.6 | [269] |
| GdAuSn | AFM | 24 | -46.7 | 8.42 | [141] |
| | SG, AFM | 15.3[g], 22.9 | -29 | 7.61 | [579, 580] |
| TbAuSn | AFM | 15 | -21.4 | 9.31 | [141] |
| DyAuSn | AFM | 8 | -10.2 | 10.33 | [141] |
| HoAuSn | AFM | 9 | -14.3 | 10.35 | [141] |
| ErAuSn[hex] | AFM | 12.1 | -0.5 | 9.53 | [140] |
| ErAuSn[cubic] | | | -11.1 | 9.83 | [141] |





| ErAuSn[cubic] | AFM | 1.3** |      |      | [445] |
|---------------|-----|-------|------|------|-------|
| TmAuSn | | | -6.6 | 8.11 | [448] |
| YbAuSn | | | -3.5 | 0.1 | [22] and ref. therein |
| EuAuAs | FM | 8 | 11 | 7.57 | [7] and ref. therein |
| EuAuSb | AFM | 4.1 | 2.6 | 8.02 | [61] |
| YbAuSb | | | -4.4 | 0.2 | [22] and ref. therein |
| EuAuBi | AFM | 8 | | | [7] and ref. therein |
| YbAuBi | | | 0.8 | | [22] and ref. therein |

g=spin glass temperature, r= reentrant spin glass temperature, # = CeFeSi-type, ## = CeScSi-type, ### = TiNiSi-type, *=along *a*-axis, ** = estimated from neutron diffraction data, *** = HTM II, **** = HTM I, + = from resistivity data, a* = $T_{max}$ ZFC

Table III. Values of $\Delta S_M$, $\Delta T_{ad}$ and RC at 5 T of some *RTX* compounds.

| Compound | Ordering temperature | -$\Delta S_M$(J/kg K) | $\Delta T_{ad}$ (K) | RC (J/Kg) | Reference |
|----------|----------------------|----------------------|---------------------|-----------|-----------|
| GdScSi | 352 | 2.5[c] | | 93.3[c] | [5] |
| GdScGe | 348 | 3.3[c] | | 88.9[c] | [5] |
| TbScGe | 250 | 3.6 | 3 | | [150] |
| TbTiGe[##] | 300 | 4.3 | 4.2 | | [18] |
| DyTiGe | 179 | 2 | | | [156] |
| TmTiGe | | ~9.5 | | | [458] |





| | | | | | |
|---|---|---|---|---|---|
| HoTiGe | 90 | 4 | | | [157] |
| GdMnAl | 274 | 0.69 | | 59 | [25] |
| GdMnSi | 314 | | 0.5$^\$$ | | [161] |
| NdFeAl[a] | 110 | -5.65 | | | [457] |
| GdFeAl | 265 | 3.7 | | 420 | [170] |
| DyFeAl | 128 | 6.4 | | 595 | [171] |
| GdFeSi[b] | 118 | 22.3 | 4.5 | 1940 | [175] |
| TbFeSi | 110 | 17.5 | 8.2 | | [176] |
| DyFeSi | 70 | 17.4 | 7.1 | | [176] |
| HoFeSi | 29, 20 | 16.2, -6 | | | [177] |
| ErFeSi | 22 | 23.1 | 5.7 | 365 | [178] |
| GdCoAl | 100 | 10.4 | | | [179] |
| TbCoAl | 70 | 10.5 | | | [179] |
| DyCoAl | 37 | 16.3 | | | [179] |
| HoCoAl | 10 | 21.5 | | | [179] |
| HoCoSi | 14 | 20.5 | 3.1[c] | 410 | [184] |
| GdNiAl | 57 | 10.6 | - | 460 | [25] |
| DyNiAl | 30 | 19 | 7 | | [195] |
| HoNiAl | 14 | 23.6 | 8.7 | 500 | [196] |
| GdNiGa | 30 | ~20 | 7.6 | 530 | [449] |
| GdNiIn | 98 | 7.1 | | 326 | [208] |
| TbNiIn | 71 | 5.3 | | 191 | [208] |
| DyNiIn | 30 | 10.4 | | 270 | [208] |
| HoNiIn | 20 | 21.7 | | 341 | [208] |
| ErNiIn | 9 | 15.1 | | 229 | [208] |
| HoNiSi | 4.2 | 12.7 | | | [260] |
| ErNiSi | 3.2 | 19.1 | 2.5 | | [213] |
| GdCuAl | 81 | 10.1 | | 460 | [456] |
| TbCuAl | 52 | 14.4 | | | [246] |
| DyCuAl | 27[d] 24[e] | 20.4[d] 14.9[e] | | 423[d] 379[e] | [455] |
| HoCuAl | 11.2 12 | 30.6 23.9 | | 486 393 | [454] [453] |
| ErCuAl | 7 | 22.9 | | 321 | [453] |
| TmCuAl | 2.8 | 24.3 | 9.4 | 371.7 | [452] |
| NdCuSi | 3.2 | 11 | | | [260] |
| GdCuSi | 14.2 | 7.2 | | | [260] |
| DyCuSi | 10 | 24 | - | 381 | [450] |
| HoCuSi | 7 | 33.1 | - | 385 | [451] |
| TbCuGe | 11.8 | 3.1 | | | [459] |
| DyCuGe | 5.2 | 15.1 | | | [459] |
| HoCuGe | 4.7 | 12.4 | | | [459] |
| ErCuGe | 4.1 | 12.7 | | | [459] |
| ErRuSi | 8 | 21.2 | | 416 | [274] |
| GdRuGe | 70 | 6* | | | [273] |
| GdRhGe | 31.8 | 1.2 | | | [295] |
| TbRhGe | 24.3 | | | | [299] |
| DyRhGe | 20.3 | | | | [299] |
| HoRhGe | 5.5 | 11.1 | | | [300] |
| ErRhGe | 10.2 | | | | [299] |
| TmRhGe | 6.8 | | | | [299] |
| PrRhSn | 2.85 | ~9.6 | | | [326] |
| GdRhSn | 16.2 | 6.5 | 4.5 | | [80] |
| TbRhSn | 18.3 | 3 | 2 | | [313] |
| DyRhSn | 7.2 | 7.2 | 5.2 | | [313] |
| HoRhSn | 6.2 | 13.2 | 6.5 | | [313] |
| ErRhSn | | 9.7 | | | [313] |
| TmRhSn | 3.1 | 8 | 5.6 | | [313] |
| GdPdAl | 46 | 10.41 | | 562 | [25] |
| TbPdAl | 43 | 11.4 | | 350 | [339] |
| HoPdAl (hex) | 12 | 20.6 | | 386 | [343] |
| HoPdAl (ortho) | 10 | 13.7 | | | [343] |
| PrPdIn | 11.2 | 6.3** | | | [354] |
| NdPdIn | 34.3 | ~6** | | | [354] |
| GdPdIn | 90 | 4.6 | | 464 | [25] |
| HoPdIn | 23 | 14.6 | 5.5 | 496 | [352] |
| TbAgAl | 59 | 68 mJ/cm$^3$K | | 3.18 J/cm$^3$ | [384] |
| HoAgGa | 7.2 | 16 | | 262 | [388] |

a=amorphous ribbon, b= ΔH=90 kOe, c= ΔH=20 kOe, d=90nm, e=38 nm, *=30 kOe, **=70 kOe, $=13.5 kOe, ## = CeScSi-type

Table IV Values of MR of some *RTX* compounds.

| Compound | T$_N$/T$_C$ (K) | Temperature (K) | Field (T) | MR (%) | Ref. |
|---|---|---|---|---|---|
| DyTiGe | 180 | 5 | 9 | -20 | [460] |
| GdMnSi | 316 | 4.2 | 14 | -17 | [461] |
| CeCoSi | 5.5 | | | ~-30[1],>70[1] | [462] |
| TbNiAl | | 5 | >0.5 | -33 | [193] |
| HoNiAl | 14, 5 | 15 | 5 | -16 | [196] |
| HoNiSi | 4.2 | 4 | 5 | -24 | [260] |
| ErNiSi | 3.2 | 3 | 5 | -34 | [213] |
| CeNiSb | 3.5 | 2 | 5 | -13 | [44] |
| NdNiSb | 23 | 22 | 4 | -18 | [44] |
| PrNiSb | - | 2 | 5 | 7 | [44] |
| LaNiSb | - | | 6.4 | 20 | [44] |
| TbNiSb | 5.5 | 2 | 5 | -20 | [44] |
| DyNiSb | 3.5 | 2 | 5 | -32 | [44] |
| HoNiSb | 2 | 2 | 4 | 27 | [44] |
| NdCuSi | 3.1 | 1.5 | 5 | -36 | [260] |
| GdCuSi | 14.2 | 1.5, 15 | 5 | 25,- 27 | [260] |
| CeRhGa | | 0.5 | 9 | 18 | [286] |
| GdRhGe | 31.8 | 2 | 5 | 48 | [295] |
| TbRhGe | 24.3 | 30,2 | 5 | -3, 19.5 | [299] |
| DyRhGe | 20.3 | 20 | 5 | -6.5 | [299] |
| HoRhGe | 5.5 | 5 | 5 | -25 | [300] |
| ErRhGe | 10.2 | 5 | 5 | -8.5 | [299] |
| TbRhSn | 18.3 | 18, 7 | 5, 3 | -8, 11.5 | [313] |
| DyRhSn | 7.2 | 2 | 5 | 14 | [313] |
| HoRhSn | 6.2 | 7, 5 | 4, 1.2 | -9, 6 | [313] |
| YbRhSn | 2 | 2 | 8 | -22 | [132] |
| CeRhSb | | 4.2 | 4 | 5 | [469] |
| CePtGa | 3.4 | 4.2 | 5 | -5 | [477] |
| NdAuAl | 10 | 10 | 5.5 | -6 | [134] |

1=estimated from Ref [462]

Table V Crystal structure and ordering temperature of *RTX* compound before and after hydrogenation.

| Parent compound | | | | Hydrogenated compound | | | |
|---|---|---|---|---|---|---|---|
| Compound | Crystal structure | Magnetic nature | $T_C$ or $T_N$ | Crystal structure | Magnetic nature | $T_C$ or $T_N$ | Ref. |
| GdTiGe (CeFeSi-type) | tetragonal | FM | 376 | tetragonal (CeScSi-type) | PM | | [16] |
| GdTiGe (CeScSi-type) | tetragonal | AFM | 412 | tetragonal (CeScSi-type) | PM | | |
| NdMnSi | tetragonal | AFM | 280 | tetragonal | AFM | 565 | [482] |
| CeMnGe | tetragonal | AFM | 41 | tetragonal | No Ce ordering | | [483] |
| CeCoGe | tetragonal | AFM | 5 | tetragonal | SF | 13.3 | [484] |
| NdCoSi | tetragonal | AFM | 7.5 | tetragonal | FM | 20.5 | [485] |
| NdCoGe | tetragonal | AFM | 8.3 | tetragonal | FM | 15.9 | [485] |
| HoNiAl | hexagonal | FM | 13 | orthorhombic | AFM | 6 | [489] |
| CeNiGa LTP HTP | hexagonal orthorhombic | IV IV | | hexagonal hexagonal | trivalent trivalent | | [38] |
| CeNiGe | hexagonal | | | orthorhombic | | | [504] |
| CeNiSn | orthorhombic | KI | | hexagonal | FM | 7 | [497,500] |
| CeCuGa | orthorhombic | | | hexagonal | | | [46] |
| CeCuSi | hexagonal | FM | 15.5 | hexagonal | AFM | 13.7, 10.6, 8.4 | [503] |
| CeCuGe | hexagonal | FM | 10 | hexagonal | PM | | [504] |
| CeRuSi | tetragonal | NM | | tetragonal | AFM | 7.5, 3.1 | [505] |
| CeRuGe | tetragonal | NM | | tetragonal | AFM | 4 | [508] |
| CeRhGe | orthorhombic | AFM | 10 | hexagonal | IV | | [510] |
| CeRhSb | orthorhombic | IV | | orthorhombic | AFM | 3.6 | [512] |
| CePdIn | hexagonal | AFM | 1.8 | hexagonal | AFM | 3 | [349, 513] |
| CePdSn | orthorhombic | AFM | 7.2 | orthorhombic | AFM | 5 | [513] |
| CeIrGa | orthorhombic | IV | | hexagonal | TV | | [514] |
| CeIrGe | orthorhombic | IV | | hexagonal | IV | | [510] |
| CeIrSb | orthorhombic | IV | | orthorhombic | AFM | 7 | [515] |
| CePtAl | orthorhombic | FM | 5.6 | unknown | FM | 11.6 | [516] |
| TbPtIn | hexagonal | AFM | 48 | hexagonal | AFM | 95 | [517] |
| ErPtIn | hexagonal | FM | 13 | hexagonal | FM | 15.5 | [517] |
| TmPtIn | hexagonal | AFM | 3.4 | hexagonal | NM | | [517] |